\documentclass[
  journal=pasa,
  manuscript=research-paper, 
  year=2021,
  volume=37,
]{cup-journal}
\usepackage{graphicx}
\usepackage{xcolor}
\usepackage{microtype,siunitx}
\usepackage{amssymb}
\usepackage{amsmath}
\usepackage{mathtools}
\usepackage{hyperref}
\definecolor{dodgerblue}{RGB}{30, 144, 255}
\definecolor{crimson}{RGB}{220, 20, 60}
\definecolor{darkerblue}{RGB}{0, 0, 139} 
\definecolor{darkred}{RGB}{150,20,20}
\definecolor{cpink}{RGB}{243, 141, 252}
\hypersetup{colorlinks,citecolor=dodgerblue,linkcolor=blue,urlcolor=magenta}
\usepackage{enumitem, xcolor}
\let\svitem\item

\usepackage{booktabs} 
\usepackage{threeparttable,longtable}  
\usepackage{makecell}

\usepackage[skip=0.5ex]{subcaption}
\usepackage{cprotect}
\renewcommand\thesubfigure{(\roman{subfigure})}

\usepackage{multirow}

\title[GLEAM-300]{GLEAM-300: The GaLactic and Extragalactic All-sky Murchison Widefield Array (GLEAM) survey at 300 MHz}

\author{S.~W.~Duchesne}
\affiliation{CSIRO Space and Astronomy, PO Box 1130, Bentley WA 6102, Australia}
\email[S.~W.~Duchesne]{Stefan.Duchesne@csiro.au}

\author{J.~H.~Cook}
\affiliation{International Centre for Radio Astronomy Research, Curtin University, Bentley, WA 6102, Australia}

\author{N.~Hurley-Walker}
\affiliation{International Centre for Radio Astronomy Research, Curtin University, Bentley, WA 6102, Australia}

\author{A.~J.~M.~Thomson}
\affiliation{CSIRO Space and Astronomy, PO Box 1130, Bentley WA 6102, Australia}

\author{S.~Paterson}
\affiliation{International Centre for Radio Astronomy Research, Curtin University, Bentley, WA 6102, Australia}

\author{C. J. Riseley}
\affiliation{Astronomisches Institut der Ruhr-Universit\"{a}t Bochum (AIRUB), Universit\"{a}tsstra{\ss}e 150, 44801 Bochum, Germany}

\author{S.~J.~McSweeney}
\affiliation{International Centre for Radio Astronomy Research, Curtin University, Bentley, WA 6102, Australia}

\author{S.~Mantovanini}
\affiliation{International Centre for Radio Astronomy Research, Curtin University, Bentley, WA 6102, Australia}

\author{G.~Heald}
\affiliation{SKA Observatory, SKA-Low Science Operations Centre, 26 Dick Perry Avenue, Kensington, WA 6151, Australia}
\alsoaffiliation{CSIRO Space and Astronomy, PO Box 1130, Bentley WA 6102, Australia}

\author{T.~M.~O.~Franzen}
\affiliation{SKA Observatory, Jodrell Bank, Lower Withington, Macclesfield SK11 9FT, UK}

\author{K.~Ross}
\affiliation{International Centre for Radio Astronomy Research, Curtin University, Bentley, WA 6102, Australia}

\author{N.~Seymour}
\affiliation{International Centre for Radio Astronomy Research, Curtin University, Bentley, WA 6102, Australia}

\author{R.~B.~Wayth}
\affiliation{International Centre for Radio Astronomy Research, Curtin University, Bentley, WA 6102, Australia}
\alsoaffiliation{SKA Observatory, Science Operations Centre, CSIRO ARRC, 26 Dick Perry Avenue, Kensington, WA 6151, Australia}

\author{T.~J.~Galvin}
\affiliation{CSIRO Space and Astronomy, PO Box 1130, Bentley WA 6102, Australia}

\doi{10.1017/pasa.20XX.XX}

\received {dd Mmm YYYY}
\revised  {dd Mmm YYYY}
\accepted {dd Mmm YYYY}
\published{22 September 2020}


\definecolor{dodgerblue}{RGB}{30, 144, 255}

\newcommand{\corrs}[1]{{\color{black} #1}}

\def\nsources{{338\,080}}
\def\psfMedianA{{128\farcs8}}
\def\psfMedianB{{112\farcs5}}
\def\psfMaxA{{275\farcs7}}
\def\psfMaxB{{222\farcs8}}
\def\psfMinA{{108\farcs8}}
\def\psfMinB{{89\farcs4}}
\def\rmsMedian{{\ensuremath{9.1_{-2.8}^{+5.5}}}}
\def\relOverall{{99.58}}
\def\relSCP{{99.94}}
\def\relHighDec{{99.71}}
\def\intPeakAll{{\ensuremath{0.99^{+0.19}_{-0.10}}}}
\def\intPeakBright{{\ensuremath{0.98^{+0.06}_{-0.02}}}}

\begin{document}

\def\gleamalpha{-0.8}
\begin{abstract}

In this paper we present a wide-field radio survey at 300\,MHz covering the sky from $-90\degr \leq \delta_\text{J2000} \lesssim {+40}\degr$ using the Murchison Widefield Array (MWA). This 300-MHz survey follows the Galactic and Extragalactic All-sky MWA (GLEAM) survey, and provides an additional comparatively high-frequency data point to existing multi-frequency (72--231\,MHz) data. With this data release we provide mosaic images and a catalogue of compact source components. We use two-minute snapshot observations covering 2015--2016, combining overlapping two-minute snapshot images to provide full-sensitivity mosaic images with a median root-mean-square noise of \rmsMedian\,mJy\,beam$^{-1}$ and median angular resolution of $\psfMedianA \times \psfMedianB$, with some position-dependent variation. We find a total of \nsources\ unique Gaussian components across the mosaic images. The survey is the first at 300\,MHz from the MWA covering the whole Southern Hemisphere. It provides a unique spectral data point that complements the existing GLEAM survey and the ongoing GLEAM-eXtended survey, and points toward results from the upcoming SKA-Low surveys. 
\end{abstract}

\section{Introduction}
\label{sec:int}

\defcitealias{Wayth2015}{GLEAM}
\defcitealias{HurleyWalker2017}{GLEAM EGC}
\defcitealias{HurleyWalker2022}{GLEAM-X}
\defcitealias{gleamx-dr2}{GLEAM-X DR2}
\defcitealias{Franzen2021a}{GLEAM SGP}

The Murchison Widefield Array \footnote{\url{https://www.mwatelescope.org/}.} \citep[MWA;][]{Tingay2013} is a low radio frequency aperture array located on Inyarrimanha Ilgari Bundara, the CSIRO \footnote{Commonwealth Scientific and Industrial Research Organisation.} Murchison Radio-astronomy Observatory. The MWA operates from 72--300\,MHz, and among its myriad science activities has been completing widefield, multi-frequency surveys in the form of the Galactic and Extragalactic All-sky MWA survey \citep[GLEAM;][]{Wayth2015}. 

GLEAM has resulted in five major data releases: the first all-sky extra-galactic images and extra-galactic catalogue \citep[GLEAM EGC;][]{HurleyWalker2017}, the Galactic plane between Galactic longitudes of $345^\circ < l  < 60^\circ$ and $180^\circ < l < 240^\circ$ \citep{HurleyWalker2019a}, the South Galactic Pole \citep[GLEAM SGP;][]{Franzen2021a}, as well as circular polarization \citep{2018MNRAS.478.2835L} 
 and linear polarization reprocessing \citep[the POlarised GLEAM Survey;][]{Riseley2018,Riseley2020}. The successor survey GLEAM-eXtended has also produced {three data releases so far \citep[GLEAM-X DR1 and DR2;][and DR3 covering the Galactic Plane; Mantovanini et al., in press.]{HurleyWalker2022,gleamx-dr2}} making use of the `extended' baseline configuration of the Phase II MWA \citep{Wayth2018}. While GLEAM and GLEAM-X cover 72--231\,MHz, the MWA bandpass extends beyond 300\,MHz. Frequencies between 242--272\,MHz are occupied by radio frequency interference (RFI) from known satellites \citep{Sokolowski2016} and are avoided. Beyond $\approx 280$\,MHz, RFI occupancy drops sufficiently to provide a usable `high'-frequency band and from 2015--2016 the Phase I MWA was used to perform another all-sky survey, based on the successful GLEAM drift-scan observing strategy, this time at 300 MHz.

Observations from the MWA at 300\,MHz have posed a challenge in generating science-ready images. This is in part due to the antenna spacing and overall antenna design producing a primary beam response with significant grating lobes, contributing to overall lower sensitivity and differences in X and Y polarizations. With the intention to determine whether it would be feasible to process Phase I MWA observations at 300\,MHz, \citet{Cook2021} showcased a processing pipeline that included careful modelling of the apparent brightness of sources for use in calibration, significant RFI flagging, and subtraction of sources in the primary beam grating lobes. These methods go beyond what is required to produce science-ready images for typical MWA observations below 230\,MHz, but are generally applicable to all observations. 

Many of the standard continuum data processing pipelines for the MWA have matured over the last few years, providing the ability to calibrate, image, and combine thousands of snapshot observations on supercomputing infrastructure. While there are still challenges that arise during processing \citep[e.g.~mysterious aliasing artefacts shown by][]{gleamx-dr2}, the current generation of MWA data processing pipelines are able to image much of the data with only minor user intervention. {Within the context of surveys, we have the GLEAM-X pipeline \footnote{\url{https://github.com/GLEAM-X/GLEAM-X-pipeline}.} \citep{HurleyWalker2022,gleamx-dr2}, which has been processing GLEAM-X on a variety of supercomputers, and the Deep Imaging Pipeline (DIP \footnote{\url{https://github.com/sjpaterson/dip}.}), which is being used to process the MWA Interestingly Deep AStrophysical (MIDAS) survey (Paterson et al., {submitted}).} On top of improvements to general calibration and imaging, the MWA All-Sky Virtual Observatory \footnote{\url{https://asvo.mwatelescope.org/}.} (ASVO) provides both a user interface to interact with MWA archives and to then perform pre-processing on observations using the \textit{Garrawarla} and \textit{Setonix} supercomputers at the Pawsey Supercomputing Research Centre \footnote{\url{https://pawsey.org.au/}. \textit{Garrawarla} was decomissioned at the beginning of 2025 after providing over four years of excellent service to the MWA Collaboration. ASVO pre-processing is now done using dedicated resources on the the Pawsey Supercomputing Research Centre cluster \textit{Setonix}.}, which can then be delivered ready for calibration to the user, or calibrated using existing bandpass and gain solutions derived from the observatory's daily calibrator scans \citep{Sokolowski2020}. With numerous upgrades to data (pre-/post-)processing since the original GLEAM survey was released, we have opted to revisit the heretofore unprocessed and unpublished 300\,MHz MWA observations.

\defcitealias{texas}{TXS}
\defcitealias{mrc}{MRC}
\defcitealias{wish}{WISH}
\defcitealias{vcss}{vcss2}

In the Southern Hemisphere to date there are no widefield surveys near 300\,MHz matching the angular resolution and sensitivity of GLEAM, motivating us to progress this work. There are only a handful of surveys near 300\,MHz, typically originating in the Northern Hemisphere, that provide some coverage. From radio interferometers these include The Texas Survey (hereinafter, \citetalias{texas}) of radio sources covering $-35.5^\circ < \delta < +71.5^\circ$ at 365\,MHz \citep{texas}, Westerbork in the Southern Hemisphere \citep[WISH;][]{wish} at 352\,MHz with $54\,\text{arcsec} \times 54 \csc \delta \,\text{arcsec}$ angular resolution, covering $-9^\circ < \delta < -26^\circ$, the Molonglo Reference Catalogue (hereinafter, \citetalias{mrc}; \citealt{mrc}) covering $-85.0^\circ < \delta_\text{B1950} <+18.5^\circ, |b| \geq 3^\circ$ at 408\,MHz. More recently, the Karl G.\ Jansky Very Large Array (VLA) Low-band Ionosphere and Transient Experiment (VLITE \footnote{\url{https://vlite.nrao.edu/}}) is being used to produce a survey at 340\,MHz as the VLITE Commensal Sky Survey \citep[VCSS;][]{vcss1,Polisensky2016,vcss2}. VCSS is a multi-epoch survey covering the sky above $\delta_\text{J2000} > -41.7\degr$, with {first epoch} data products {\citep[VCSS1;][]{vcss-memo}} showing an average angular resolution of $\approx 15\arcsec \times 15\arcsec$ and root-mean-square (rms) noise of $\approx 3$\,mJy\,beam$^{-1}$. In addition to the interferometric surveys, the `Low-Band South' component of the Global Magneto-Ionic Medium Survey \citep[GMIMS-LBS,][]{Wolleben2019} covers 286.25 to 487.75\,MHz and $-90 < \delta_\text{J2000} < +20\degr$, using Murriyang, CSIRO's Parkes radio telescope. Being a single dish survey, its angular resolution is coarse (81--45\arcmin) but is sensitive to all angular scales down to the resolution element.

This paper describes the GLEAM 300 MHz survey in detail, named GLEAM-300, detailing the data processing for the individual observations (Section~\ref{sec:processing}), combining the data products to produce science-ready images (Section~\ref{sec:mosaic}) and a compact source catalogue (Section~\ref{sec:catalogue}), and finally a discussion about applications and future work (Section~\ref{sec:future}). The catalogue and images are publicly released alongside this paper. We define a spectral index, $\alpha$, via $S_\nu \propto \nu^\alpha$ for a flux density $S_\nu$ at frequency $\nu$. Asymmetric uncertainties are reported at the 16-th and 84-th percentile unless otherwise noted. 

\section{300 MHz observations and data processing}\label{sec:processing}

\begin{table*}[tp]
    \centering
    \begin{threeparttable}
    \caption{\label{tab:strips}Declination strips and observation information.}
    \begin{tabular}{rcc c c c}
    \toprule
        Declination & Dates & $N_\text{snapshots}$ & Solutions type \tnote{a} & Median rms \tnote{b} & Good snapshots \\
        ($^\circ$) & & & ( C / N / S) & (mJy\,beam$^{-1}$) & (\%) \\\midrule
        $+32.0$ &  2015 Sep 15, Nov 06, Dec 12; 2016 Feb 15, Apr 19, Jun 23 & 1877 & 39\% / 34\% / 27\%  & $128_{-30}^{+43}$ & 84\% \\
         $+23.0$ &  2015 Sep 16, Nov 07,  Dec 13; 2016 Feb 16, Apr 20, Jun 24 & 939 & 0\% / 55\% / 45\% & $75_{-17}^{+37}$ & 84\% \\\midrule
        $+18.6$ &  2015 Sep 21, Nov 12, Dec 19; 2016 Feb 21, Apr 26, Jun 30 & 922 & 34\% / 48\% / 18\% & $64_{-14}^{+29}$ & 85\% \\
        $+9.7$ &  2015 Sep 21, Nov 12, Dec 19; 2016 Feb 21, Apr 26, Jun 30 & 921 & 23\% / 71\% / 7\% & $55_{-10}^{+19}$ & 86\%\\
        $+1.6$ &  2015 Sep 20, Nov 11, Dec 18; 2016 Feb 20, Apr 25, Jun 28 & 936 & 62\% / 10\% / 29\% & $57_{-14}^{+31}$ & 77\% \\
        $-5.9$ &  2015 Sep 20, Nov 11, Dec 18; 2016 Feb 20, Apr 25, Jun 28 & 933  & 48\% / 42\% / 10\% & $60_{-19}^{+26}$ & 82\% \\
        $-13.0$ &  2015 Sep 19, Nov 10, Dec 22; 2016 Feb 19, Apr 24, 27, Jun 27 & 1026 & 72\% / 13\% / 15\% & $51_{-9}^{+26}$ & 66\% \\

        $-19.9$ &  2015 Sep 19, Nov 10, Dec 22; 2016 Feb 19, Apr 24, 27, Jun 27 & 1025 & 7\% / 53\% / 40\% & $55_{-11}^{+19}$ & 86\% \\

        $-26.7$ &  2015 Sep 15, Nov 06, Dec 12; 2016 Feb 15, Apr 19, Jun 23 & 1877 & 40\% / 33\% / 27\% & $56_{-13}^{+13}$ & 85\% \\
        
        $-33.5$ &  2015 Sep 16, Nov 07,  Dec 13; 2016 Feb 16, Apr 20, Jun 24 & 939 & 0\% / 51\% / 49\% & $49_{-9}^{+23}$ & 72\% \\
        
        $-40.4$ &  2015 Sep 16, Nov 07,  Dec 13; 2016 Feb 16, Apr 20, Jun 24 & 940 & 8\% / 59\% / 33\% & $52_{-11}^{+21}$ & 86\% \\
        
        $-47.5$ &  2015 Sep 17, Nov 08; 2016 Feb 17, Apr 21, Jun 22 & 800 & 0\% / 87\% / 13\% &  $59_{-17}^{+20}$ & 84\% \\
        
        $-55.0$ &  2015 Sep 17, Nov 08; 2016 Feb 17, Apr 21, Jun 22 & 799 & 4\% / 77\% / 19\% & $66_{-25}^{+29}$ & 87\% \\
        
        $-63.1$ & 2015 Sep 18, Nov 9, Dec 16; 2016 Feb 18, Apr 23, Jun 26   & 837 &  9\% / 75\% / 16\% & $62_{-21}^{+35}$ & 78\% \\
        
         $-72.0$ & 2015 Sep 18, Nov 9, Dec 16; 2016 Feb 18, Apr 23, Jun 26  & 837 & 8\% / 70\% / 22\% & $72_{-21}^{+15}$ & 79\% \\\midrule
         
         $-76.0$ &  2015 Sep 19, Nov 10, Dec 22; 2016 Feb 19, Apr 24, 27, Jun 27 & 1025 & 6\% / 54\% / 39\% & $90_{-23}^{+34}$ & 86\% \\
         
        $-86.0$ & 2015 Nov 06, Dec 12; 2016 Feb 15, Apr 19, Jun 23 & 1877 & 39\% / 32\% / 29\%& $150_{-29}^{+56}$ & 82\% \\\midrule
        All & - & & 26\% / 48\% / 26\% & $68_{-22}^{+59}$ & 83\% \\\bottomrule
    \end{tabular}
    \begin{tablenotes}[flushleft]
{
\footnotesize \item[a] `C': solutions derived from a dedicated
calibrator scan; `N': nearest-in-time best solutions from other observations; `S': solutions derived from the observation itself. \item[b] Median rms noise in a single two-minute snapshot image {after applying the brightness scale correction described in Section~\ref{sec:post:scale}}.
}
    \end{tablenotes}
    \end{threeparttable}
\end{table*}

\begin{figure*}[pth!]
    \centering
    \begin{subfigure}[b]{0.49\linewidth}
    \includegraphics[width=1\linewidth]{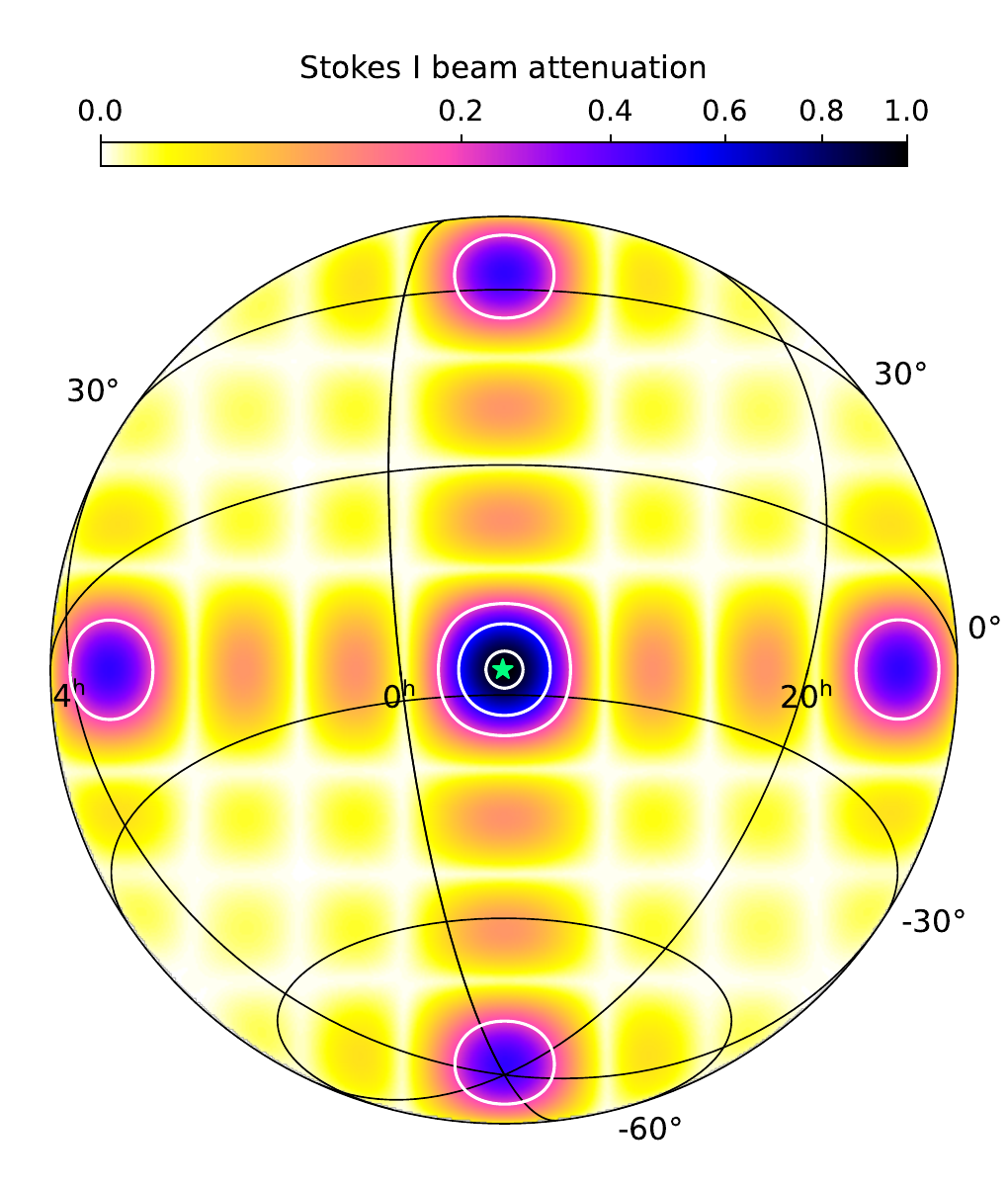}
    \caption{\label{fig:beam:m26} ObsID 1150751128.}
    \end{subfigure}%
    \begin{subfigure}[b]{0.49\linewidth}
    \includegraphics[width=1\linewidth]{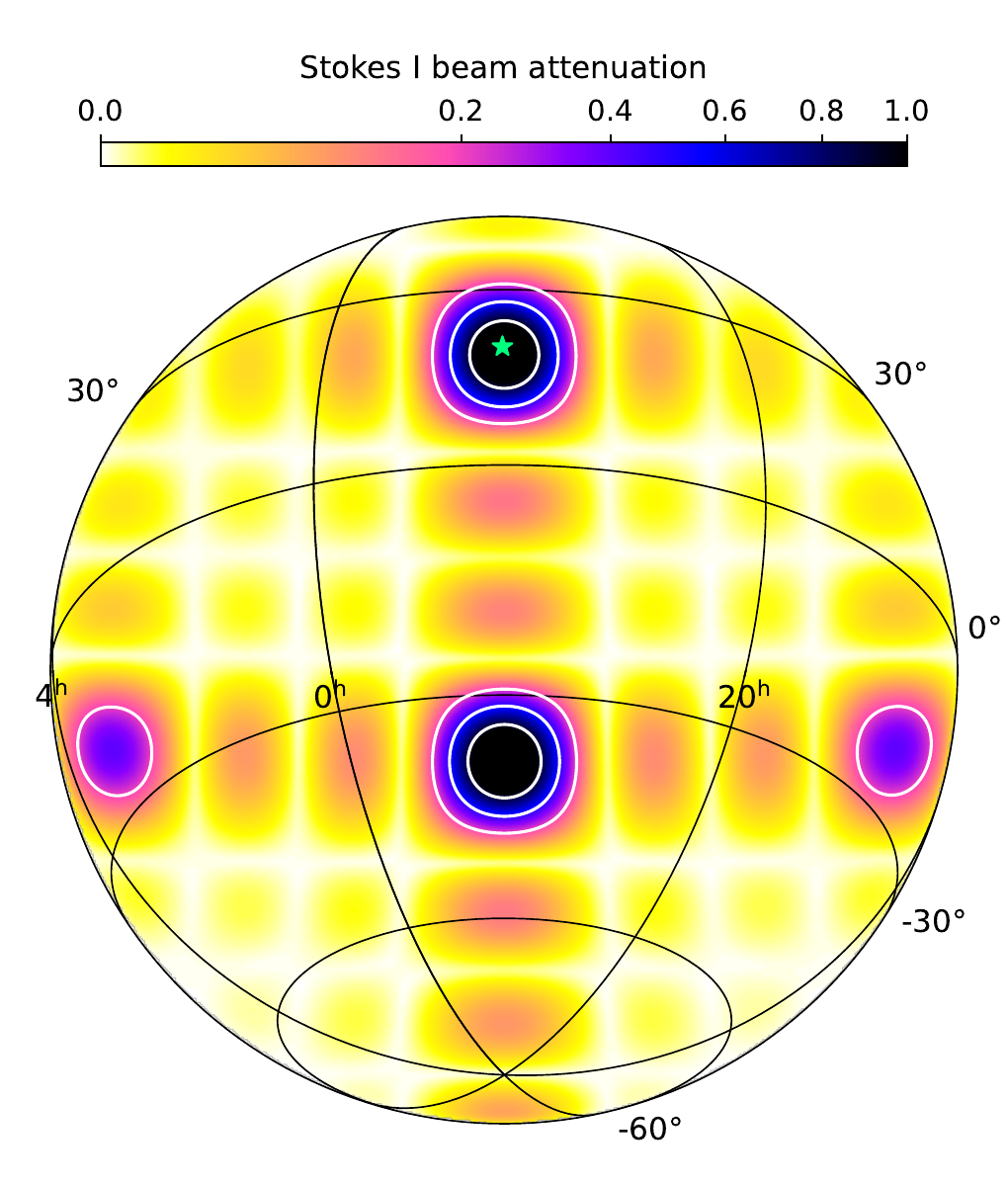}
    \caption{\label{fig:beam:p18} ObsID 1151351872.}
    \end{subfigure}
    \caption{\label{fig:beam} Example primary beam response for a zenith pointing [declination $-26.7^\circ$, \subref{fig:beam:m26}] and a low-elevation pointing [declination $+18.6^\circ$, \subref{fig:beam:p18}]. The attenuation is displayed with square-root stretch, and the white contours trace $[0.2, 0.5, 0.9]$. The green stars indicate the main lobe, in the pointing direction, and significant grating lobes are those above the 0.2 contour.}
\end{figure*}

Observations for the 300-MHz GLEAM survey follow the same drift-scan observing strategy used for GLEAM and GLEAM-X. This observing mode has the telescope form a primary beam near the meridian towards a given declination, and takes two-minute snapshots as the sky drifts above the observatory over the course of an observing night. Due to the smaller field-of-view (FoV) of the MWA at 300\,MHz, extra declinations are used to ensure uniform sensitivity, resulting in 13~declination strips from $-72.0\degr \lesssim \delta_\text{J2000} \lesssim +18.6\degr$. These observations provide primary beam mainlobe sensitivity from $-80^\circ \lesssim \delta_\text{J2000} \lesssim +24^\circ$. The declination strips and observation dates are summarised in Table~\ref{tab:strips}. {The observations have 30.72\,MHz bandwidth (covering 284.8--315.5\,MHz), and were recorded at 40\,kHz/0.5\,s frequency/time resolution, following the original GLEAM. Individual observations are referred to by unique observation IDs (ObsIDs) defined by the start time in GPS time format.}

GLEAM 300 MHz data processing follows standard continuum snapshot processing strategies. In this case, we follow the processing strategy used by \citet{Duchesne2020} \footnote{Using the `Phase II Pipeline' (\texttt{piip}): \url{https://gitlab.com/Sunmish/piip}, though note despite the name the pipeline is also suitable for processing MWA Phase I data as is the case here.}. {Data are pre-processed through the ASVO, which includes automated flagging with \texttt{AOFlagger} \citep{Offringa2012,Offringa2015} and averaging to 4\,s time resolution.} In the following sections, we briefly describe the processing with particular emphasis on differences between GLEAM/GLEAM-X and GLEAM 300 MHz. All two-min snapshots are pre-processed, calibrated, imaged, and post-processed individually prior to mosaicking and final source-finding activities described in later sections.  

\subsection{Calibration and flagging}\label{sec:calibration}

\begin{figure*}[t]
    \centering
    \includegraphics[width=1\linewidth]{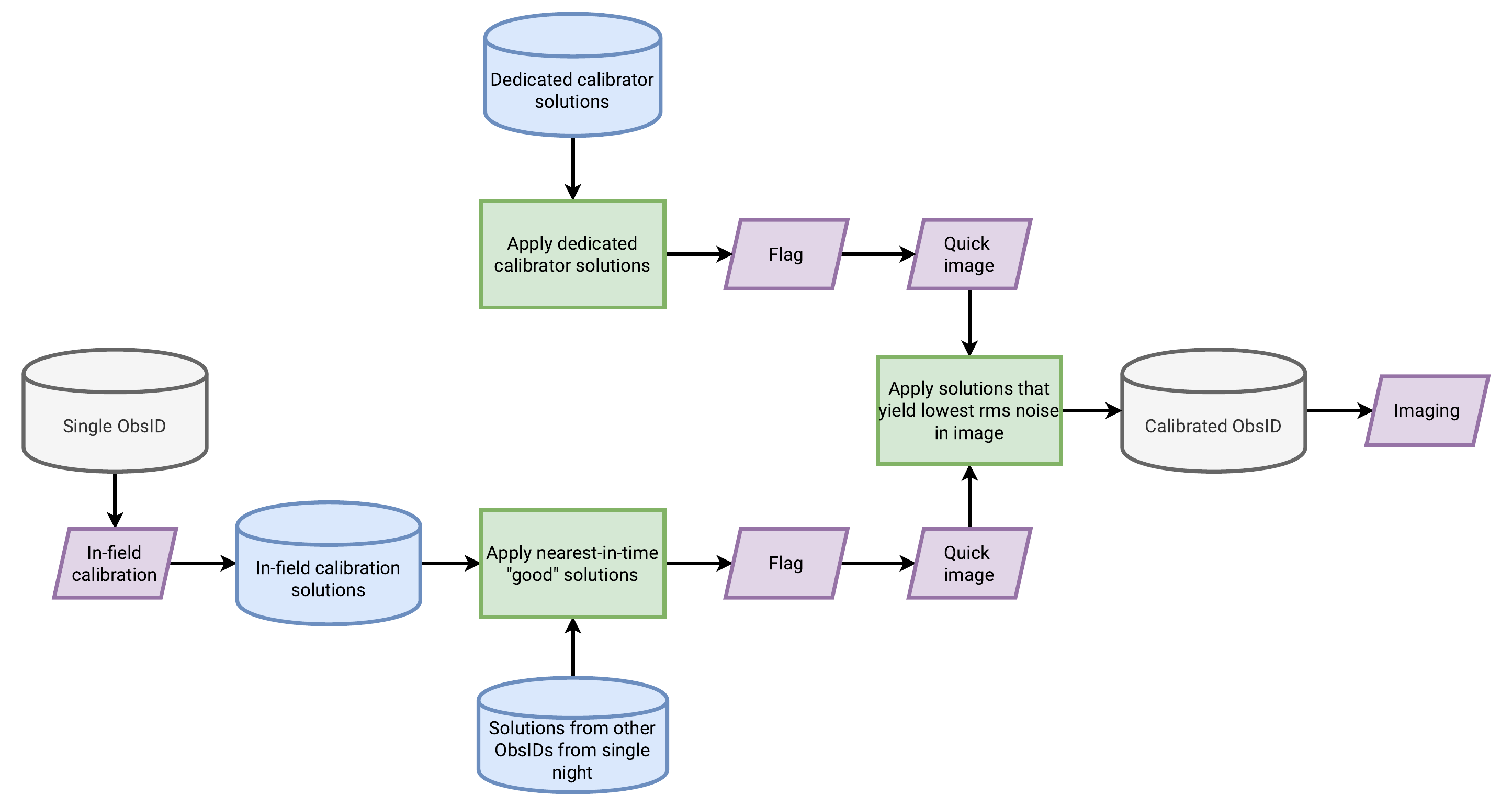}
    \caption{\label{fig:calibration} Flow diagram of the calibration procedure. The diagram shows the processing steps to assign good calibration solutions to each ObsID after initial pre-processing, and prior to peeling/outlier source subtraction, and imaging.}
\end{figure*}

Figure~\ref{fig:beam} shows example primary beam attenuation patterns (with main, grating, and side lobes) for a zenith scan [pointed towards declination $-26.7^\circ$, \ref{fig:beam:m26}] and a low-elevation scan [pointed towards declination $+18.6^\circ$, \ref{fig:beam:p18}] with the pointing directions indicated by green stars. The main lobe is closest to the pointing direction, but at this frequency and particularly for low elevations, the grating lobes approach the sensitivity of the main lobe. \citet{Cook2021} used observations of bright `calibrator' sources to ensure bandpass and direction-independent complex gains can be solved when the primary beam grating lobes contribute significant amounts of power.  We found that following this approach presented two issues: (1) {we were unable to achieve good calibration solutions for some dedicated calibrator scans for unknown reasons}, and (2) the gains drift over the course of the night and good calibration solutions become less effective as the night progresses. The alternate approach being used in other MWA processing strategies is to perform in-field calibration for every snapshot observation \citep[e.g.][but see also \citealt{Duchesne2020,Lynch2021,HurleyWalker2022}]{Franzen2021a}. For these 300-MHz observations, the in-field approach suffers a higher failure rate than the dedicated calibrator scans, reducing to only $26$\% of observations arriving at good bandpass and gain solutions. {In the context of these 300-MHz observations, `in-field' refers to the main lobe as well as grating lobes above an attenuation of 20\%. A {local} sky model is constructed for each observation from a subset of the {GLEAM global sky model \citet[][]{HurleyWalker2022} \footnote{See \url{https://github.com/GLEAM-X/GLEAM-X-pipeline}.}. We take the 1\,000~sources with the highest attenuated brightness using their frequency-dependent models. We note that while the sky model is largely based on GLEAM, the sky above $\delta_\text{J2000} \gtrsim +30\degr$ is filled in with sources from the NRAO \footnote{National Radio Astronomy Observatory.} VLA Sky Survey catalogue \citep[NVSS;][]{ccg+98} at 1\,400\,MHz. These northern sources have a two-point spectral index that is derived after cross-matching to the VLA Low-frequency Sky Survey redux catalogue \citep[VLSSr;][]{Lane2014} \footnote{\url{https://github.com/johnsmorgan/marco/tree/master/gleam_sky_model}.} at 74\,MHz.} 

{Our calibration process makes use of a mixture of in-field solutions and dedicated calibrator solutions, selecting solutions that reduce the image rms noise. This process is outlined in a flow diagram in Figure~\ref{fig:calibration} and summarised as follows.} {We initially obtain calibration solutions from each nightly calibrator scan, using the same in-field calibration approach on the calibrator observation. We then apply the solutions to the survey ObsIDs, perform additional flagging, then create a shallow quick-look image.} 

{Then, on uncalibrated data we perform in-field calibration with all ObsIDs.} We then consider in-field solutions `good' if they satisfy \begin{enumerate}
    \item $\sigma_\text{amplitude} < 40$\,Jy,
    \item $\sigma_\text{phase} < 2\degr$,
    \item $<35$\% solutions flagged from data that are not initially flagged,
\end{enumerate}
where $\sigma_\text{amplitude}$ and $\sigma_\text{phase}$ are the standard deviations of the amplitude and phase of the complex gains, checking both the XX and YY polarizations separately. We ignore XY and YX instrumental polarizations as we do not perform polarization calibration and assume all calibrators are unpolarized. Solutions identified as good are applied directly to their host snapshots, and for ObsIDs with `bad' solutions the nearest-in-time good solution is applied with no interpolation. Another set of shallow quick-look images is then made after further flagging.

{The two quick-look images are compared for each ObsID, and we record the set of solutions that produced the lowest rms noise in the image and use those solutions to calibrate the data. In Table~\ref{tab:strips} we note the fraction of ObsIDs for each night that use either dedicated calibrator solutions (`C'), the in-field solutions derived from the ObsID itself (`S'), or the nearest-in-time in-field solutions (`N').}

{At multiple stages in the calibration process, prior to imaging, we perform a significant amount of RFI flagging. Alongside \texttt{AOFlagger}, we also follow \citet{Cook2021} and make heavy use of the automated flagging tasks \texttt{rflag} and \texttt{tfcrop} within the Common Astronomy Software Applications \citep[CASA;][]{CASA2022} \footnote{\url{https://casa.nrao.edu/}} package. This is done before and after calibration. We also use the gain solutions to identify poorly-performing MWA tiles that are typically flagged for a given observing night. While the total percentage of flagged data varies between observations, we see approximately 60\% of data flagged, which is a reasonable increase from the 20--30\% normally seen at lower frequencies from other GLEAM observations processed with the same pipeline.} 

\subsection{Imaging}\label{sec:imaging}

\subsubsection{Outlier source subtraction}
Prior to imaging, we opt to peel and/or subtract a selection of sources outside of the primary beam mainlobe. We begin with peeling and subtraction of specific bright sources (e.g. Pictor~A, Cygnus~A---see Appendix~\ref{app:subtraction} for a full list). This peeling process uses a model of the primary beam to estimate the apparent brightness of sources in this collection of bright sources, and if above 77\,Jy at 300\,MHz they are peeled and if above 2.3\,Jy directly subtracted (without gain calibration) from the data \footnote{The default behaviour of the pipeline is to scale flux densities from 215\,MHz reference values, so 77\,Jy is scaled from 100\,Jy at 215\,MHz assuming a spectral index of $-0.77$, and 2.3\,Jy is scaled from 3\,Jy.}. We use the peeling pipeline \texttt{PotatoPeel} \footnote{\url{https://gitlab.com/Sunmish/potato}.} \citep[see][for a description of the process]{racs-mid}, which uses the CASA package and \texttt{WSClean} \citep{wsclean1} \footnote{\url{https://gitlab.com/aroffringa/wsclean/}.} to generate image-based CLEAN component models of the apparent brightness of the source, which are used for amplitude and phase gain calibration before subtraction of the offending source as done in peeling \citep[e.g.][]{Smirnov2011b}. {After removal of individual bright sources, we also image a selection of primary beam grating and side lobes in the order of attenuation (i.e. brightest lobe first). The CLEAN component models from the lobe images are subtracted to remove {any} additional contamination from off-axis sources. For most observations, a majority of the grating lobes are subtracted alongside a small number of side lobes.}

\subsubsection{Snapshot images}
After outlier sources and grating/side lobes are subtracted, we make images of the main lobe covering approximately $21\degr \times 21\degr$. We make use of multi-scale deconvolution from \texttt{WSClean} \citep{wsclean2} and the $w$-gridder algorithm \citep{Arras2021,Ye2022}. For image weighting, we use the `Briggs' weighting scheme \citep{Briggs95} with a robust parameter of 0. This weighting is chosen as it optimises sensitivity without significant cost to angular resolution for these data. While some baselines end up flagged prior to imaging due to RFI and poor calibration, there is no explicit baseline or $(u,v)$ cut added. 

{Initial imaging runs used automatic masking within \texttt{WSClean}, generating a mask once CLEANing reaches an initial global threshold of $3\sigma_\text{rms}$, and then continuing down to $1\sigma_\text{rms}$ within the derived mask. We found that this approach introduced a bias in flux density measurements of sources below the initial masking threshold. We suspect that too many noise components were deconvolved and may have redistributed flux away from un-deconvolved sources. We detail an exploration of this problem in Appendix~\ref{app:faint}.}

{Our imaging setup instead makes use of two imaging modes: the first uses no mask, with a global $3\sigma_\text{rms}$ threshold and up to five major iterations. $\sigma_\text{rms}$ is calculated as a position-independent value for each snapshot. This comparatively shallow CLEAN reduces the aforementioned bias, but results in up to $\approx 30\%$ higher rms noise in the output images. This is similar to the CLEAN threshold used for GLEAM and GLEAM-SGP, though we note some noise pixels may still be deconvolved in this case, depending on the observation. The second method retains the original masking and thresholding, but adds an additional mask. This additional mask restricts CLEAN to traditional boxes around sources from the sky model. The box size is set to $5\,\text{pixel} \times 5\,\text{pixel}$. As this model is derived from deeper surveys than an individual snapshot image (namely GLEAM), almost all sources in a snapshot are included. We also restrict the mask further to sources that have $>25$\,mJy apparent flux density at 300\,MHz, corresponding to approximately the lowest rms noise in any of the snapshots. We note that the automatic mask computed by \texttt{WSClean} is used within the regions covered by the sky model mask.}

{In general, the masking approach provides the best imaging quality and largely removes the aforementioned bias, however, around certain bright extended sources like Fornax~A and in the Galactic Plane the masks are not adequate to capture the extent of the source so results in poorer images. For snapshots within Galactic latitudes of $|b|<7.5\degr$ and snapshots within $7.5\degr$ of specific bright extended sources \footnote{Orion~A, Taurus~A, Fornax~A, Centaurus~A, and the radio galaxy associated with IC~4296, though note some of these overlap with the Galactic Plane selection.}, we select the non-masked image, and select the masked image in all other cases where CLEAN did not diverge. We therefore use the masked image for 88\% for the survey.}
 
 As in other GLEAM surveys, the snapshot data are split into four subbands during imaging by \texttt{WSclean} (\texttt{-channels-out 4}) to account for instrumental and physical spectral effects, with CLEAN components derived from the averaged full-band 30.72-MHz image. However, we do not retain the comparatively lower-sensitivity 7.68-MHz narrowband images. Sensitivity at 300\,MHz is reduced compared to the lower part of the MWA band used for the rest of GLEAM (i.e.\ 72--231 MHz), and with more of the band flagged due to RFI---particularly below 300\,MHz---the narrow band images are significantly poorer quality than the equivalent images from the other GLEAM datasets.

\begin{figure*}[p]
    
    \begin{subfigure}[b]{1\linewidth}
        \includegraphics[width=1\linewidth]{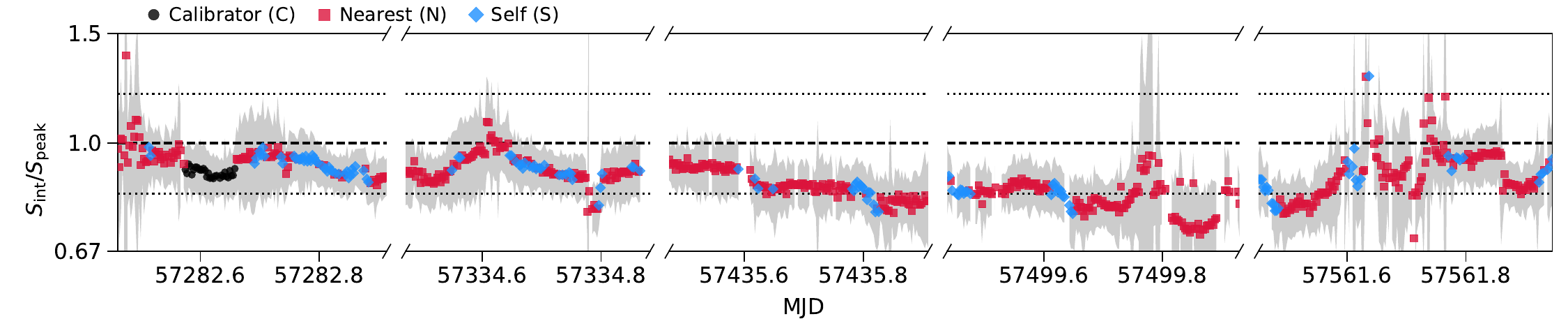}
        \caption{\label{fig:night:intpeak} Median $S_\text{int}/S_\text{peak}$ per snapshot.}
    \end{subfigure}
    \begin{subfigure}[b]{1\linewidth}
        \includegraphics[width=1\linewidth]{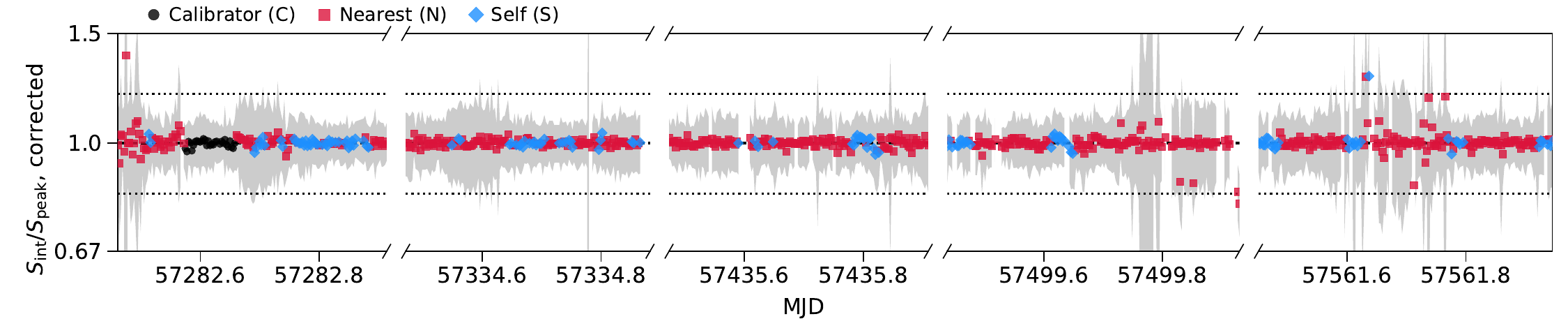}
        \caption{\label{fig:night:intpeak2} Median $S_\text{int}/S_\text{peak}$ per snapshot after applying $f_\text{PSF}$.}
    \end{subfigure}
    \begin{subfigure}[b]{1\linewidth}
        \includegraphics[width=1\linewidth]{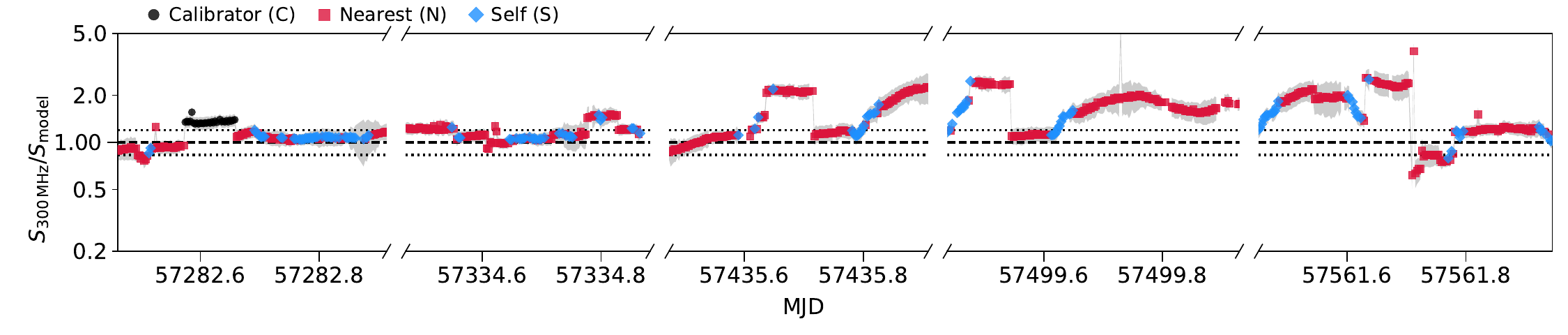}
        \caption{\label{fig:night:fluxratio} Median $S_\text{300 MHz} / S_\text{model}$ per snapshot.}
    \end{subfigure}
    
    \begin{subfigure}[b]{1\linewidth}
        \includegraphics[width=1\linewidth]{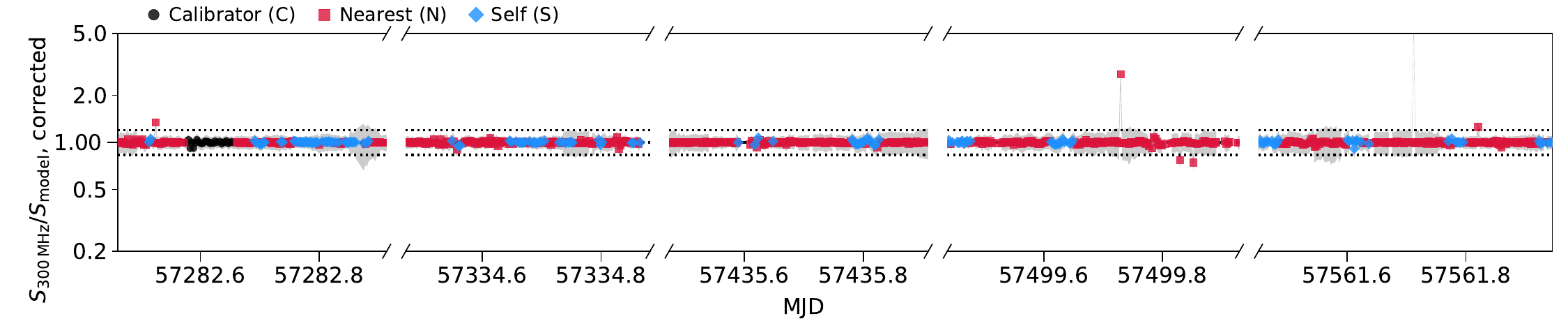}
        \caption{\label{fig:night:fluxratio2} Median $S_\text{300 MHz} / S_\text{model}$ per snapshot after applying $f_\text{PSF}$ and $f_\text{scale}$.}
    
    \end{subfigure}
\begin{subfigure}[b]{1\linewidth}
        \includegraphics[width=1\linewidth]{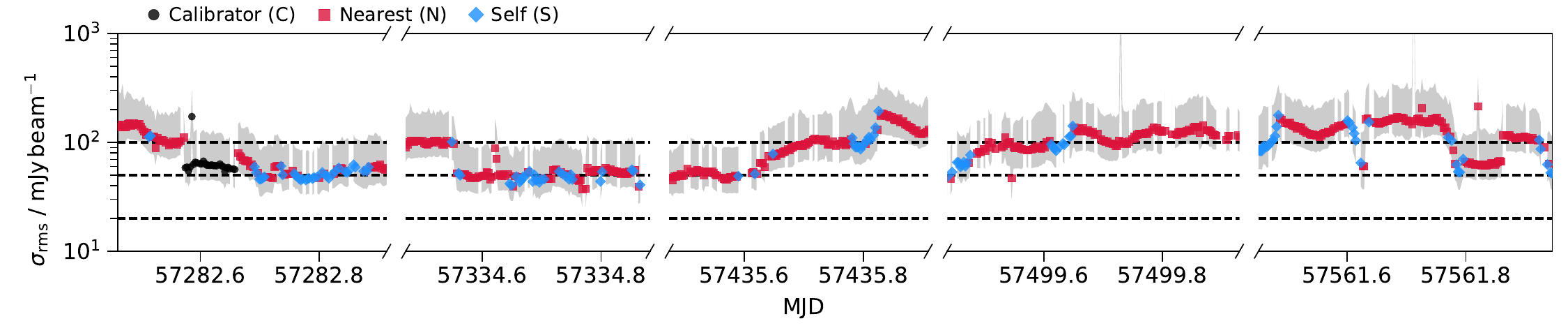}
        \caption{\label{fig:night:rms} Median $\sigma_\text{rms}$ per snapshot {after corrections}.}
    \end{subfigure}
    
    \caption{\label{fig:night} Basic image properties per snapshot for declination strip $-55.0\degr$. \subref{fig:night:intpeak}--\subref{fig:night:intpeak2} {SNR-weighted mean} $S_\text{int}/S_\text{peak}$ per snapshot after applying $f_\text{PSF}$.  The dashed horizontal line is drawn at 1, and the dotted horizontal lines are drawn at a ratio difference of 20\%. \subref{fig:night:fluxratio}--\subref{fig:night:fluxratio2} {SNR-weighted mean} flux density ratios, $S_\text{300 MHz} / S_\text{model}$, comparing to {the GGSM} measurement, before and after applying $f_\text{PSF}$ and $f_\text{scale}$. The horizontal lines are as in \subref{fig:night:intpeak}. \subref{fig:night:rms} median rms noise ($\sigma_\text{rms}$) per snapshot {after applying the brightness scale correction factor}. Dashed horizontal lines indicate 20, 50, and 100 mJy\,beam$^{-1}$.  In {\subref{fig:night:intpeak}--\subref{fig:night:fluxratio2}, the grey shaded region is drawn between $\pm1\sigma$ and  in \subref{fig:night:rms} between the 16-th and 84-th percentiles. Snapshots} are coloured by their calibration solutions. Black circles: solutions derived from a dedicated calibrator scan; red squares: nearest-in-time best solutions from other observations; blue diamonds: solutions derived from the observation itself. Note for this declination strip, only a small number of snapshots used the dedicated calibrator solutions in the first night of observing. Note the y-axis in all panels is logarithmically scaled.}
\end{figure*}

\subsection{Patching gaps with grating lobes}\label{sec:scp}
The lowest declination strip is at $-72^\circ$ and the mainlobe full-width at half-maximum (FWHM) is only $\approx 15\degr$ so does not cover the South Celestial Pole (SCP). While there are no pointings towards the SCP, most of the observations in the declination {$-26.7\degr$ and $-19.9\degr$} strips have significant primary beam grating lobes that {provide coverage below $\delta_\text{J2000} < -72\degr$, including the SCP, with maximum sensitivity at $\delta_\text{J2000} \approx -86\degr$ and $\approx -76\degr$, respectively.} Figure~\ref{fig:beam:m26} shows an example pointing with a grating lobe with a peak in sensitivity at $\delta_\text{J2000} \approx -86\degr$, covering the SCP. To ensure complete coverage of the Southern Sky we image {these lobes for all the relevant observations}, using the same calibration solutions for the given snapshot observation. We also repeat the peeling process, assuming the grating lobe is the pointing centre (as opposed to the main lobe) and continue with imaging and post-processing in an otherwise identical fashion.

{In addition to the grating lobes covering the SCP region, we also use the equivalent grating lobes at $\delta_\text{J2000} \approx +32\degr$ and $\approx +23\degr$ from the declination $-26.7\degr$ and $-33.5\degr$ strips to provide survey coverage up to $\approx +40\degr$. We refer to these additional data as separate declination strips in Table~\ref{tab:strips}.}

\subsection{Post-imaging corrections}

\begin{figure}[t]
    \centering
    \includegraphics[width=1\linewidth]{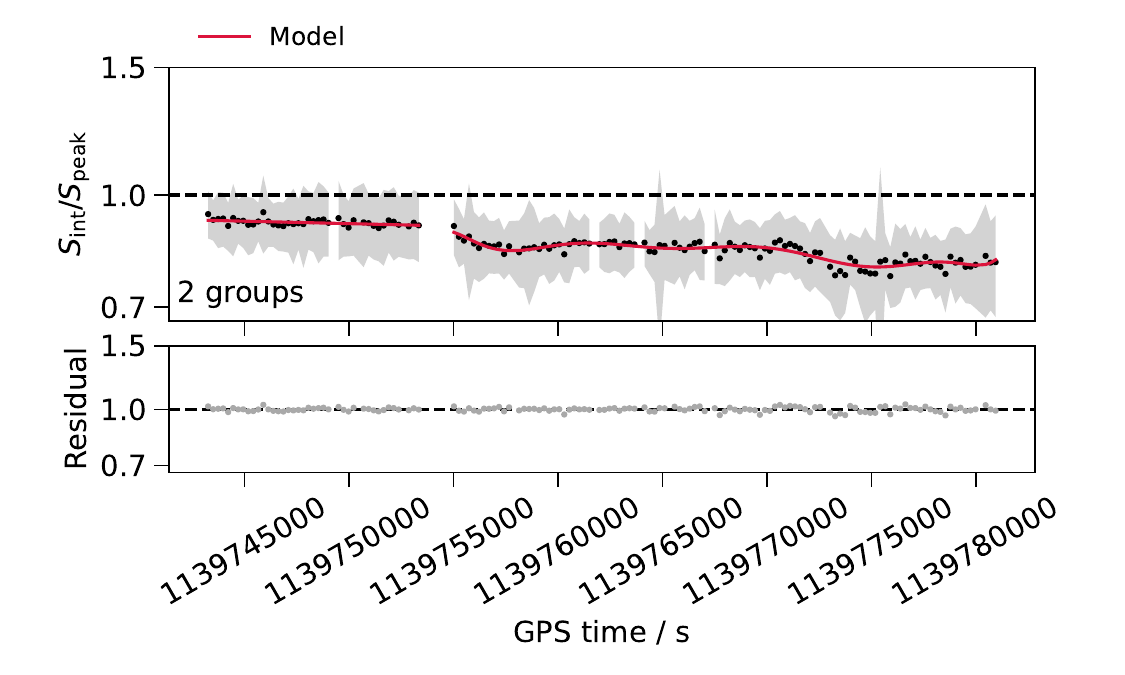} \\
    \includegraphics[width=1\linewidth]{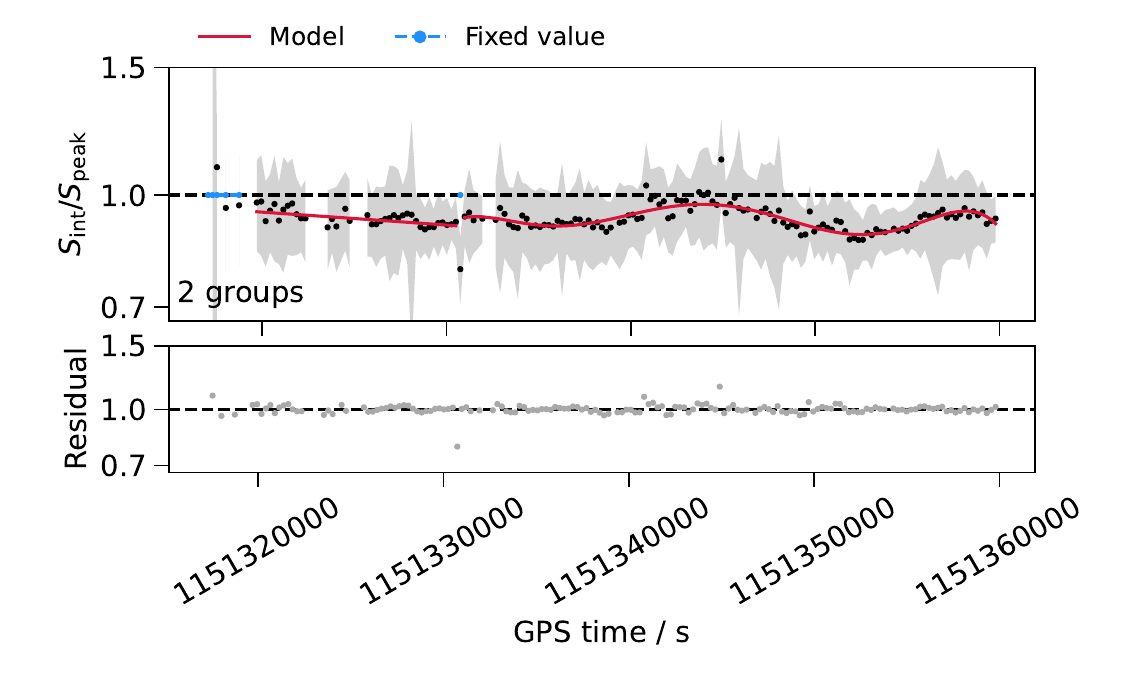}
    \caption{\label{fig:corrections:ip} {SNR-weighted mean} $S_\text{int}/S_\text{peak}$ {of sources cross-matched to the `unresolved' RACS-SUMSS/MGPS-2 catalogue} as a function of time for two example nights (\emph{top panels}) and the residual ratio (\emph{bottom panels}).  The black points indicate {SNR-weighted mean} values for a given ObsID, and {the red solid lines indicate fitted polynomial models to each group as described in the text and the blue dashed lines with blue markers indicate a fixed value was assigned (only relevant for the bottom panel in this case).} The gray shaded regions show $\pm 1\sigma$ for each snapshot. The residuals after applying the model values are shown in the bottom panels.}
\end{figure}

\begin{figure}[t]
    \centering
    \includegraphics[width=1\linewidth]{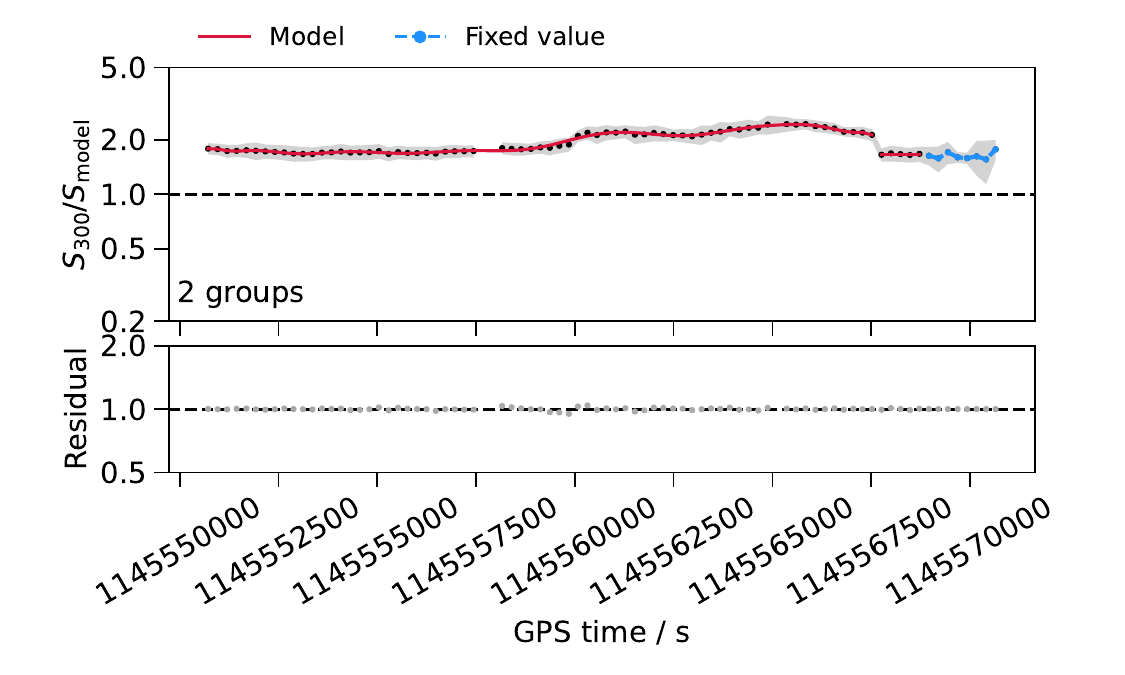}
    \includegraphics[width=1\linewidth]{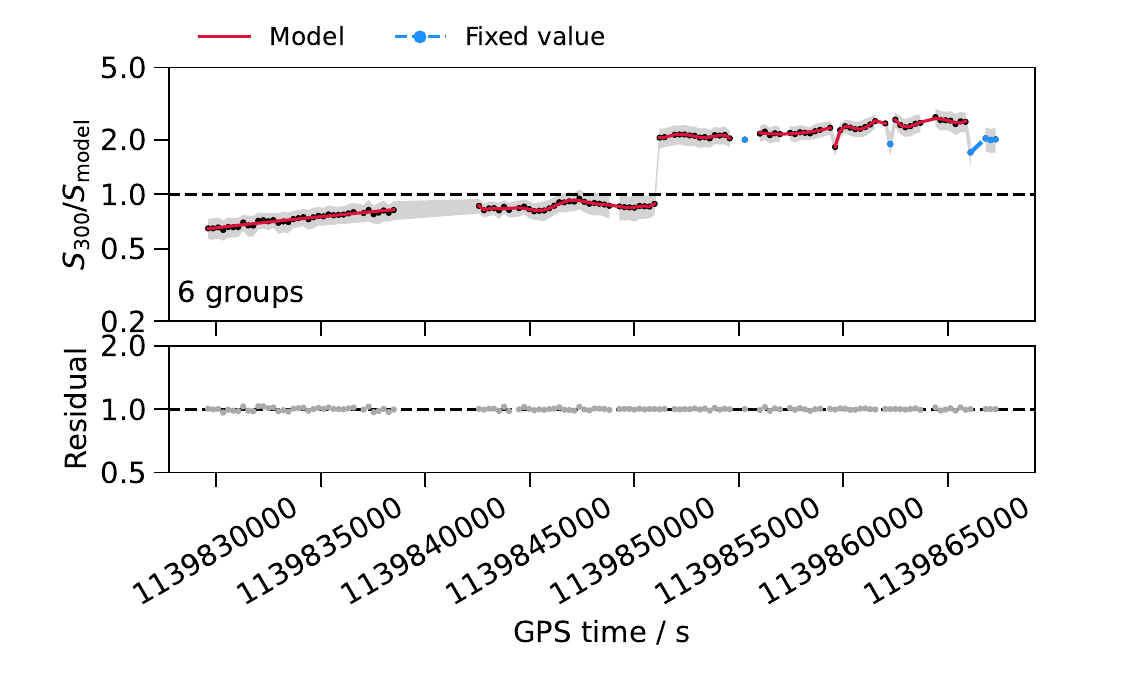}
    \caption{\label{fig:brightness_scaling} Models of the per-ObsID brightness scale correction factors for two example nights (\emph{top panels}) and the residual ratio (\emph{bottom panels}). The black points indicate weighted mean values for a given ObsID, and {the red solid lines} indicate fitted polynomial models to each group as described in the text. The shaded, gray regions show the weighted standard deviation. {The blue dashed lines with blue markers indicate a fixed value was assigned based on the weighted mean of the individual ObsID.}}
\end{figure}

\begin{figure*}[t]
    \centering
    \includegraphics[width=1\linewidth]{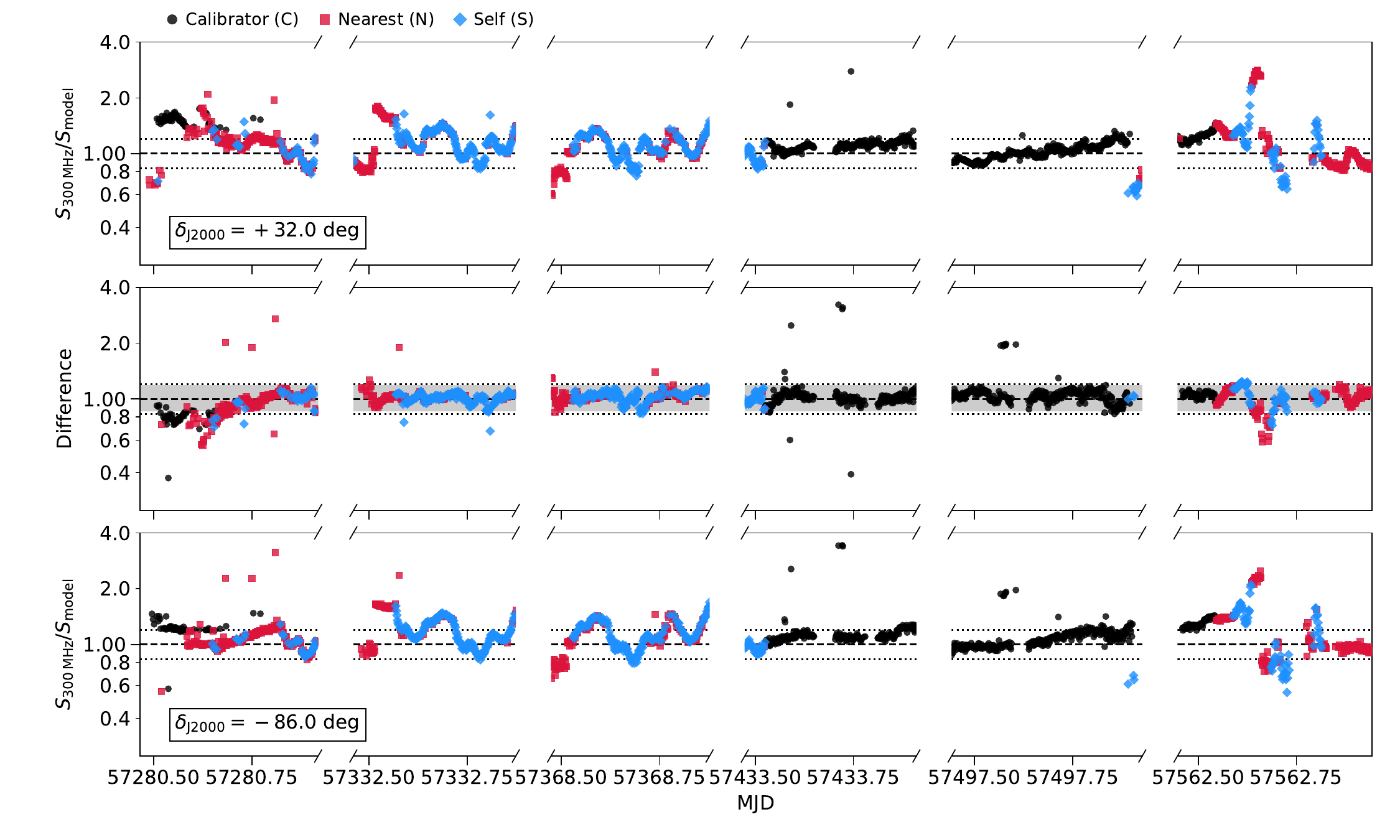}
    \caption{\label{fig:brightness_scaling:lobes} {SNR-weighted mean flux density ratios, $S_\text{300 MHz} / S_\text{model}$, comparing to the extrapolated sky model measurement for grating lobe images that make up the $\delta_\text{J2000} = +32.0\degr$ (\textit{top panel}) and $-86.0\degr$ (\textit{bottom panel}) declination strips. The middle panel shows the difference in the SNR-weighted mean ratios between the two strips for each ObsID. The shaded region in the middle panel is drawn at $\pm 1\sigma$ (16\%), and the dashed and dotted lines in each panel are drawn at a ratio of 1 and $\pm 20\%$, respectively. The points are coloured by their calibration solutions type, as in Figure~\ref{fig:night}. The y-axis scale is logarithmic, though has a reduced range compared to the flux density ratios on Figure~\ref{fig:night}.}}
\end{figure*}

\subsubsection{Primary beam and astrometry}
After imaging we correct for the direction-dependent sensitivity pattern by applying the Full Embedded Element (FEE) model of the primary beam \citep{Sokolowski2017}. The images are clipped where they are attenuated by $\geq 90\%$ (i.e. the apparent brightness is $<10$\%). After primary beam correction, we use the source-finder \texttt{aegean} \footnote{\url{https://github.com/PaulHancock/Aegean}.} \citep[][]{Hancock2012,Hancock2018} to create a source list for each snapshot, keeping sources with peak brightness above $5\sigma_\text{rms}$, where $\sigma_\text{rms}$ is the local rms noise in the snapshot as calculated by the Background And Noise Estimation (\texttt{BANE}) tool packaged with \texttt{aegean} \citep{Hancock2018}. The source lists for each snapshot are cross-matched to the sky model used for initial calibration, and we use these cross-matched source lists to correct for position-dependent astrometric offsets using \texttt{fits\_warp.py} \footnote{\url{https://github.com/nhurleywalker/fits_warp}.} \citep{HurleyWalker2018}. This process uses a position-dependent set of interpolated pixel shifts to align the image data with the global sky model.

\subsubsection{The point-spread function}\label{sec:post:psf}

After imaging and source-finding with \texttt{aegean}, we inspect the ratio of integrated and peak flux densities ($S_\text{int}/S_\text{peak}$), comparing the median for each snapshot and each declination strip to check that the point-spread function (PSF) is appropriately estimated for each snapshot. To avoid extended sources, we {construct a reference point source catalogue by combining the Sydney University Molonglo Sky Survey \citep[SUMSS;][at 843 MHz, $\approx 45\times 45$\,arcsec$^{2}$ angular resolution]{bls99,mmb+03}, Molonglo Galactic Plane Survey \citep[MGPS-2;][]{mmg+07}, and the Rapid ASKAP \footnote{Australian SKA Pathfinder \citep{Hotan2021}.} Continuum Survey \citep[RACS;][]{racs1} 888-MHz catalogue \citep[RACS-low;][at $25 \times 25$ arcsec$^{2}$ angular resolution]{racs2}. We use SUMSS/MGPS-2 for the sky south of $\delta_\text{J2000}<-80\degr$ and RACS-low elsewhere. For RACS-low sources, we use the unresolved definition from \citet[][see their section 5.2.1]{racs2}, and for SUMSS/MGPS-2 we consider a source unresolved if $S_\text{int}/S_\text{peak} < 1.2$. We also remove sources in the SUMSS/MGPS-2 and RACS-low catalogues that have neighbours within 4 and 3\,arcmin, respectively. We cross-match this unresolved and isolated RACS-SUMSS/MGPS-2 catalogue to the GLEAM-300 snapshot source list to identify point sources, but we do not enforce any similar selection criteria on the GLEAM-300 data.}

Figure~\ref{fig:night:intpeak} shows the declination $-55.0\degr$ {signal-to-noise (SNR) weighted mean $S_\text{int}/S_\text{peak}$} ratios for cross-matched unresolved sources as a function of ObsID. In general, this ratio should tend towards $>1$ ($=1$ for an ideal point source), where residual extended sources and ionospheric blurring push the ratio above 1. With median ratios $<1$ there are likely calibration, deconvolution, and/or a mismatched PSF issues. {We find that for most snapshots this ratio is less than 1.} {The declination $-55.0\degr$ strip has a median across all sources/snapshots of $0.91_{-0.06}^{+0.05}$}, which reduces to {$0.88_{-0.05}^{+0.03}$} for sources with peak flux densities above $100\sigma_\text{rms}$. The median across all declination strips is {$0.96_{-0.06}^{+0.07}$} (reducing to {$0.90_{-0.05}^{+0.08}$} for $100\sigma_\text{rms}$ sources). For comparison, the {equivalent GLEAM cross-match to the unresolved RACS-SUMSS/MGPS-2 catalogue has a median of $0.99_{-0.03}^{+0.02}$ at 200 MHz (with $0.96_{-0.06}^{+0.11}$ above $100\sigma_\text{rms}$). The median ratios $<1$ are likely related to calibration errors in this case due to mismatch between sky model and true sky (i.e.\ point sources being treated as extended or vice versa).}

We attempt to correct for this by scaling the PSF (as the image restoring beam) which will leave the surface brightness in the images untouched, but allow for more accurate integrated flux density measurements. To avoid over-correcting individual snapshots, particularly those with a small {number of sources}, we opt to fit generic polynomial models to contiguous groups of ObsIDs for each {night} and for each declination strip separately. The groups are defined by ObsIDs that are not separated by more than {60} mins, and where the logarithmic ratios do not change by more than 0.05 to avoid jumps that require high-order polynomial models. This effectively creates a piece-wise polynomial model over the night. For fitting, we also use a sliding window filter across 10 ObsIDs and remove those with ratios greater than {three} times the standard deviation of the window. {We also remove snapshots from fitting if they had less than 100 sources in the original source lists prior to cross-matching.} For each group with a number of snapshots $N_\text{obs}\geq10$, we trial polynomial models up to degree $\lceil{N_\text{obs} / 3}\rceil$, using the Bayesian Information Criterion \citep{Schwarz1978} to select the model from the range of fitted models. Snapshots that do not lie within a fitted group will {take the median value for the night, or a value of 1 if no groups were fit for that particular night}, and those within the group range will take a fitted value from the selected model. These PSF correction factors, $f_\text{PSF}$, are applied to the FWHM of both the PSF major and minor axes as {$\theta_\text{corrected} = \theta \sqrt{f_\text{PSF}}$}. Figure~\ref{fig:corrections:ip} shows two example nights of $S_\text{int}/S_\text{peak}$ highlighting the results of the piece-wise polynomial fitting.

{Due to the lack of coverage of the RACS-SUMSS/MGPS-2 catalogue, we also cross-match the $\delta_\text{J2000} = +32.0\degr$ grating lobe strip to the NVSS and check the results against the $\delta_\text{J2000} = -26.7\degr$ mainlobe strip. We find similar ratios of $S_\text{int}/S_\text{peak}$ within the standard deviation for the ObsIDs, but with more significant scatter. For this reason, we assume the $\delta_\text{J2000} = +32.0\degr$ grating lobe declination strip mirrors the main lobe in terms of the resultant PSF correction and use the corrections derived from the mainlobe images.} {After application, the median ratio for unresolved sources across all snapshots is {$1.02_{-0.03}^{+0.04}$ ($0.98_{-0.02}^{+0.02}$} for $100\sigma_\text{rms}$ sources), bringing the 300 MHz data closer to the expected ratio of 1.}

\subsubsection{Snapshot brightness scale}\label{sec:post:scale}

\begin{figure}[t]
    \centering
    \includegraphics[width=1\linewidth]{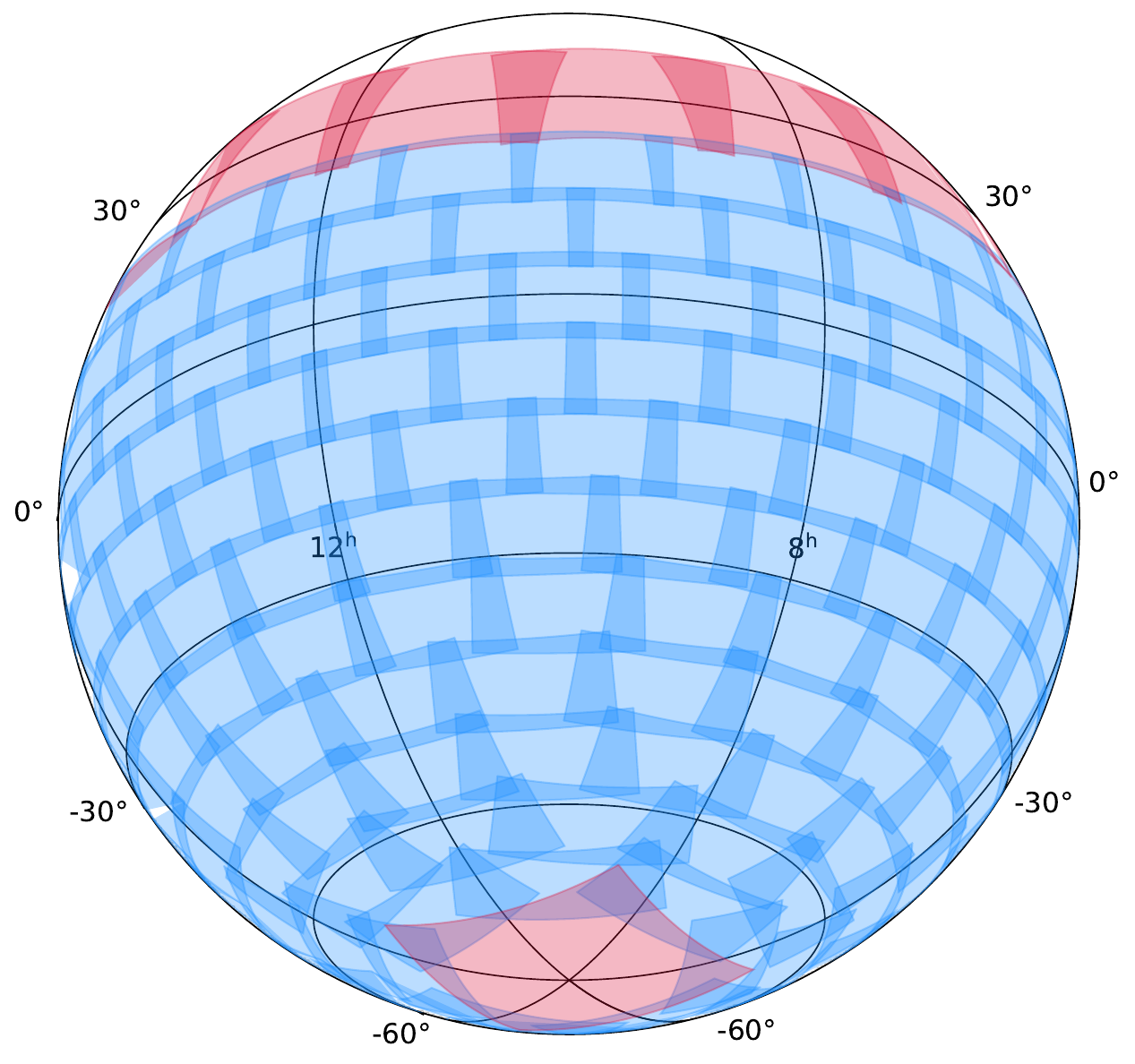}
    \caption{\label{fig:regions} Mosaic and source-finding regions in orthographic projection. {The grating lobe regions are shown in red (SCP cap region and high-declination regions).} Only regions on the front side of the sphere are shown.}
\end{figure}

\begin{figure*}[t]
    \centering
    \includegraphics[width=1\linewidth]{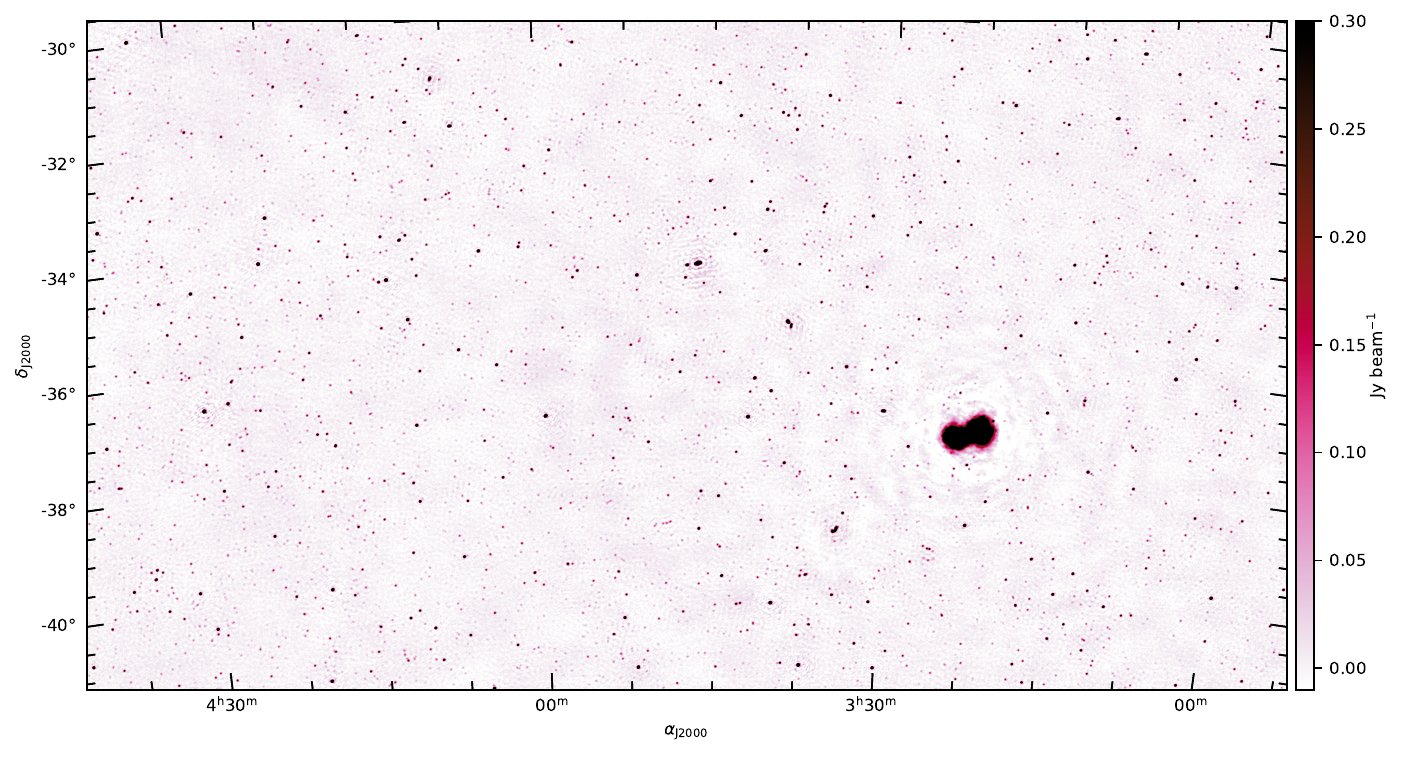}
    \caption{\label{fig:example:fornaxA} Example mosaic covering {$21.0\degr \times 11.7\degr$} containing the radio galaxy Fornax~A. The image represents the region of sky close to lowest rms noise ({$5.6_{-0.6}^{+0.8}$}\,mJy\,beam$^{-1}$) in the survey, partly due to the strong and well-modelled in-field calibrator.}
\end{figure*}

\begin{figure*}[p]
    \centering
    \includegraphics[width=1\linewidth]{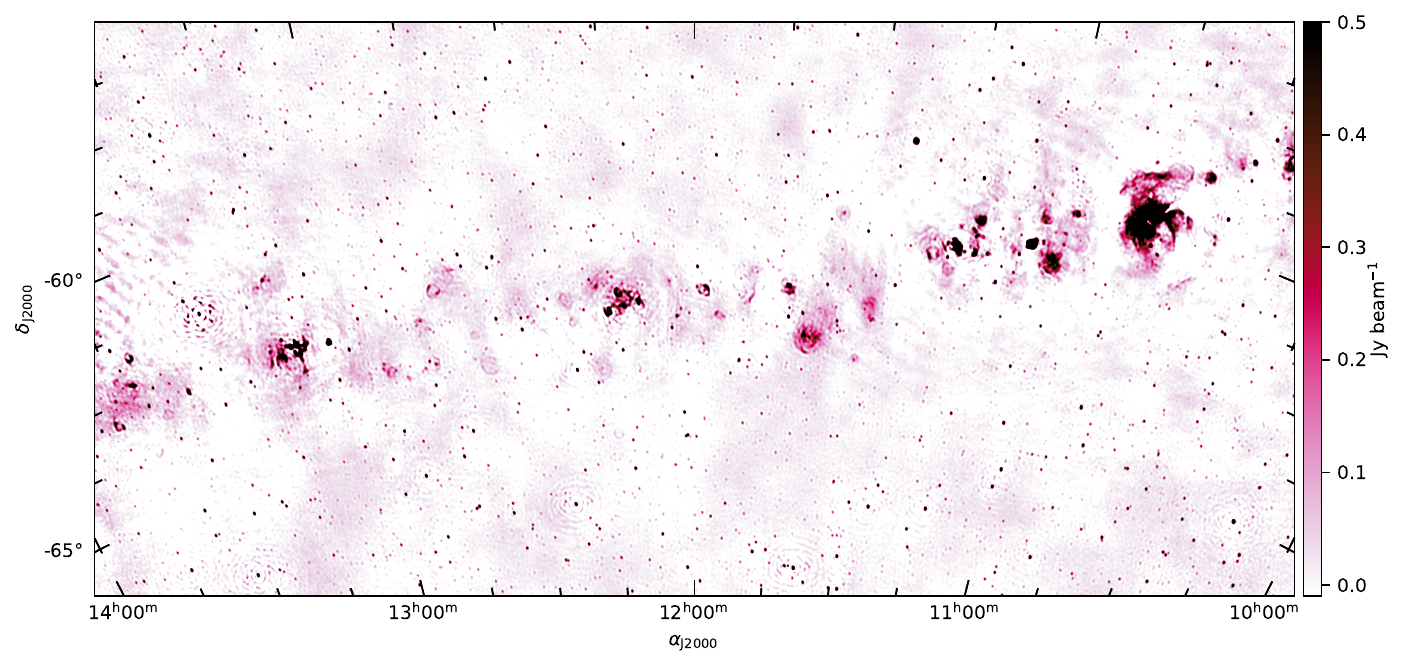}
    \caption{\label{fig:example:gp} Example mosaic covering {$24.5\degr \times 11.7\degr$} covering Galactic longitudes $285\degr \lesssim l \lesssim 308\degr$, showing the background ripples due to the extended sources. The median rms noise of the {mosaic} is {$14.1_{-3.1}^{+10.9}$}\,mJy\,beam$^{-1}$.}
\end{figure*}

\begin{figure*}[p]
\centering
\includegraphics[width=1\linewidth]{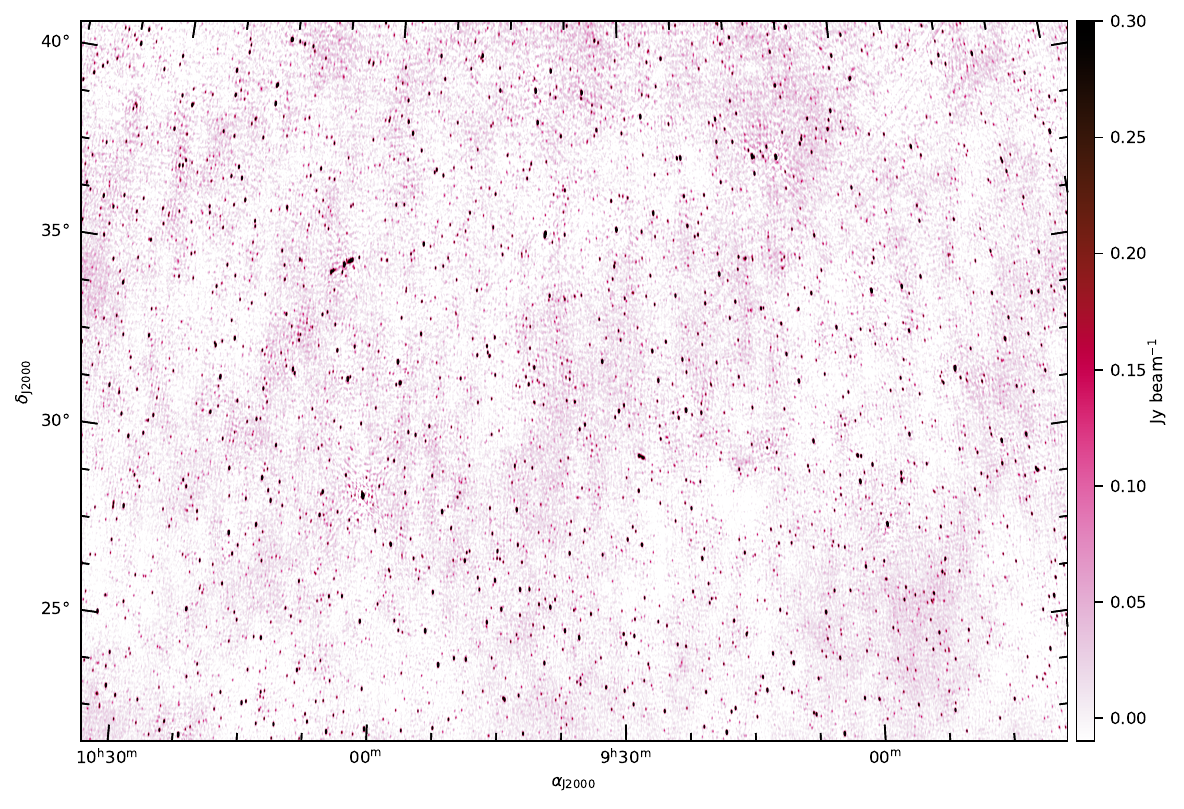}
\caption{\label{fig:example:highdec} Example mosaic at high declination covering {$26.3\degr \times 19.2\degr$}, representing the lowest noise ({$16.5_{-2.7}^{+3.7}$}\,mJy\,beam$^{-1}$) for the northern part of the survey. }
\end{figure*}

\begin{figure*}[t]
    \centering
    \includegraphics[width=1\linewidth]{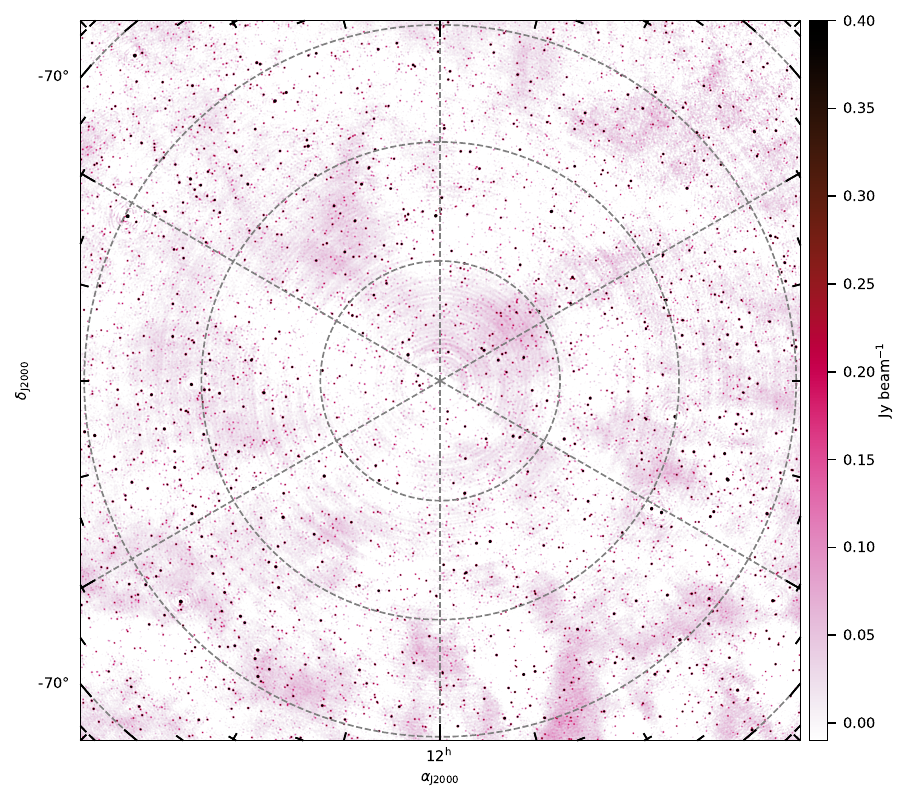}
    \caption{\label{fig:example:scp} {The full SCP mosaic, with median noise $12.2_{-1.9}^{+4.0}$\,mJy\,beam$^{-1}$. The colourscale stretch is chosen to highlight the large-scale background features.}}
\end{figure*}

After application of the primary beam model, we also compare the measured flux densities of sources in the 300 MHz images compared to {the {sky model} to check the overall brightness scale of the snapshot images. To compare, we {cross-match sources detected above $10\sigma_\text{rms}$ and with $S_\text{int}/S_\text{peak} < 1.2$ in the 300-MHz snapshots to the sky model and scale the sky model flux densities to 300\,MHz using the reported spectral indices.} This is done after application of the PSF correction factors. We noticed residual variation to the brightness scale as a function of ObsID. Figure~\ref{fig:night:fluxratio} shows the {SNR-weighted mean} measured to model flux density ratio ($S_\text{300 MHz} / S_\text{model}$) per ObsID as a function of time for the declination $-55.0\degr$ strip. We also see variation as a function of time/ObsID for all declination strips, with some link to specific calibration solutions (i.e.\ the ratio between observations that share a set of solutions shows less variation generally). While not shown in the {SNR-weighted mean} values, we also see no significant variation in brightness scale as function of position across the images, but the residual variation as a function of ObsID is likely related {to} variation in the local sky model for each ObsID and how nearest-in-time `good' solutions are applied to observations. 

To correct this residual brightness scale variation, we perform a similar polynomial fitting procedure for the ratio $S_\text{300 MHz} / S_\text{model}$, using the SNR-weighted mean ratio per snapshot, with model-fitting weights determined by the sum of weights from flux density measurement uncertainties at 300 MHz. We restrict the ratios to 0.1--10, with a slightly larger logarithmic jump of 0.1 to determine where fitting group boundaries are defined. After obtaining the models, we apply the brightness scale correction factor, $f_\text{scale}$, to snapshots {by evaluating the models for snapshots within a group that was fit. For snapshots that were not part of a fitted group, we take the measured SNR-weighted mean ratio as $f_\text{scale}$ if over 25 sources were in the cross-match, otherwise snapshots not in a group and with few cross-matches are discarded}. Figure~\ref{fig:brightness_scaling} shows the SNR-weighted mean flux density ratios along with the fitted models for two {example} nights from the declination $-72.0^\circ$ strip (top panel) the declination $-19.9^\circ$ strip (bottom panel), highlighting the general shape and variance of the polynomial models.

{The GLEAM flux density scale uncertainty is reported to be 80\% for $\delta_\text{J2000} < -83.5\degr$ and our sky model makes use of NVSS-VLSSr sources above $\delta_\text{J2000} \gtrsim +30\degr$ so it is important to check the validity of the brightness scaling process for the grating lobe images in those regions. We compare the flux density ratios calculated for the $\delta_\text{J2000} -86.0\degr$ and $+32.0\degr$ images, which are derived from the same observations and calibration solutions. 

Figure~\ref{fig:brightness_scaling:lobes} shows the flux density ratios of the ObsIDs with the sky model in the far grating lobe images (top panel $\delta_\text{J2000} = +32.0\degr$ and bottom panel  $\delta_\text{J2000} = -86.0\degr$) along with the difference in the ratios for each ObsID (middle panel). The general form of the flux density ratios is similar between the two declination strips except around complex regions (e.g.\ ObsIDs where the main lobe covers the Galactic Centre), but with some residual small-scale features. The standard deviation of the difference between the ObsIDs is $\approx 16\%$, which we take as an intrinsic brightness scale error for these regions (see Section~\ref{sec:brightness_scale} for details on the final brightness scale uncertainty). This reduces to $\approx 6$\% after application of $f_\text{PSF}$ and $f_\text{scale}$. We note that comparisons of the far grating lobe data with the main lobe data for those ObsIDs returns almost identical standard deviations, though with an additional time-independent offset likely related to the primary beam models which is implicitly corrected for in this post-image scaling.}

\section{Stacked images}\label{sec:mosaic}

\subsection{Mosaics}

After post-imaging corrections, snapshots are combined via linear mosaicking to form full-sensitivity images at 300 MHz. Because of the smaller FoV at 300 MHz, observations typically have less overlap compared to the GLEAM/GLEAM-X observations in the lower part of the MWA band even with the additional declination strips. A smaller FoV allows us to mosaic smaller regions than the full declination strip mosaics used for GLEAM, though there is still some redundancy in the mosaic processing. Therefore, we follow a similar mosaicking process used for the RACS-low \citep{racs2} to create full-sensitivity images in regions covering \begin{equation} f_\text{pad}[15\degr \times {9}\degr] \, .
\end{equation}
The regions are placed {with equal spacing between $-72\degr \leq \delta_\text{J2000} \leq +18\degr$}, with dimensions in $\alpha_\text{J2000}$ scaled by $\cos \delta_\text{J2000}$. These regions are shown in blue on Figure~\ref{fig:regions}. We include a padding factor, $f_\text{pad} = {1.2}$, in the region size to ensure overlap between the regions so that extended sources are less likely to be cut off during source-finding. 

{We include additional regions covering the SCP and high-declination grating lobe images (highlighted red in Figure~\ref{fig:regions}). The SCP mosaic is $30 \times 30$\,deg$^2$ and the large high-declination mosaic regions are centered on $\delta_\text{J2000} = +32.0\degr$, with dimensions of $26.3 \times 19.2$\,deg$^2$, following a similar padding and spacing as the main lobe regions.}   

There are a total of {217 mosaic regions, covering $-90\degr \leq \delta_\text{J2000} \lesssim +40\degr$.} These regions are formed from the linear mosaic of all snapshots with image centres within 5 degrees of region boundary or {with centres within the region boundary}. {Due to the generally lower quality of the grating lobe images, we only use the grating lobe images in the specific grating lobe regions.} This results in {87--461 snapshots per non-SCP region, and {2\,046} for the SCP region (which draws also from the $\delta_\text{J2000} = -72.0\degr$ snapshots)}. We use a position-dependent rms map as weights to stack the snapshots via a weighted average. The rms map is generated by \texttt{BANE}, where we increase the grid and box size (100 and 500 pixels) used for calculating the rms to ensure a smoothly varying rms estimate that largely follows the primary beam sensitivity.

Prior to forming the individual mosaics, we also convolve all snapshots within a mosaic group to the lowest common angular resolution of that particular group. We make use of the convolution tools from \texttt{RACS-tools} \footnote{\url{https://github.com/AlecThomson/RACS-tools}.} which makes use of the \texttt{common\_beam} function from the \texttt{radio-beam} Python package \footnote{\url{https://github.com/radio-astro-tools/radio-beam}.} to find the smallest PSF that the group can be deconvolved by, ensuring all snapshots can be convolved to a matching resolution. To ensure that the resolution does not become too degraded, we remove snapshots that have a PSF major axis ($\theta_\text{major}$) that is $> \overline{\theta_\text{major}} + 2\sigma_{\theta_\text{major}}$ where  $\overline{\theta_\text{major}}$ is the mean PSF major axis of the snapshot group and $\sigma_{\theta_\text{major}}$ is the standard deviation. For the SCP, we restrict this further to a {$0.5\sigma_{\theta_\text{major}}$ variation to avoid an overly-large PSF in some of the SCP images and because there is a significant amount of overlap in images for this region.} In addition to removal of snapshots with large PSFs, we also filter snapshots based on the median root-mean-square noise and we remove {99} snapshots where the Moon was present within the image boundaries. 

Figure~\ref{fig:example:fornaxA} shows the example mosaic containing the bright radio galaxy Fornax~A. That particular mosaic presents a `best-case' for imaging quality as Fornax~A provides a strong model for in-field calibration and most snapshot images in this region arrive at good solutions. {The median rms noise for the mosaic in Figure~\ref{fig:example:fornaxA} is {$5.6_{-0.6}^{+0.8}$}\,mJy\,beam$^{-1}$. Figure~\ref{fig:example:gp} highlights the imaging quality in the Galactic Plane, showing a mosaic covering the plane and Galactic longitudes of $285^\circ \lesssim l \lesssim 308^\circ$. The median noise in this region is higher than the Fornax~A example, at {$14.1_{-3.1}^{+10.9}$}\,mJy\,beam$^{-1}$, owing to the number of complex and bright extended sources that are not as well-modelled or well-calibrated as Fornax~A.} 

{A region with low rms noise at high-declination is shown in Figure~\ref{fig:example:highdec}---a median rms noise of $16.5_{-2.7}^{+3.7}$\,mJy\,beam$^{-1}$ is a best-case scenario for the most northern part of the survey. High-declination mosaics are constructed from low-elevation pointings {and grating lobes} which have lower sensitivity. Finally, Figure~\ref{fig:example:scp} shows the full mosaic covering the SCP. Due to the significant overlap in images here the median rms noise is $12.2_{-1.9}^{+4.0}$\,mJy\,beam$^{-1}$. This region features noticeable ripples radiating from the SCP, likely residual RFI in the declination $-26.7\degr$ scans. The same ripples can be seen in the mainlobe images, though at a much lower level. In addition, all of the grating lobe regions (SCP and high-declination) tend to feature more significant large-scale background undulations than elsewhere.} 

\subsection{Source finding and source lists}

Once mosaics are made, we use \texttt{aegean} to generate lists of 2-D Gaussian components. We use \texttt{BANE} again to create a position-dependent rms noise and background map for each mosaic, used by \texttt{aegean} for source-finding thresholds and measurements. For the mosaics, we use a $4\sigma_\text{rms}$ threshold for source detection (the `\texttt{seedclip}' parameter in \texttt{aegean}), and $3\sigma_\text{rms}$ for growing source detections (`\texttt{floodclip}'). This process is consistent with source-finding from other GLEAM data releases, focusing on compact sources. While our initial source-finding threshold is {$4\sigma_\text{rms}$ we remove any} components with a peak flux density $<5\sigma_\text{rms}$ from  each source list. In addition, we remove sources with reported integrated flux density uncertainties greater than 100\% of the measured value, with $\approx 1$--20 sources removed from a median $\approx 2\,500$ sources per mosaic. {For example, the mosaic region with Fornax~A in Figure~\ref{fig:example:fornaxA} contains {3\,861} components above $5\sigma_\text{rms}$.}

\subsection{Completeness}\label{sec:completeness}

\begin{figure}[t]
    \centering
    \begin{subfigure}[b]{1\linewidth}
    \includegraphics[width=1\linewidth]{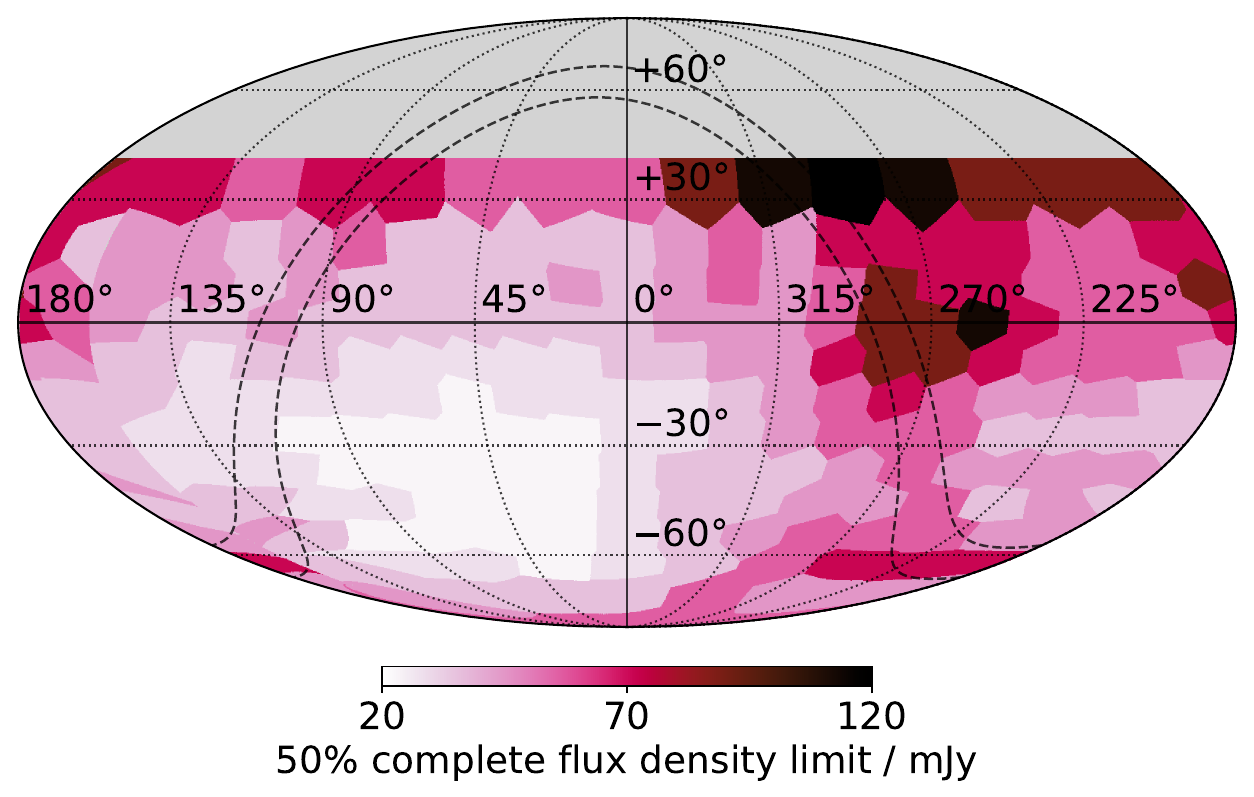}
    \end{subfigure}\\%
    \begin{subfigure}[b]{1\linewidth}
    \includegraphics[width=1\linewidth]{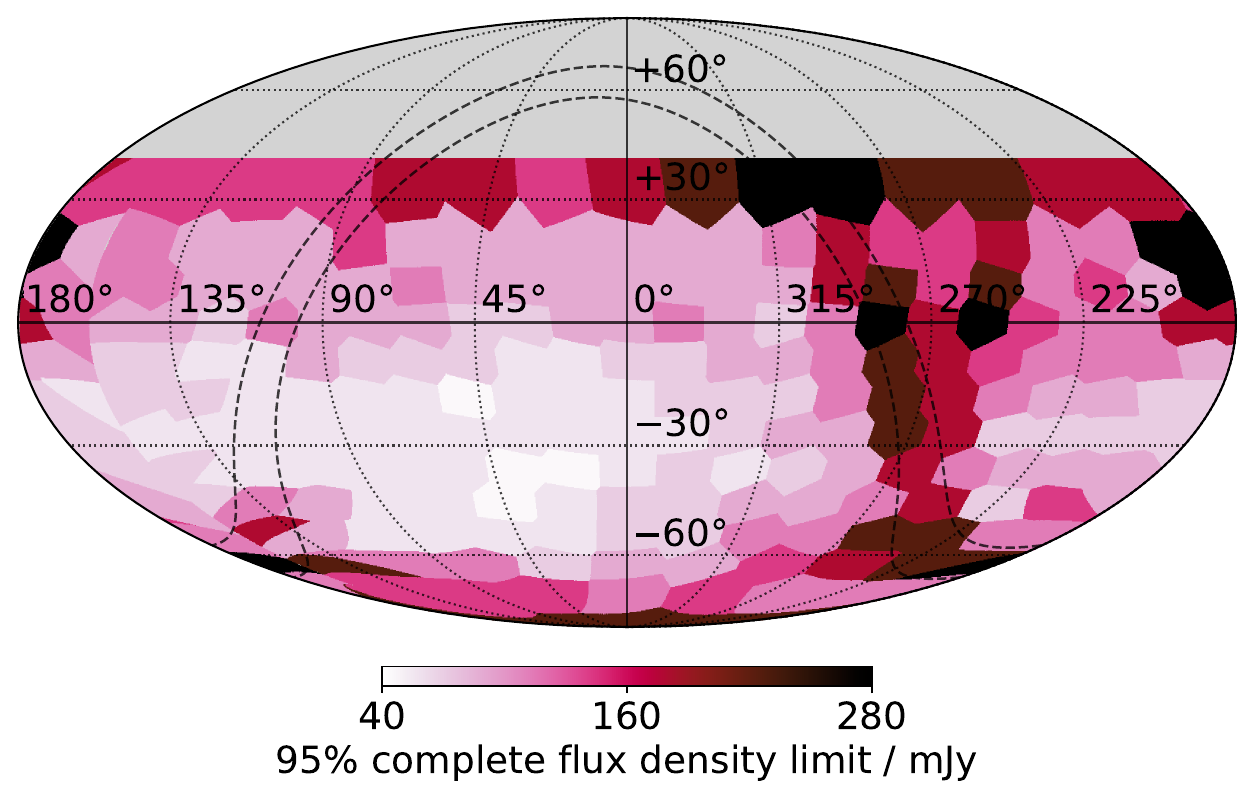}
    \end{subfigure}
    \caption{{\label{fig:compl} 50\% (\emph{top}) and 95\% (\emph{bottom}) flux density completeness limits across the survey. The limits are represented by an average over each mosaic region, with nearest-neighbour interpolation. Note each panel uses a different colourscale.}}
\end{figure}

{Due to difference in image properties, we investigate how complete the source lists for each mosaic region are expected to be prior to combining them into a single catalogue. We follow the process used for other GLEAM data releases \citep[][but see also \citealt{Franzen2021a,HurleyWalker2022,gleamx-dr2}]{HurleyWalker2017}. We inject point sources onto our images using \texttt{AeRes} \citep{Hancock2012,Hancock2018} with pseudo-random positions (requiring no neighbours within 4~arcmin) and for multiple realisations at a range of logarithmically-spaced flux densities from 0.012--1\,Jy. The number of sources varies per mosaic region, corresponding to $\approx 10$ sources per square degree matching the overall source density in the final catalogue (Section~\ref{sec:catalogue}). The completeness is defined by the number of injected sources that are recovered at each injected flux density level after repeating the same source-finding process used for the normal images. We note that this method largely assesses the source-finding approach and does not take into account how other instrumental effects such as $(u,v)$ coverage would affect recovery of sources.}

{The median flux density limit at 50\% completeness is $45_{-16}^{+12}$\,mJy (ranging from 21--141 mJy) and at 95\% is $89_{-33}^{+89}$ (ranging from 45--891\,mJy) across the mosaic regions. We show the 50\% and 95\% complete flux density limits as a function of position in Figure~\ref{fig:compl}. The data are shown as a average for each tile, with nearest-neighbour interpolation to generate a contiguous map. Generally, completeness decreases in the Galactic Plane where large, extended Galactic radio emission is present, and around areas with significant artefacts following positional variation of the rms noise across the survey (see Section~\ref{sec:noise}). Completeness metrics are included in the FITS headers for each mosaic image.}

\section{The 300-MHz catalogue}\label{sec:catalogue}

\begin{figure}[t]
    \centering
    \includegraphics[width=1\linewidth]{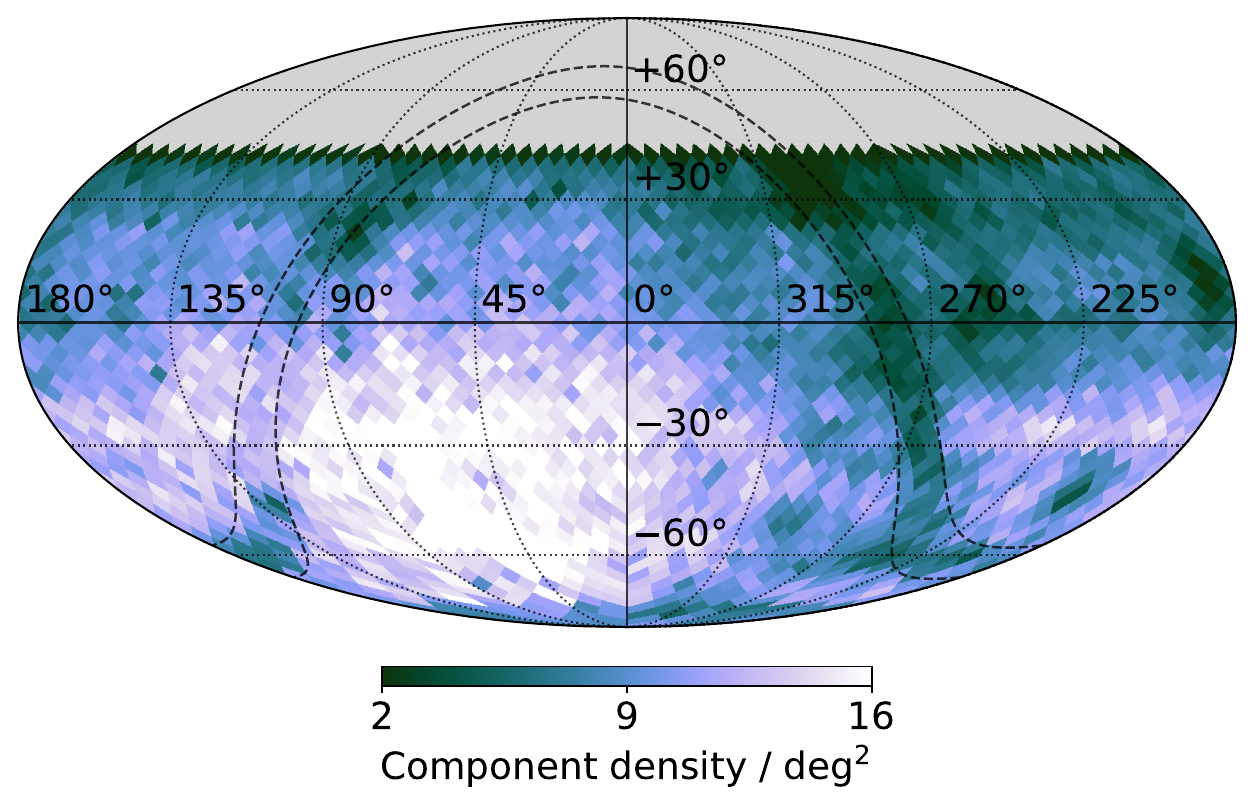}
    \caption{\label{fig:density} The density of Gaussian components across the survey in HEALPix bins of {$13.4$}\,deg$^2$.}
\end{figure}

\subsection{Merging source lists}

The mosaic regions have a significant overlap by construction so any combined source catalogue needs to have duplicate measurements from sources detected across adjacent mosaics removed. We follow the method of \citet{racs-mid2} and construct the full catalogue by concatenating the source lists one-by-one. As each source list is added, we cross-match the incoming source list to the partially constructed catalogue with a separation of $0.5 \times \langle\Theta_\text{major}\rangle$, where $\Theta_\text{major}$ is the fitted major axis FWHM of the Gaussian component. Any sources cross matched within half their averaged size are considered duplicate measurements of the same source. For duplicated measurements, we opt to keep the measurement from the mosaic lowest rms noise at the location of the source. The resultant merged catalogue contains \nsources\ components, with a median density of $\approx 10$\,deg$^{-2}$. Figure~\ref{fig:density} shows the component density across the whole survey using HEALPix\,\footnote{Hierarchical Equal Area isoLatitude Pixelation \citep{Gorski2005}.} binning, generally highlighting regions of lower/higher sensitivity (see Section~\ref{sec:noise}). 

\subsection{Catalogue columns}\label{sec:columns}

The information included in the catalogue is similar to a single frequency from GLEAM-X, with component coordinates, measured sizes and flux densities, and additional metadata useful for further measurements or comparisons to other catalogues. The columns are summarised as:

\begin{enumerate}
    \item \texttt{name}: Component name following IAU convention: GLEAM-300 JHHMMSS$\pm$DDMMSS. 
    \item \texttt{RAJ2000}: J2000 right ascension of the component in decimal degrees.
    \item \texttt{err\_RAJ2000}: Uncertainty on right ascension from fitting the component position in decimal degrees.
    \item \texttt{DEJ2000}: J2000 declination of the component in decimal degrees.
    \item \texttt{err\_DEJ2000}: Uncertainty in declination from fitting the component position in decimal degrees.
    \item \texttt{local\_rms}: Estimate of the local root-mean-square noise in Jy\,beam$^{-1}$. 
    \item \texttt{background}: Estimate of the local background in Jy\,beam$^{-1}$.
    \item \texttt{peak\_flux}: Peak flux density of the component in Jy\,beam$^{-1}$.
    \item \texttt{err\_peak\_flux}: Uncertainty in the peak flux density in Jy\,beam$^{-1}$.
    \item \texttt{int\_flux}: Integrated flux density of the component in Jy.
    \item \texttt{err\_int\_flux}: Uncertainty in the integrated flux density of the component in Jy.
    \item \texttt{a}: FWHM of the major axis of the component in arcsec.
    \item \texttt{err\_a}: Uncertainty in the FWHM of the major axis in arcsec.
    \item \texttt{b}: FWHM of the minor axis of the component in arcsec.
    \item \texttt{err\_b}: Uncertainty in the FWHM of the minor axis in arcsec.
    \item \texttt{pa}: Position angle of the component in degrees.
    \item \texttt{err\_pa}: Uncertainty in the position angle in degrees.
    \item \texttt{bmaj}: Local major axis of the beam in arcsec.
    \item \texttt{bmin}: Local minor axis of the beam in arcsec.
    \item \texttt{bpa}: Local position angle of the beam in degrees.
    \item \texttt{flags}: \texttt{Aegean} fitting flags.
    \item \texttt{flux\_scale\_err}: {Declination-dependent brightness scale fractional uncertainty (Section~\ref{sec:brightness_scale})}.
\end{enumerate}

\subsection{Noise}\label{sec:noise}

\begin{figure*}[t]
    \centering
    \includegraphics[width=1\linewidth]{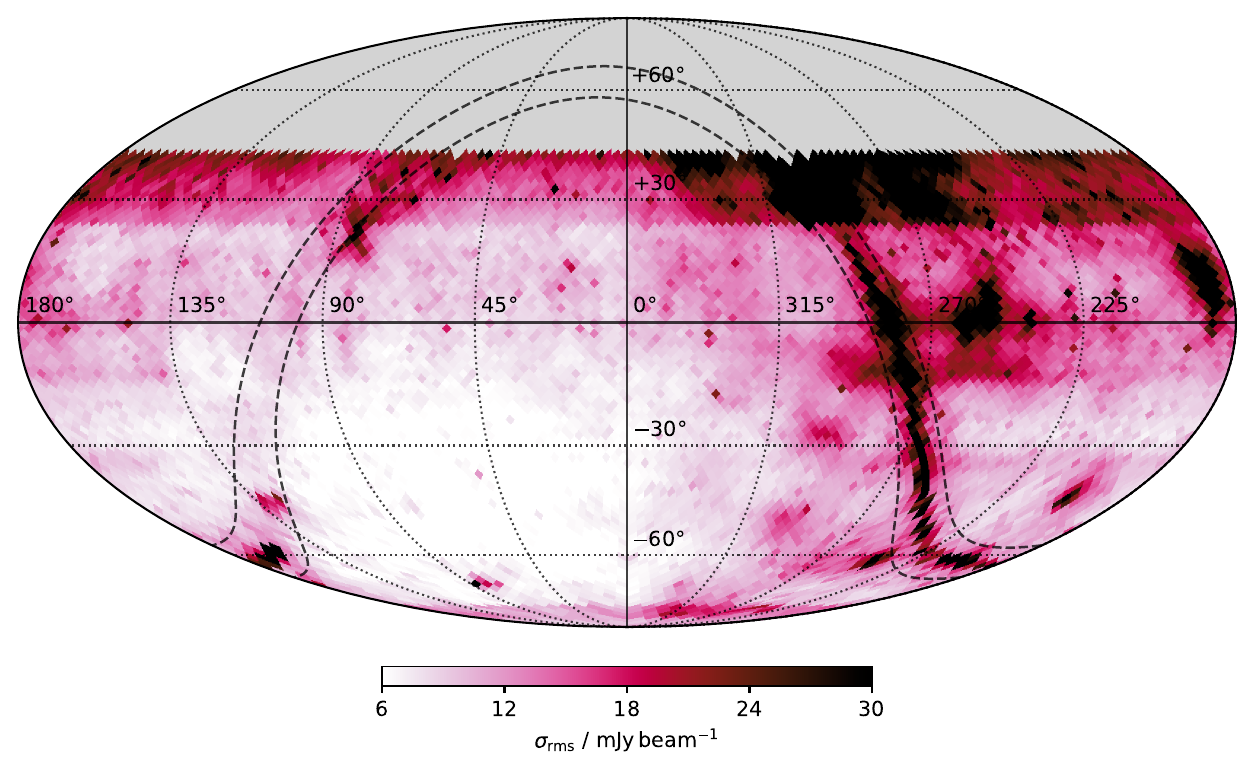}
    \caption{\label{fig:rms} HEALPix representation of the local rms noise at the position of sources in the constructed catalogue.}
\end{figure*}

\begin{figure}[t]
\includegraphics[width=1\linewidth]{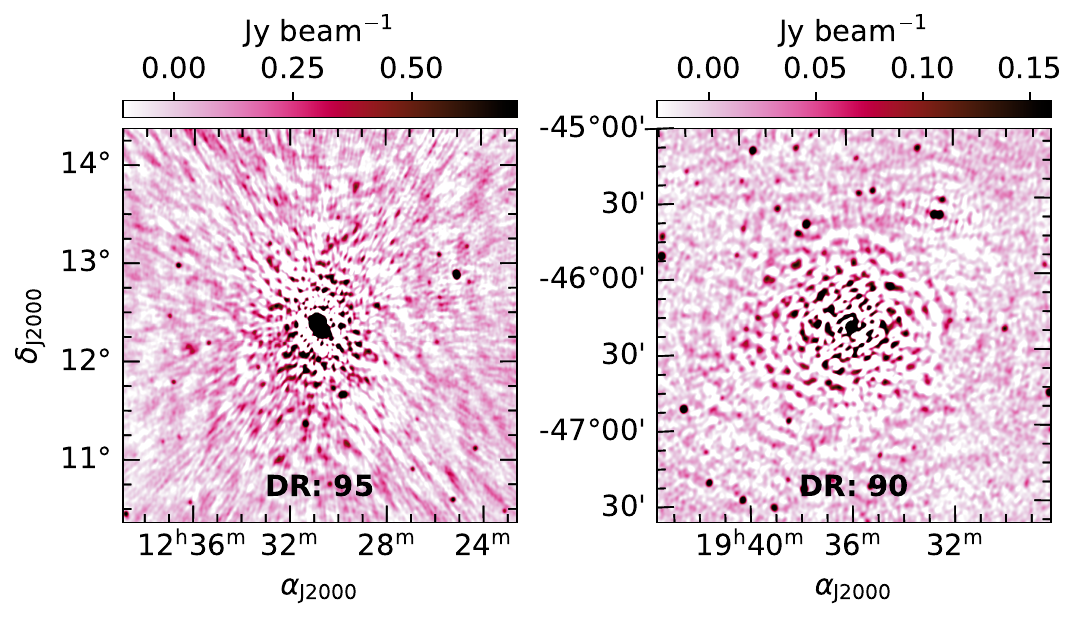}
\caption{\label{fig:artefacts} Common artefacts around bright sources for declination $\gtrsim 0\degr$ (Virgo~A, left) and declination $\ll 0\degr$ (PKS~1932$-$46, right). The dynamic range (DR) is shown, estimated based on peak values of the sources and the most significant nearby artefacts.}
\end{figure}

\begin{figure}[t]
    \centering
    \includegraphics[width=1\linewidth]{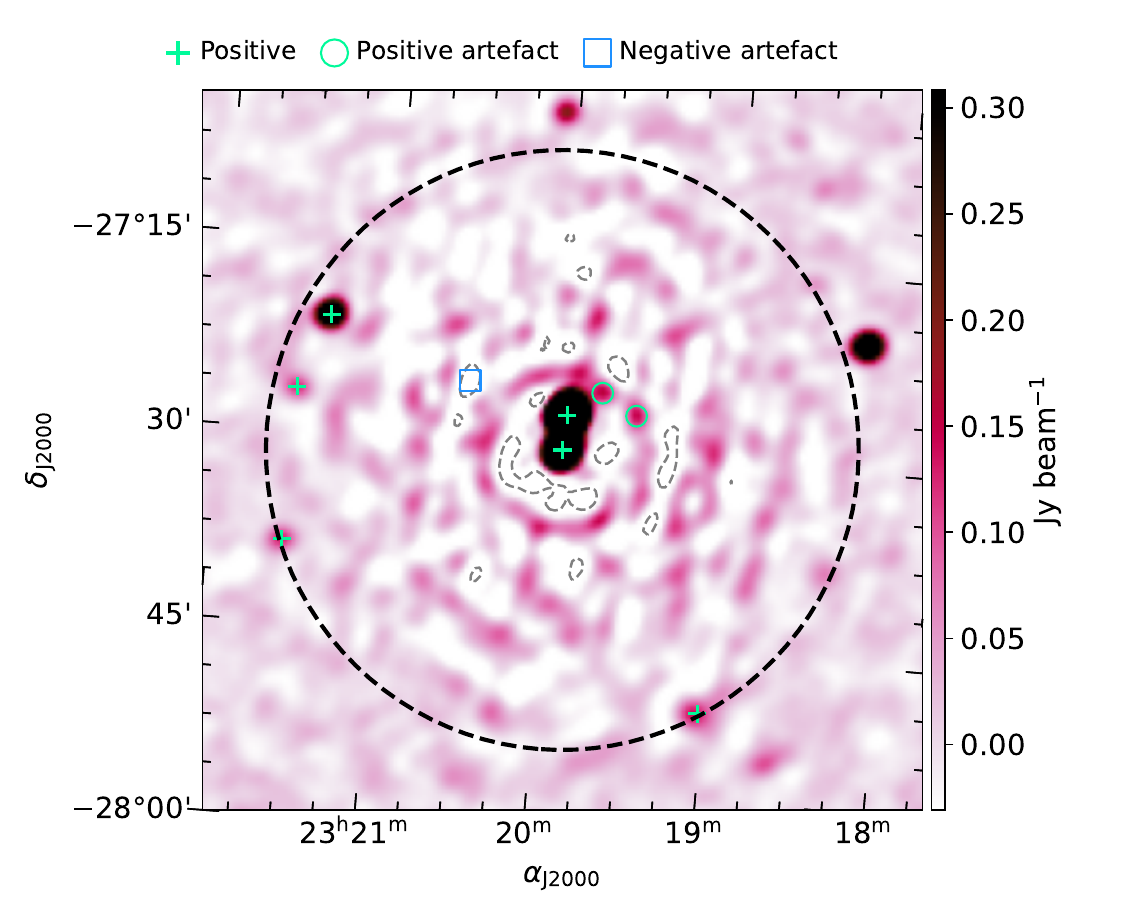}
    \caption{\label{fig:artefact_removal}Example of the artefact filtering process. The panel is centered on a {2.4}\,Jy component. Positive and negative components within the initial filter radius are shown, with markers indicating whether they are considered artefacts or not (see legend). The single dashed-grey contour is drawn at $-3\sigma_\text{rms}$.}
\end{figure}

\begin{figure}[t]
    \centering
    \includegraphics[width=1\linewidth]{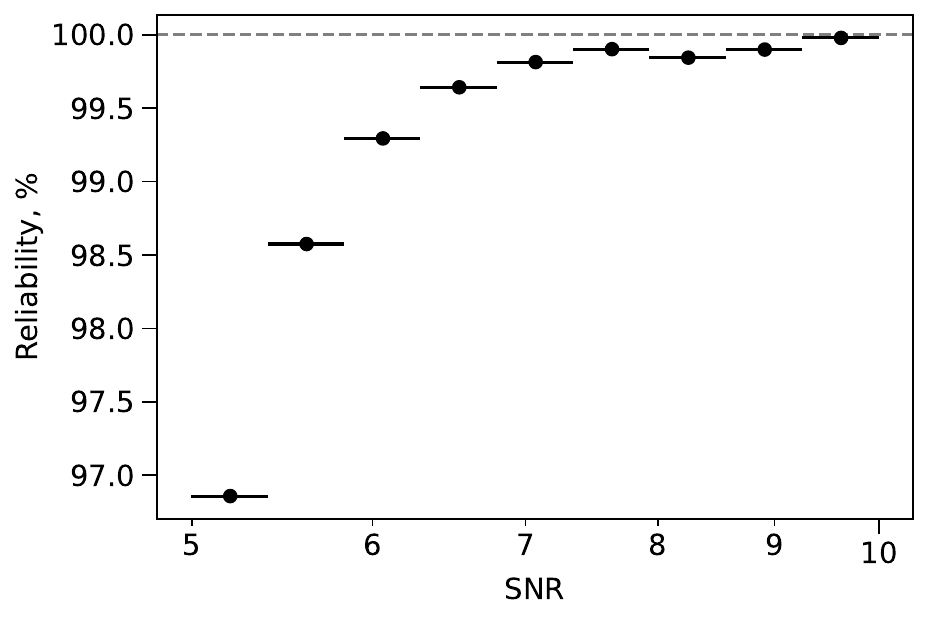}
    \caption{\label{fig:reliability} Reliability as a function of SNR. The dashed horizontal line shows 100\% reliability, and the bars on each point show the SNR bin width. Note the minimum SNR in source lists is confined to $5\sigma_\text{rms}$. }
\end{figure}


The rms noise in the final mosaic images varies across the survey, with a median value of \rmsMedian\,mJy\,beam$^{-1}$. Figure~\ref{fig:rms} shows the HEALPix-binned map of the rms noise as reported in the catalogue. The noise is raised around some bright sources and in regions of the Galactic Plane, with some additional increase in noise around $\alpha_\text{J2000} \approx 270\degr$ and $\delta_\text{J2000} \approx -13\degr$ to $\approx +1.6\degr$ due to calibration challenges in this region. {Some of the high-declination mosaics feature higher noise due to complex bright sources within the images (Cygnus~A) and within the main lobe (the Galactic Centre)---these are peeled/removed when outside of the FoV, but significant artefacts still remain in some cases.} Figure~\ref{fig:example:gp} also shows typical large-scale background features present near the extended sources in the Galactic Plane.

\subsection{Image artefacts}\label{sec:artefacts}

 In common with most all-sky surveys, the 300-MHz images feature noticeable artefacts at the ${\approx} 1$\% level, from both position-dependent and position-independent errors around bright sources. These artefacts can be mitigated with self-calibration, {and are reduced when using in-field calibration}, although with our processing strategy self-calibration is not done due to a high failure rate and in-field calibration is only done for a subset of the snapshots. Common artefacts following the general shape of the PSF around bright sources are shown in Figure~\ref{fig:artefacts}, highlighting the form of the errors for $\delta_\text{J2000} \gtrsim 0\degr$ (Virgo~A, left) and for $\delta_\text{J2000} \ll 0\degr$ (PKS~1932$-$46, right). Despite the lack of self-calibration, we are able to achieve a dynamic range of at least $\approx {90}$ with typical errors ${\approx} 1$\%. 

\subsection{Catalogue reliability}\label{sec:reliability}

A common method of determining the number of false positive detections around {bright} sources (hence the reliability of the source finding) is to repeat source-finding on the same images multiplied by $-1$ (i.e.\,the `negative' image, or `inverted' image; e.g.\,\citealt{Intema2017,racs2,HurleyWalker2022,gleamx-dr2}). Assuming the noise and artefacts are Gaussian and symmetric, a measure of source-finding reliability, $r$, is then defined as \begin{equation}
r = 100 \times \left(1 - N_\text{negative} / N_\text{positive}\right) \, \% ,
\end{equation}
for $N_\text{negative}$ sources found in the negative image and $N_\text{positive}$ sources found in the original image. 

{To help reduce the number of artefacts that are included in the catalogue, we filter catalogued Gaussian components around bright ($S_\text{bright} = S_\text{300} > 1$\,Jy) sources. This process is modified from the artefact filtering done for GLEAM-X DR1/DR2 \citep{HurleyWalker2022} with the inclusion of a simplified flux-dependent radial filter \citep[see e.g.][for a similar method]{Knowles2022}. We do this prior to merging source lists. We define a beam- and flux-dependent filter around sources that scales linearly between ${5}$--${20}\times\theta_\text{major}$ for bright sources between $1$--$5$\,Jy. Sources with flux densities $>5$\,Jy use a fixed ${20}\times\theta_\text{major}$ radius. All catalogued components within the defined radius of a bright source are considered artefacts if they are sufficiently fainter than the bright source by $S_\text{faint}\times 100 < S_\text{bright}$. In addition, if a `negative' source lies within the filter radius, we reduce the filter radius to match the maximum {angular separation} between the bright source and any negative sources. We then consider any positive source within the same absolute flux limit as the negative source to also be an artefact. Negative sources satisfying these conditions are also removed from their respective source lists. An example {2.4}-Jy source is shown in Figure~\ref{fig:artefact_removal}, with sources within the ${12.1} \times \theta_\text{major}$ radius shown, with markers indicating whether they are artefacts or not. The positive artefacts either side of the source are common source sidelobe artefacts seen in that region.}

We perform the assessment of the reliability after removing artefacts on the final catalogue and negative catalogue. The overall reliability in the catalogue is \relOverall\%, and Figure~\ref{fig:reliability} shows the reliability of all mosaics combined as a function of SNR ($S_\text{peak}/\sigma_\text{rms}$). Reliability generally increases with increasing SNR, with $\approx 99$\% reliability achieved at {$\approx 6.0\sigma_\text{rms}$}, and 100\% reliability above {$\approx 10\sigma_\text{rms}$}. We note \citet{Franzen2021a} reports an overall higher reliability for the GLEAM SGP catalogue, likely a result of self-calibration reducing artefacts around bright sources. {We note that the SCP region has comparatively higher overall reliability at \relSCP\%, while the high-declination regions more closely match the main survey at \relHighDec\%.}

\subsection{Photometry}\label{sec:catalogue:flux}

\subsubsection{The point spread function}\label{sec:catalogue:psf}

\begin{figure}[t]
    \centering
    \includegraphics[width=1\linewidth]{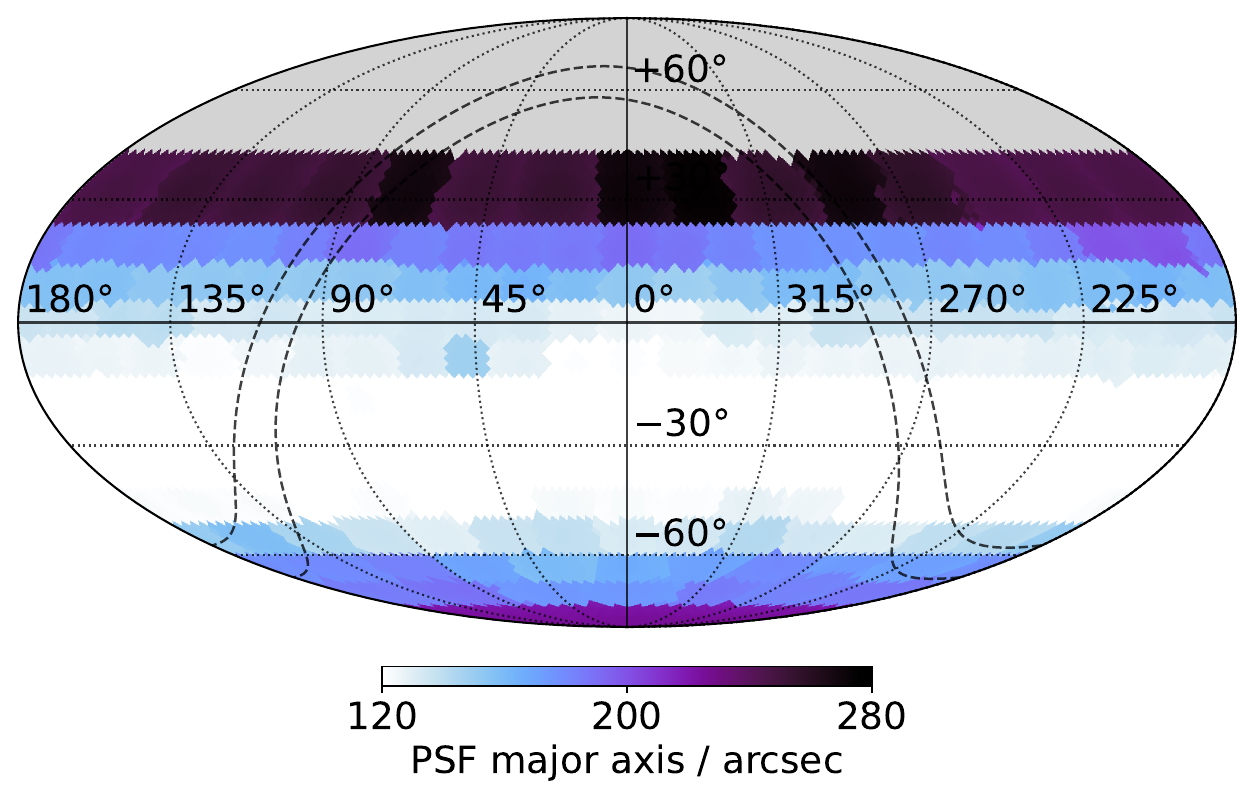}\\
    \includegraphics[width=1\linewidth]{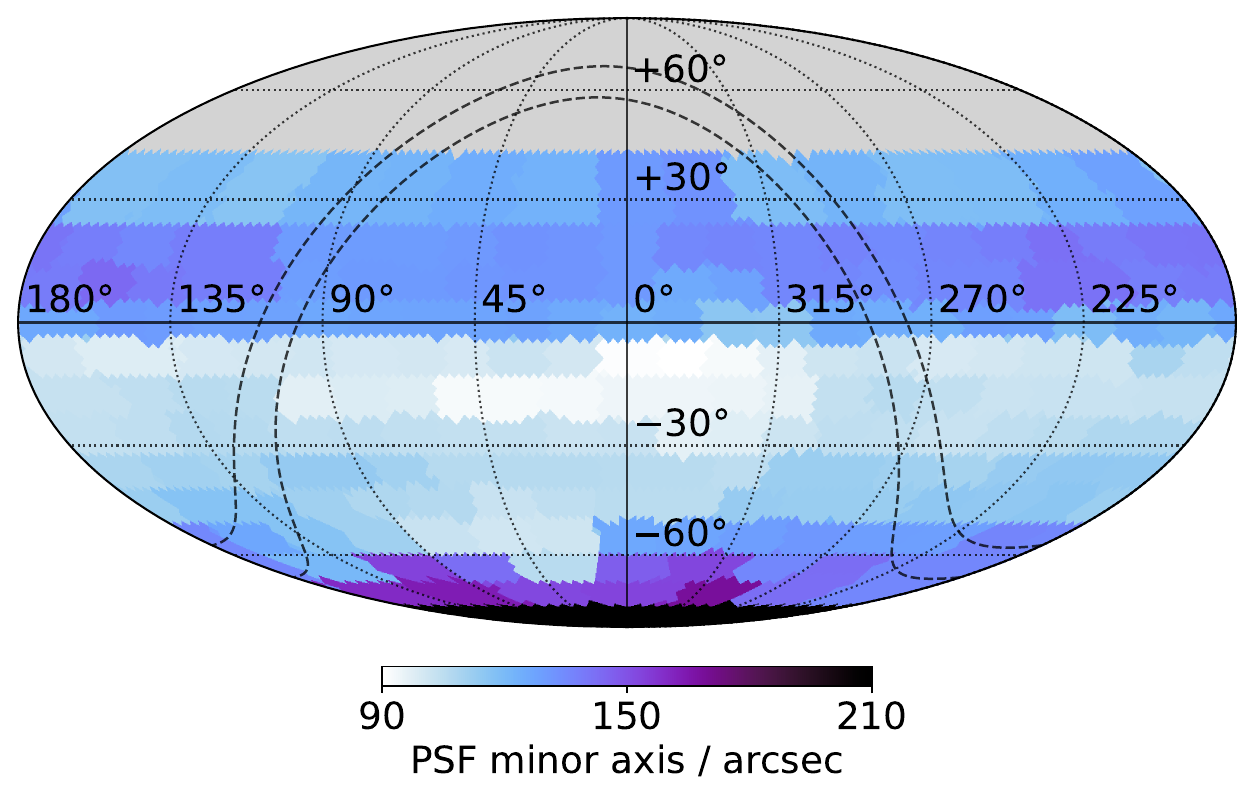}
    \caption{\label{fig:psf} HEALPix representation of the PSF major (\emph{top}) and minor (\emph{bottom}) axes across the sky as recorded at source positions in the catalogue.}
\end{figure}

\begin{figure}[t]
    \centering
    \includegraphics[width=1\linewidth]{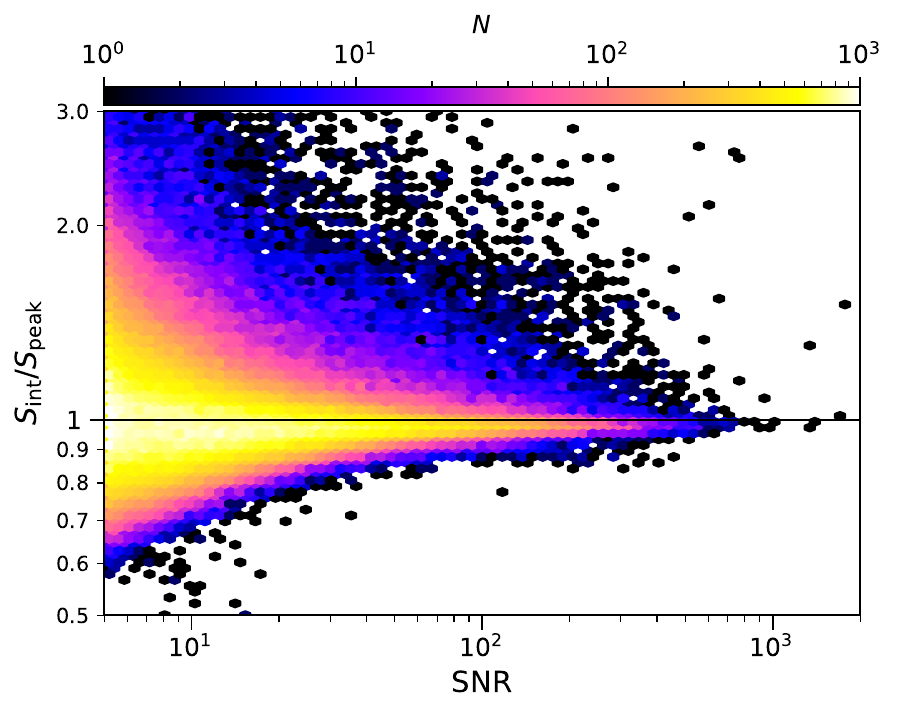}
    \caption{\label{fig:peakint} The ratio of $S_\text{int}/S_\text{peak}$ as a function of SNR for all sources in the GLEAM-300 catalogue, represented in hexagonal bins. The horizontal line indicates a ratio of 1.}
\end{figure}

As the individual mosaics have different angular resolutions, the PSF for sources in {the} catalogue also varies across the sky. Figure~\ref{fig:psf} shows the position-dependent major and minor axes of the image PSF as recorded in the catalogue using HEALPix binning. The median major and minor axes are $\psfMedianA \times \psfMedianB$, and  range from \psfMinA--\psfMaxA\ and \psfMinB--\psfMaxB, respectively. The PSF major and minor axes largely vary as a function of declination, with some additional variation around specific regions, relating to areas with either fewer snapshots, more RFI flagging, or poorer calibration. In particular, the {high-declination and SCP regions feature the largest PSFs, corresponding to the lower elevation of the grating lobe directions}.

{Figure~\ref{fig:peakint} shows the ratio of $S_\text{int}/S_\text{peak}$ as a function of source SNR for all sources in the catalogue. The overall median ratio is \intPeakAll, reducing slightly to \intPeakBright\ for sources with $\text{SNR}>100$, consistent with results from the individual snapshots prior to mosaicking (Section~\ref{sec:post:psf}).}

\subsubsection{The brightness scale}\label{sec:brightness_scale}

\begin{figure*}[p]
    \centering
    \begin{subfigure}[b]{0.5\linewidth}

        \includegraphics[width=1\linewidth]{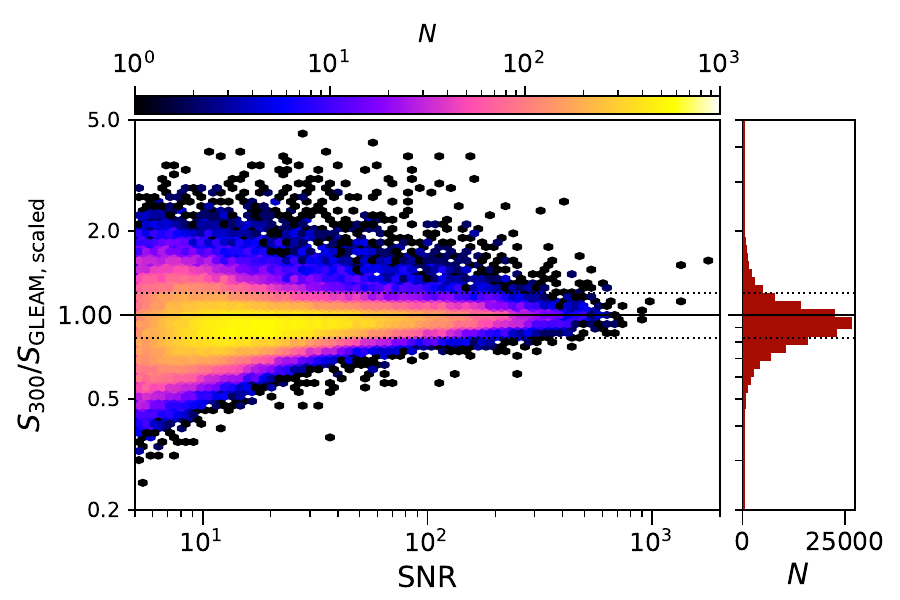}
        \caption{\label{fig:flux:snr:GLEAM} GLEAM-300 and GLEAM.}
    \end{subfigure}%
        \begin{subfigure}[b]{0.5\linewidth}
        \includegraphics[width=1\linewidth]{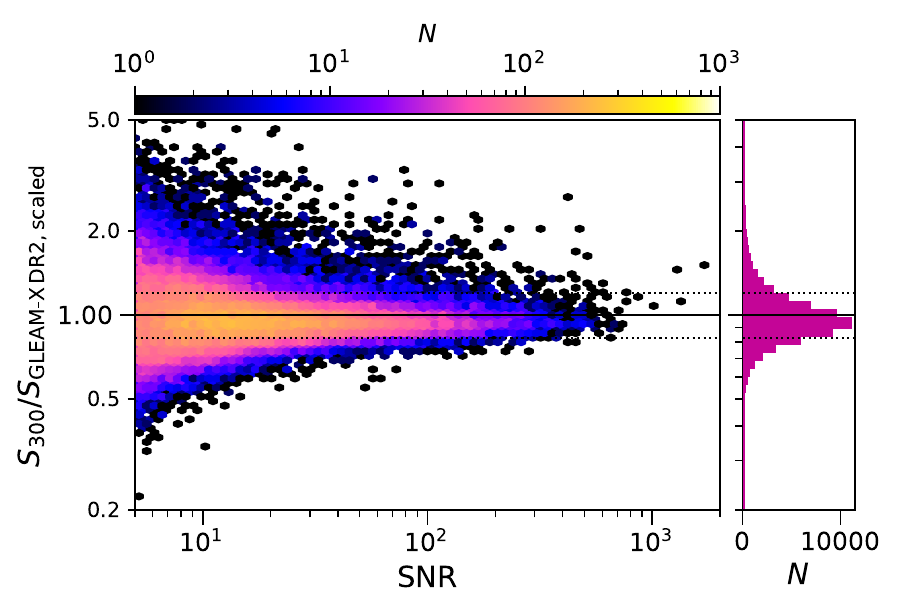}
        \caption{\label{fig:flux:snr:GLEAMX} GLEAM-300 and GLEAM-X DR2.}
    \end{subfigure}\\%
    \begin{subfigure}[b]{0.5\linewidth}
        \includegraphics[width=1\linewidth]{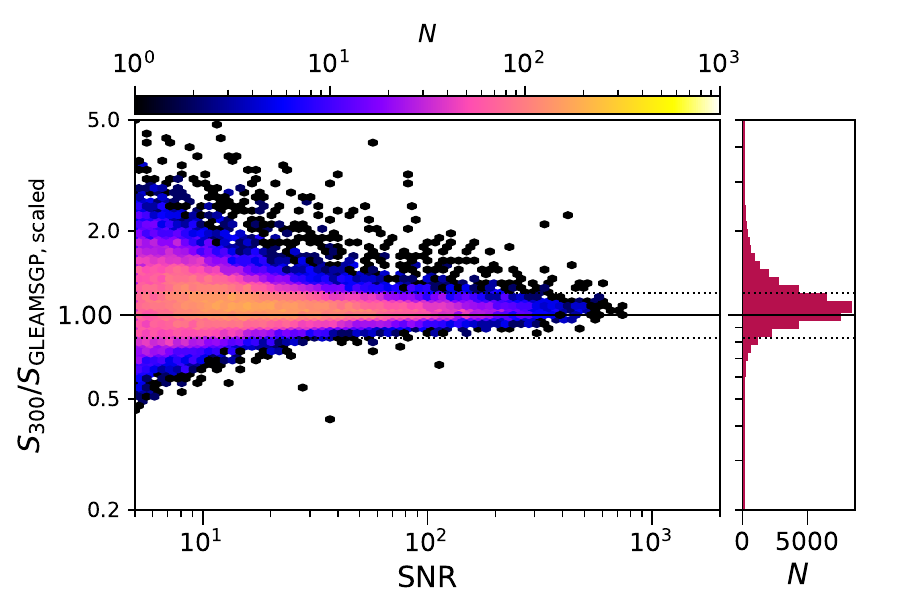}
        \caption{\label{fig:flux:snr:SGP} GLEAM-300 and GLEAM SGP.}
    \end{subfigure}%
        \begin{subfigure}[b]{0.5\linewidth}
        \includegraphics[width=1\linewidth]{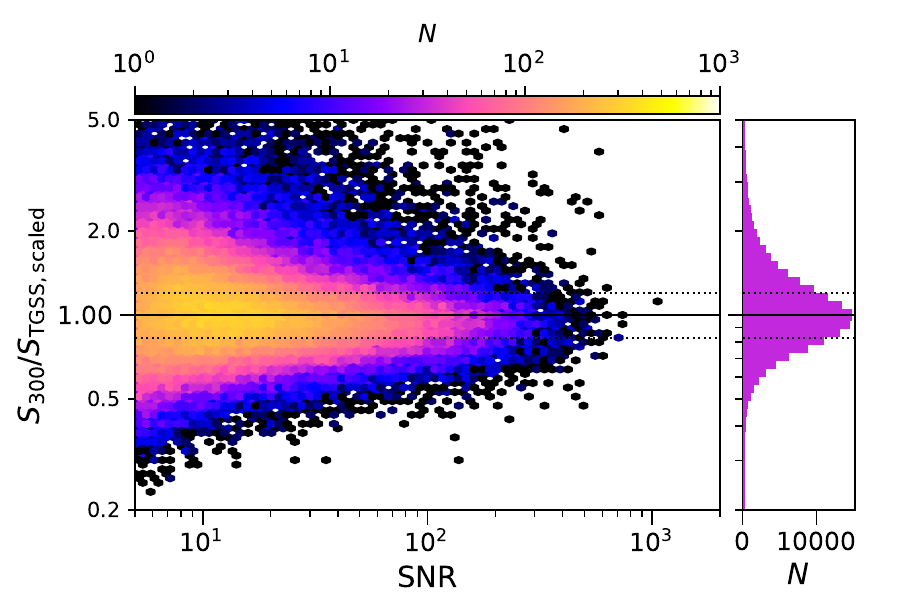}
        \caption{\label{fig:flux:snr:TXS} GLEAM-300 and the TGSS ADR1.}
    \end{subfigure}\\%
        \begin{subfigure}[b]{0.5\linewidth}\includegraphics[width=1\linewidth]{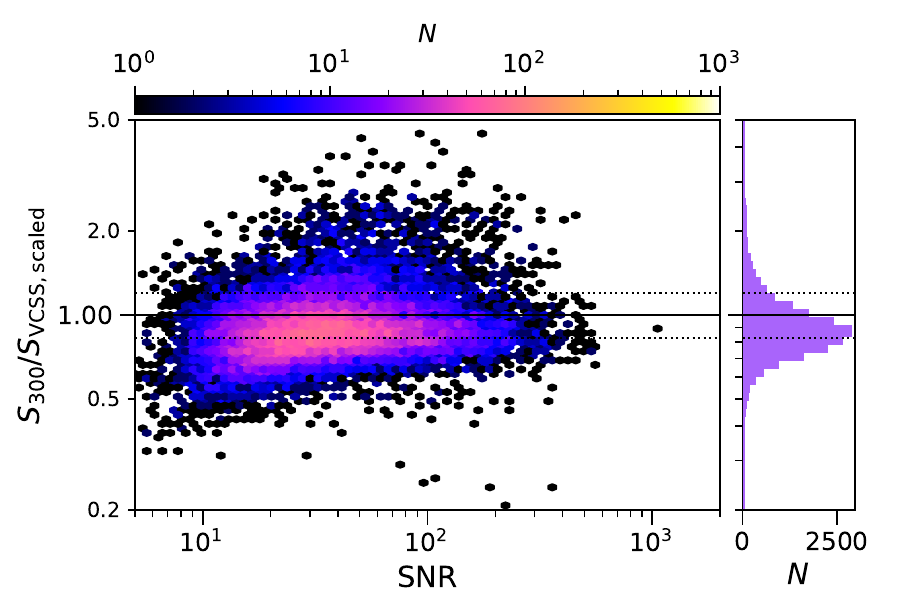}
        \caption{\label{fig:flux:snr:VCSS} GLEAM-300 and the {VCSS1}.}
    \end{subfigure}%
        \begin{subfigure}[b]{0.5\linewidth}
        \includegraphics[width=1\linewidth]{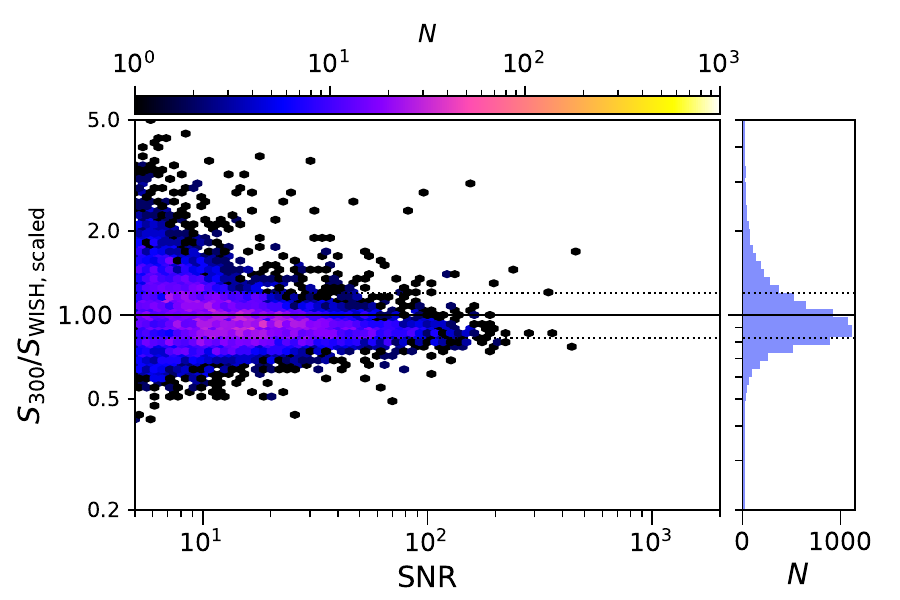}
        \caption{\label{fig:flux:snr:WISH} GLEAM-300 and WISH.}
    \end{subfigure}%

    \caption{\label{fig:flux:snr} Flux density ratios of sources cross-matched between GLEAM-300 and GLEAM \subref{fig:flux:snr:GLEAM}, GLEAM-X DR2 \subref{fig:flux:snr:GLEAMX}, GLEAM SGP \subref{fig:flux:snr:SGP}, {VCSS1} \subref{fig:flux:snr:VCSS}, WISH \subref{fig:flux:snr:WISH}, and TXS \subref{fig:flux:snr:TXS}, after scaling flux densities to 300\,MHz and correcting for Eddington bias, as a function of SNR in the GLEAM-300 catalogue. Solid horizontal line is drawn at 1, and the dashed lines are drawn at $\pm 20\%$. Note the y-axis is scaled logarithmically.}
\end{figure*}

The brightness scale is set by the input sky model we use for calibration and post-processing so we expect the final mosaics to be consistent with GLEAM and GLEAM-X by extension. To assess any residual brightness scale errors we compare the GLEAM-300 catalogue measurements to scaled flux densities from GLEAM, GLEAM-X DR2, and GLEAM SGP (scaled from 200\,MHz measurements), the {VCSS1} (scaled from 340\,MHz), WISH (scaled from 352\,MHz), {and the} TGSS~ADR1 (scaled from 147.5\,MHz). {These surveys provide data products that are the reasonably close in sensitivity, resolution, and/or frequency to GLEAM-300.} We note that GLEAM-X DR2 and GLEAM SGP inherits a flux density scale from GLEAM, and GLEAM overall is set to the \citet[][hereinafter B77]{Baars1977} flux density scale. WISH is flux-corrected to the NVSS, which is largely set to the B77 scale as well}. {For TGSS ADR1, the brightness scale is set to the low-frequency models from \citet{Scaife2012}, tied to the \citet[RCB;][]{rcb} brightness scale. There is an expected $\approx 4$\% difference between B77 and RCB at these frequencies. VCSS1 is flux-calibrated using \citet{Perley2017} models for 3C~286 and 3C~138, which again differ by a few percent from the B77 and RCB scales.}

For consistency, we take the median $\alpha$ from all cross-matched GLEAM sources ($\gleamalpha$) to scale all surveys to 300\,MHz. During cross-matching we only consider cross-matches within 30\,arcsec and only take reasonably compact sources ($S_\text{int}/S_\text{peak} < 1.2$ in both catalogues, where appropriate) and with no neighbours within 240\,arcsec. An Eddington bias  \citep{Eddington1913} correction is also applied to each catalogue, following equation~4 from \citet{Hogg1998}. Figure~\ref{fig:flux:snr} shows the scaled flux density ratio $(S_\text{300} / S_\text{survey}$) between the catalogues as a function of the signal-to-noise ratio in the GLEAM-300 catalogue. For comparisons with GLEAM, GLEAM-X DR2, GLEAM SGP, {and TGSS~ADR1} we see GLEAM-300 flux density measurements are {largely in agreement, with average offsets within $\approx 5\%$ for sources above $100\sigma_\text{rms}$. {VCSS1} and WISH flux density comparisons show GLEAM-300 flux densities are $\approx 10\%$ lower, with some additional structure in their distributions.}

\begin{table}[t]
    \centering
    \caption{\label{tab:scale} {Brightness scale uncertainty as a function of declination.}}
 \begin{tabular}{l c}\toprule
 Declination range  & Brightness scale uncertainty \\\midrule
      $\delta_\text{J2000} < -78.0 \degr$  & 21\% \\
      $-78.0\degr \leq \delta_\text{J2000} < -72.0 \degr$   & 16\% \\
      $-72.0\degr \leq \delta_\text{J2000} < +18.5 \degr$&  12\%\\
     $ +18.5\degr \leq \delta_\text{J2000} \leq +24.0 \degr$ & 16\% \\
      $ \delta_\text{J2000} > +24.0\degr$ &  21\% \\\bottomrule
    \end{tabular}
\end{table}

{To estimate the uncertainty in the brightness scale we also cross-match the catalogue to the sky model that was used for calibration and post-imaging brightness scale corrections. From the 84-th percentile of the $>100\sigma_\text{rms}$ cross-matched sources, we suggest an initial brightness scale uncertainty of 9\% for declinations covered by the main lobe images ($-78\degr \leq \delta_\text{J2000} \leq +24\degr$). For the grating lobe regions ($\delta_\text{J2000} < -78\degr$ and $\delta_\text{J0000}>+24\degr$) we get $\approx 13$\% for the 84-th percentile, though we noted in Section~\ref{sec:post:scale} that the two regions had differences represented by a standard deviation of $\approx 16$\%. To be conservative, we suggest the 16\% represents the uncertainty in the brightness scale calibration of the individual snapshots, so consider this the initial brightness scale uncertainty for these regions. These initial uncertainties should be added in quadrature to the GLEAM external brightness scale uncertainty for the relevant declination (see table 4 from \citealt{HurleyWalker2017}), except we do not consider the 80\% brightness scale uncertainty reported for $\delta_\text{J2000} < -83.5\degr$ to be necessary here. This yields the full declination-dependent brightness scale uncertainty reported in Table~\ref{tab:scale}, and is included in the catalogue for each source.}

\subsubsection{Spectral energy distributions}\label{sec:seds}
\begin{figure*}[p]
    \centering
    \begin{subfigure}[b]{0.32\linewidth}
    \includegraphics[width=1\linewidth]{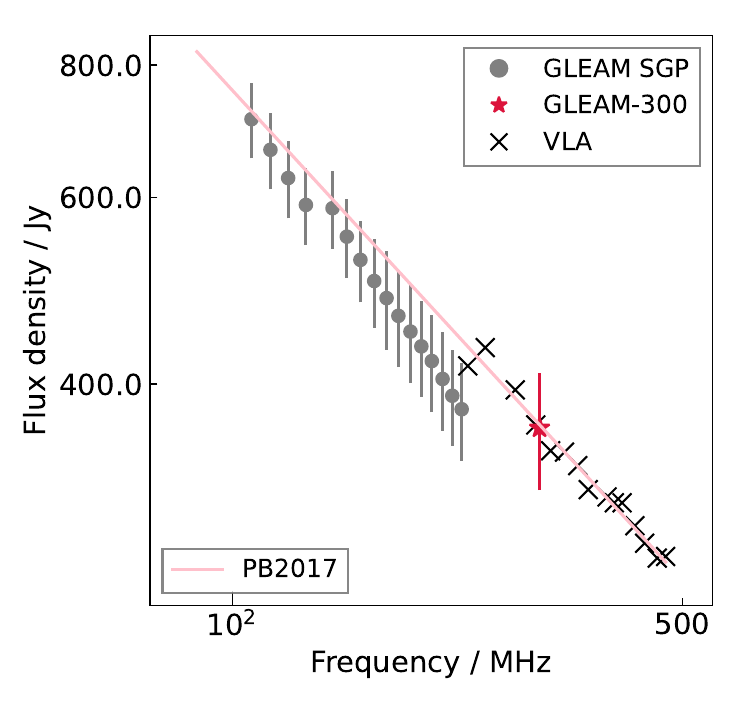}
    \caption{\label{fig:sed:fornax} Fornax A.}
    \end{subfigure}%
    \begin{subfigure}[b]{0.32\linewidth}
    \includegraphics[width=1\linewidth]{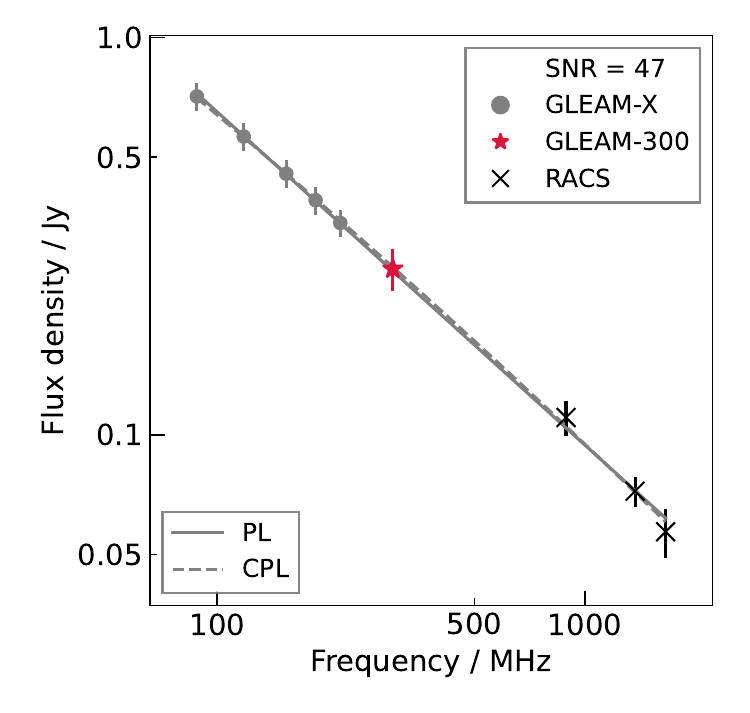}
    \caption{\label{fig:sed:example1} GLEAM-300 J011238$-$384332.}
    \end{subfigure}%
    \begin{subfigure}[b]{0.32\linewidth}
    \includegraphics[width=1\linewidth]{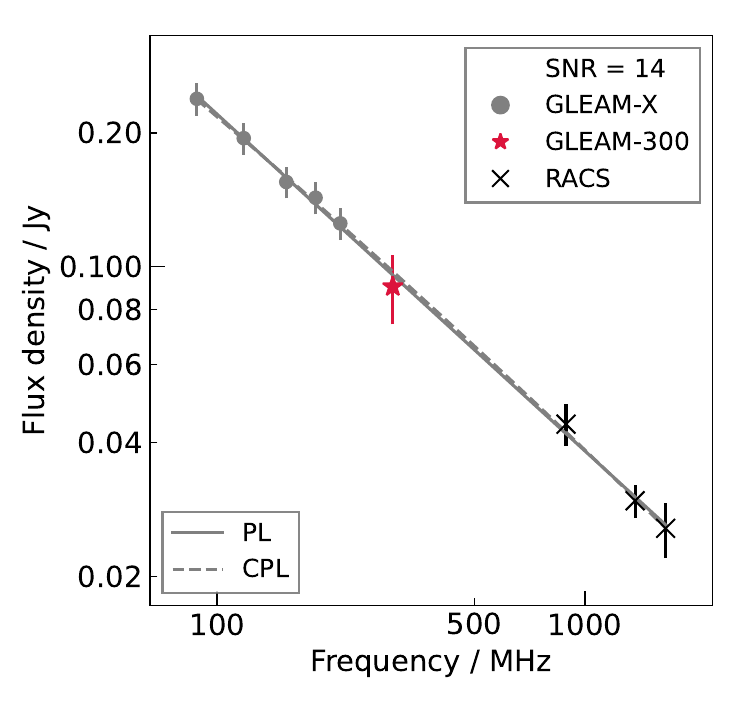}
    \caption{\label{fig:sed:example2} GLEAM-300 J005657$-$752918.}
    \end{subfigure}\\%
        \begin{subfigure}[b]{0.32\linewidth}
    \includegraphics[width=1\linewidth]{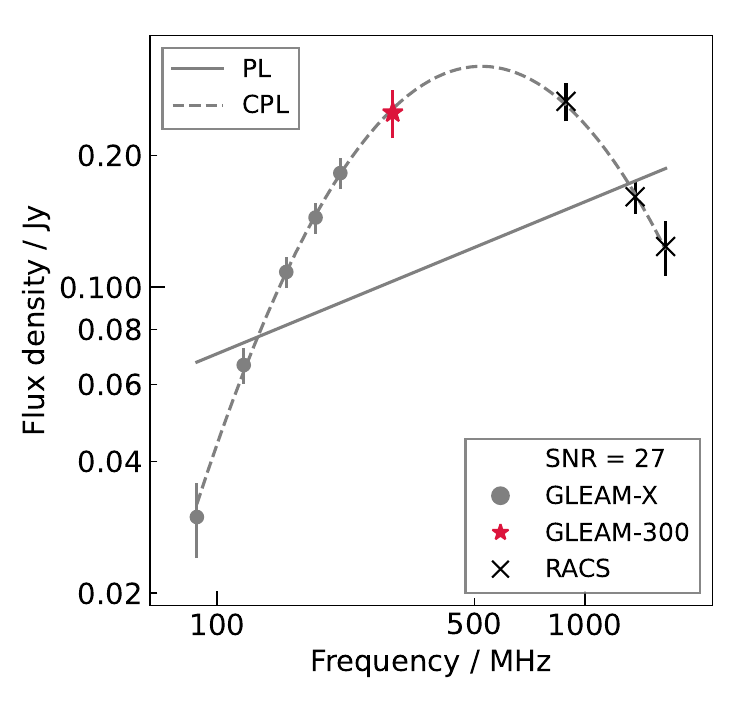}
    \caption{\label{fig:sed:example3} GLEAM-300 J000234$+$040649.}
    \end{subfigure}%
    \begin{subfigure}[b]{0.32\linewidth}
    \includegraphics[width=1\linewidth]{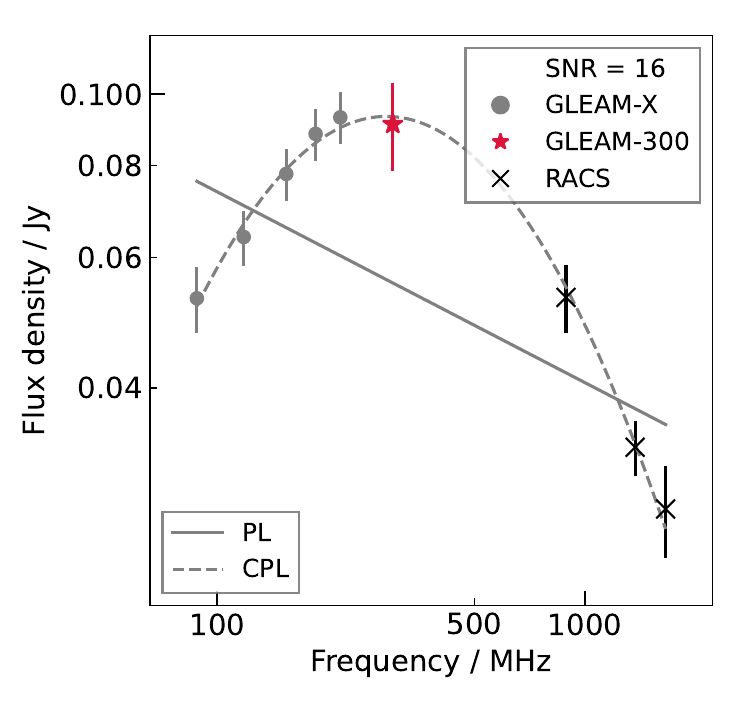}
    \caption{\label{fig:sed:example4} GLEAM-300 J001551$-$322921.}
    \end{subfigure}%
    \begin{subfigure}[b]{0.32\linewidth}
    \includegraphics[width=1\linewidth]{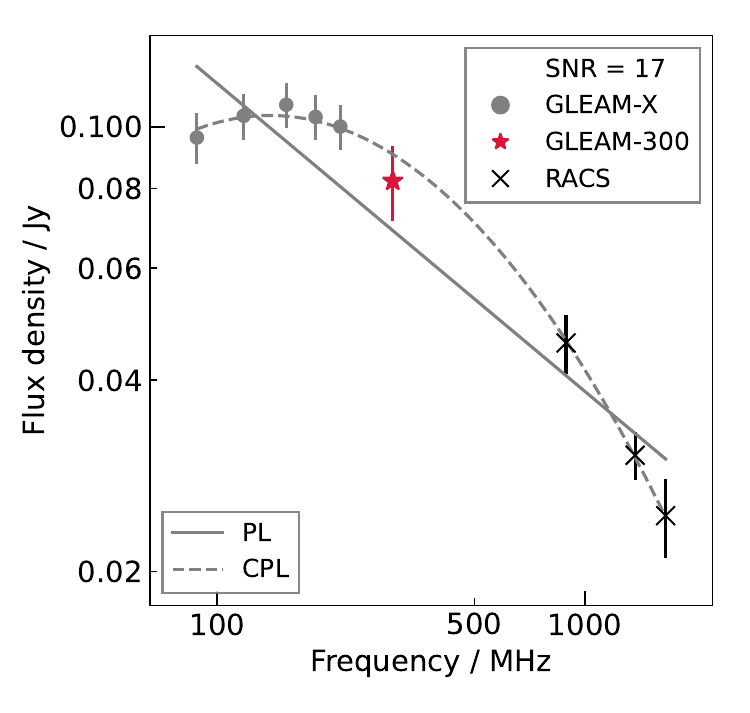}
    \caption{\label{fig:sed:example5} GLEAM-300 J045234$-$400225.}
    \end{subfigure}\\%
    \begin{subfigure}[b]{0.32\linewidth}
    \includegraphics[width=1\linewidth]{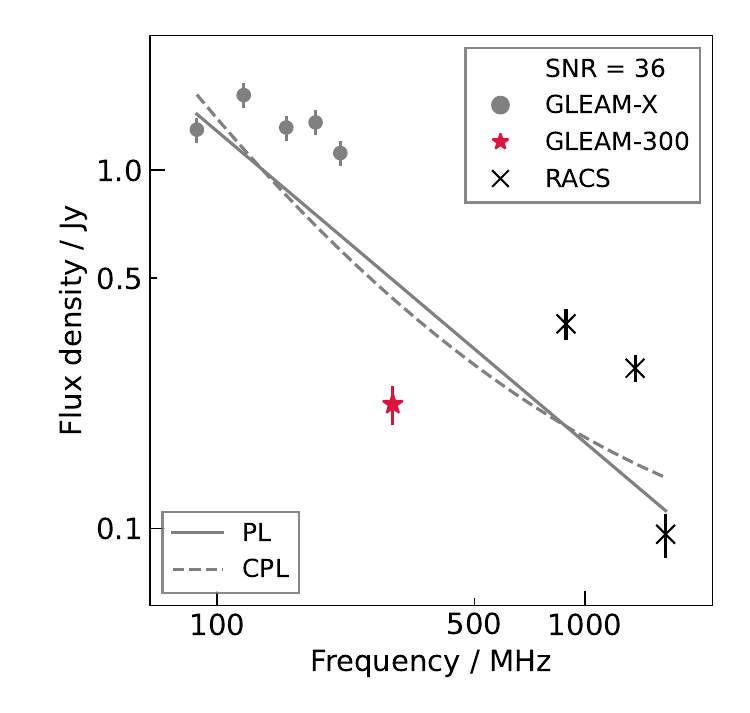}
    \caption{\label{fig:sed:example6} GLEAM-300 J043715$-$471507.}
    \end{subfigure}%
    \centering
    \begin{subfigure}[b]{0.32\linewidth}
\includegraphics[width=1\linewidth]{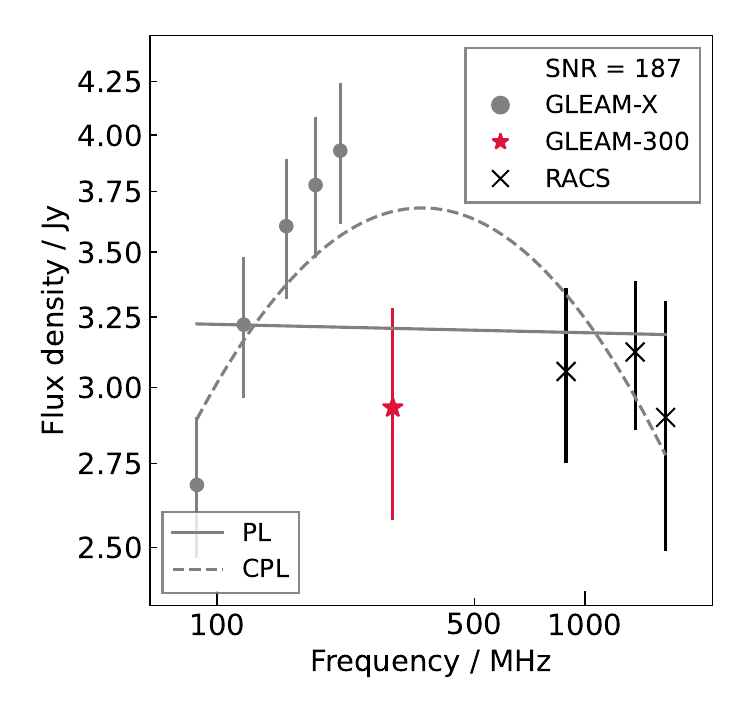}
    \caption{\label{fig:sed:example7} GLEAM-300 214805$+$065745.}
    \end{subfigure}%
    \begin{subfigure}[b]{0.32\linewidth}
    \includegraphics[width=1\linewidth]{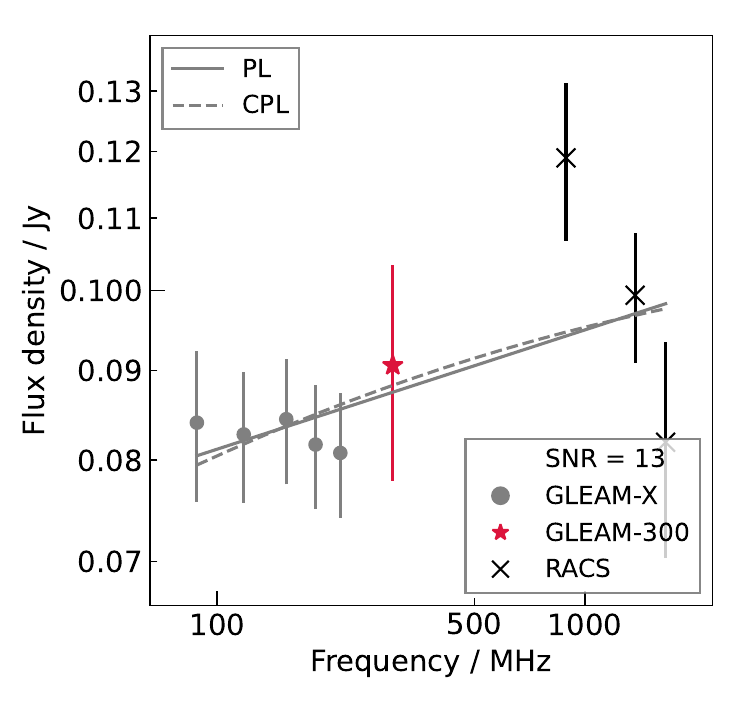}
    \caption{\label{fig:sed:example8} GLEAM-300 J051043$-$094807.}
    \end{subfigure}\\%

    \caption{\label{fig:sed} Example SEDs including a 300-MHz measurement. \subref{fig:sed:fornax} shows the SED of Fornax~A including the GLEAM SGP measurements and VLA measurements from \citet{Perley2017}. The fitted logarithmic polynomial model from \citet{Perley2017} is also shown. \subref{fig:sed:example1}--\subref{fig:sed:example2} show unresolved point sources with standard power law spectra, \subref{fig:sed:example3}--\subref{fig:sed:example5} show curved power law spectra, and \subref{fig:sed:example6}--\subref{fig:sed:example8} comprises variable{/flat spectrum} sources, all selected after cross-matching the GLEAM-300 catalogue with GLEAM-X DR2 and the RACS catalogues. Power law (PL) and curved power law (CPL) models are fit for illustrative purposes. Note both the x- and y-axes are scaled logarithmically.}
\end{figure*}

\begin{figure*}[p]
    \centering
    \begin{subfigure}[b]{0.33\linewidth}
        \includegraphics[width=1\linewidth]{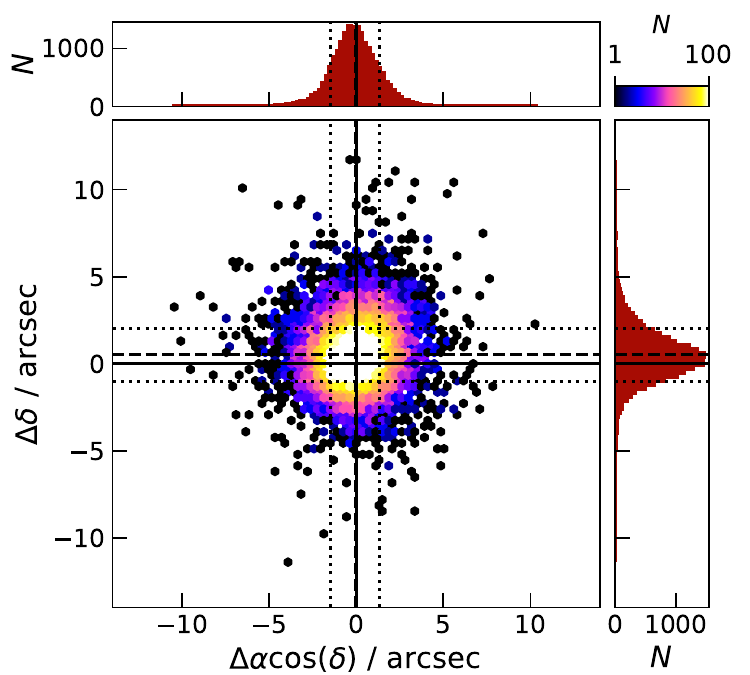}
        \caption{GLEAM-300 $-$ GLEAM\label{fig:astrometry:gleam}}
    \end{subfigure}%
    \begin{subfigure}[b]{0.33\linewidth}
        \includegraphics[width=1\linewidth]{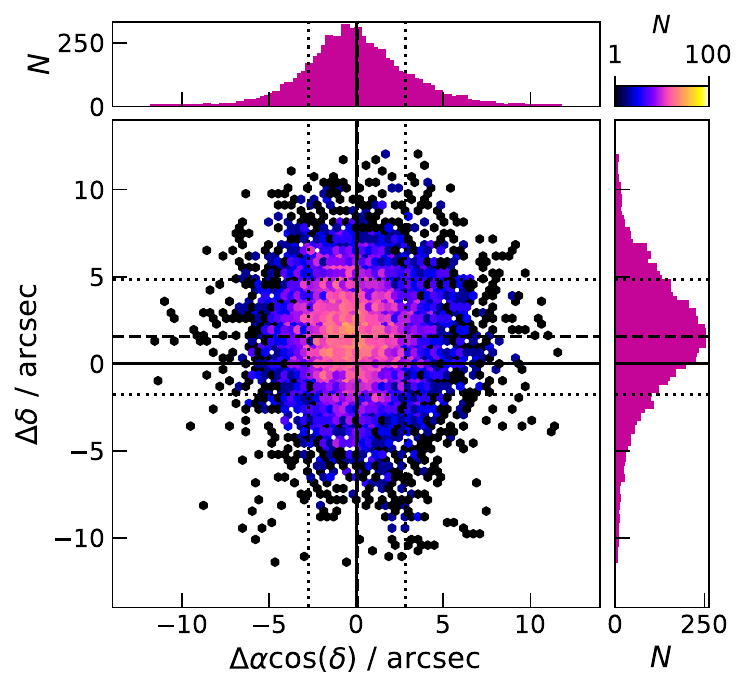}
        \caption{GLEAM-300 $-$ GLEAM-X DR2\label{fig:astrometry:gleamxdr2}}
    \end{subfigure}%
    \begin{subfigure}[b]{0.33\linewidth}
        \includegraphics[width=1\linewidth]{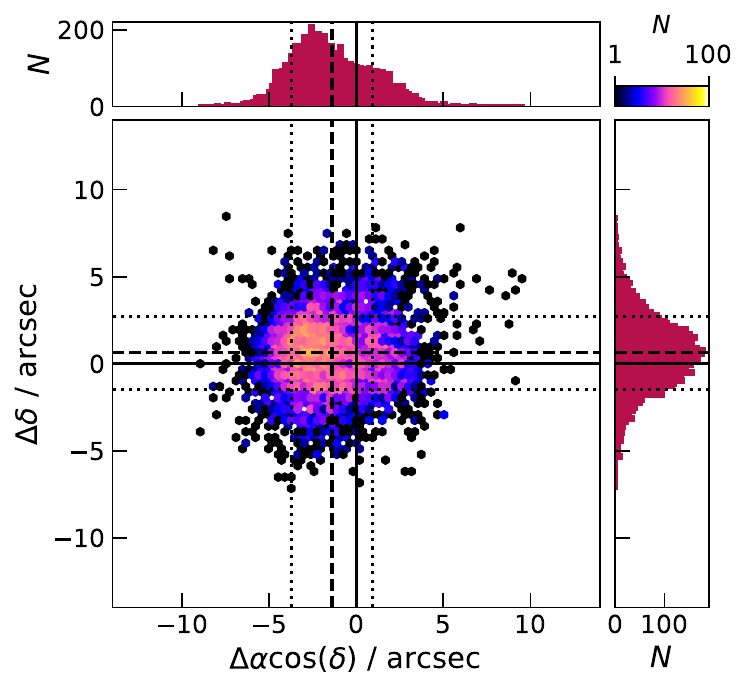}
        \caption{GLEAM-300 $-$ GLEAM SGP\label{fig:astrometry:sgp}}
    \end{subfigure}\\[1em]%
    \begin{subfigure}[b]{0.33\linewidth}
        \includegraphics[width=1\linewidth]{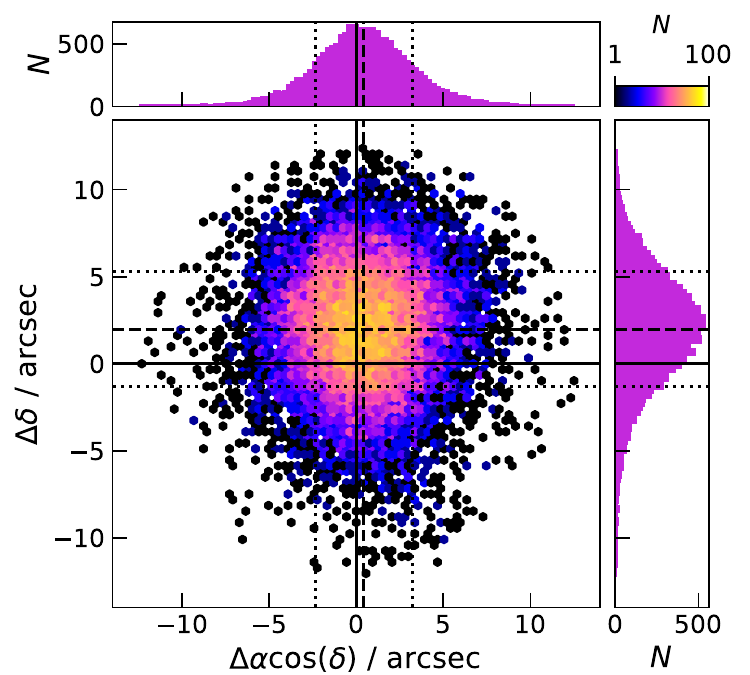}
        \caption{GLEAM-300 $-$ TGSS ADR1\label{fig:astrometry:tgss}}
    \end{subfigure}%
    \begin{subfigure}[b]{0.33\linewidth}
        \includegraphics[width=1\linewidth]{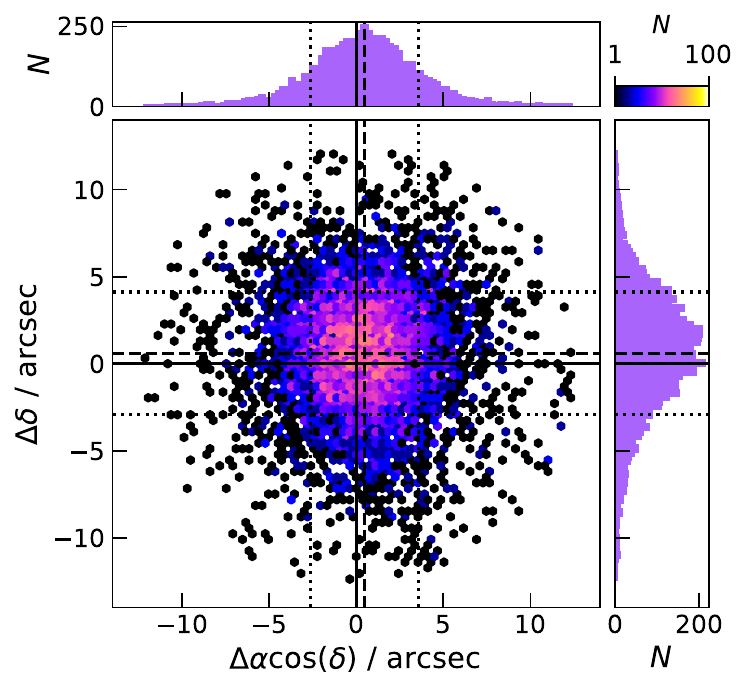}
        \caption{GLEAM-300 $-$ {VCSS1}\label{fig:astrometry:vcss}}
    \end{subfigure}%
    \begin{subfigure}[b]{0.33\linewidth}
        \includegraphics[width=1\linewidth]{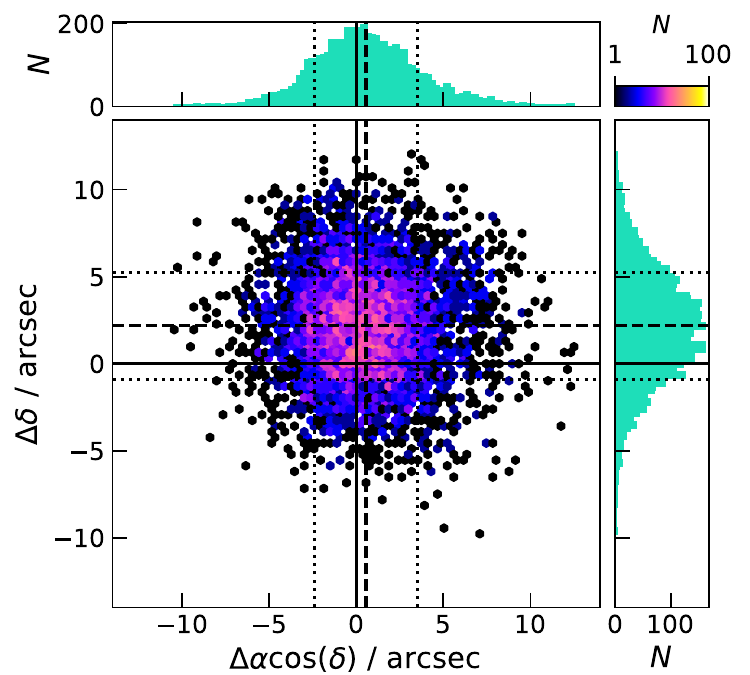}
        \caption{GLEAM-300 $-$ SUMSS\label{fig:astrometry:sumss}}
    \end{subfigure}\\%
    \begin{subfigure}[b]{0.33\linewidth}
        \includegraphics[width=1\linewidth]{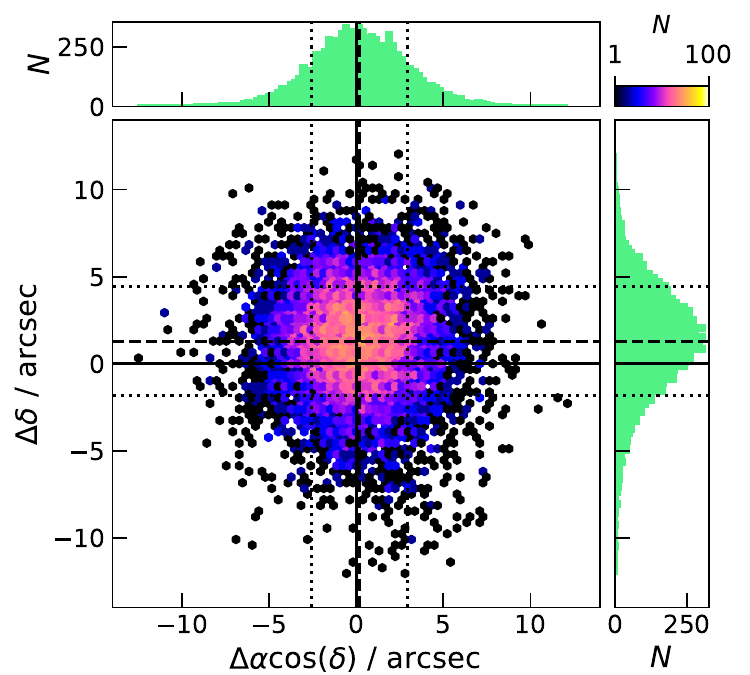}
        \caption{GLEAM-300 $-$ NVSS\label{fig:astrometry:nvss}}
    \end{subfigure}%
    \caption{\label{fig:astrometry} Right ascension ($\alpha$) and declination ($\delta$) offsets between GLEAM-300 and GLEAM \subref{fig:astrometry:gleam}, GLEAM-X DR2 \subref{fig:astrometry:gleamxdr2}, GLEAM SGP \subref{fig:astrometry:sgp}, TGSS~ADR1 \subref{fig:astrometry:tgss}, {VCSS1} \subref{fig:astrometry:vcss}, SUMSS \subref{fig:astrometry:sumss}, and the NVSS \subref{fig:astrometry:nvss}. Only sources with an SNR $>50$ in the GLEAM-300 catalogue are shown. The solid black lines are drawn at 0 offset, the dashed black lines are drawn at the mean offset value, and the dotted black lines are drawn at $\pm 1\sigma$ about the mean offset.}
\end{figure*}

\begin{figure}[t]
    \centering
    \includegraphics[width=1\linewidth]{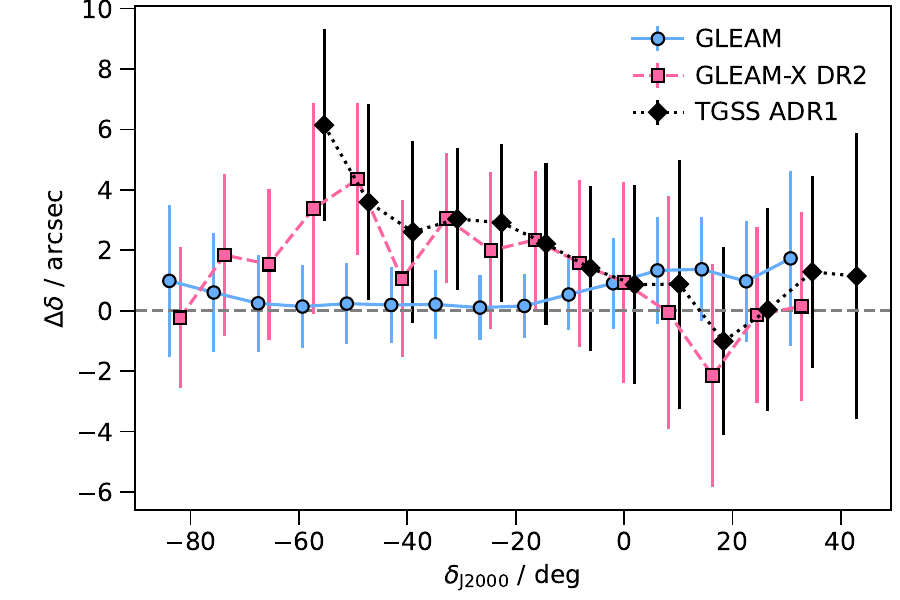}
    \caption{\label{fig:astrometry:dec} Median-binned declination offsets as a function of declination for cross-matches to GLEAM, GLEAM-X DR2, and TGSS ADR1. The error bars are drawn at $\pm 1\sigma$ for each bin. {Bins are offset by $2\degr$ for each survey for clarity.}}
\end{figure}

One of the goals of this work is to provide an additional `high'-frequency data point to GLEAM and/or GLEAM-X measurements to constrain the shapes of source spectral energy distributions (SEDs), particularly in the presence of spectral curvature. As a point of comparison we measure the integrated flux density of Fornax~A within a polygon region containing the radio galaxy. We also make the equivalent measurements from the 16 GLEAM SGP mosaics between 107--227\,MHz. We compare these GLEAM-300 and GLEAM SGP measurements to the VLA measurements and spectral model from \citet{Perley2017}. The spectral data are shown in Figure~\ref{fig:sed:fornax} with {the}  \citet{Perley2017} logarithmic polynomial model overlaid. The GLEAM-300 data agrees with the VLA measurements and model, though we see some discrepancy between the VLA data and GLEAM SGP data. The SGP data appear to show a steeper overall integrated spectrum for Fornax~A. This is likely due to use of robust image weighting combined with a 30-$\lambda$ $(u,v)$ cut and 15-$\lambda$ Tukey taper (corresponding to sensitivity up to $\approx 1.5$\,deg), so artificial steepening of extended sources is expected, and care should be taken when interpreting spectra in cases of extended sources. Overall, the agreement with the \citet{Perley2017} suggests the flux density scale (and associated uncertainty) is sensible for the GLEAM-300 data. We do not show a comparison to the original GLEAM as the Fornax~A images at some frequencies contain artefacts from the linear mosaic process.

As an example of where the GLEAM-300 datapoint sits within the ecosystem of widefield Southern Sky surveys, we also cross-match GLEAM-300 to GLEAM-X DR2 and the three RACS bands: RACS-low at 887.5 MHz \citep{racs1,racs2}, RACS-mid at 1\,367.5 MHz \citep{racs-mid,racs-mid2}, and RACS-high at 1\,655.5\,MHz \citep{Duchesne2025}. We repeat a similar strict multi-matching process employed by \citet{Duchesne2025}, which focuses on isolated and compact point sources detected above $10\sigma_\text{rms}$ in all five catalogues under consideration. This choice is to avoid the aforementioned concern with extended source spectra, and to avoid issues arising from differing angular resolution and sensitivities between the surveys. This limits the resultant cross-matched catalogue to {9\,880} sources. We then fit generic power law and curved power law models as in e.g.\ \citet[][]{gleamx-dr2}. For the GLEAM-X DR2 data, we opt to use 30-MHz wideband measurements as opposed to 7.68-MHz narrowband measurements for consistency with the GLEAM-300 30-MHz measurement.  

Figure~\ref{fig:sed:example1}--\subref{fig:sed:example8} show a set of example sources with fitted SEDs that include the GLEAM-X, GLEAM-300, and RACS measurements. The sources are selected from the cross-matched sources to showcase a range of spectral morphologies. {Figures~\ref{fig:sed:example1}--\subref{fig:sed:example2} show a selection of sources with standard power law spectra, and Figures~\ref{fig:sed:example3}--\subref{fig:sed:example5} show example sources with well-modelled curved power law spectra. In addition to the power law and curved spectra, we also show a few examples of more exotic spectra arising from source variability. Figure~\ref{fig:sed:example6} shows the millisecond pulsar PSR~J0437$-$4715 \citep{Johnston1993} which was also detected in the MWA circular polarization survey \citep{Lenc2017}. Figure~\ref{fig:sed:example7} shows a `flat-spectrum' {blazar source, 4C~$+$06.69 \citep[e.g.][]{Healey2007}, and } Figure~\ref{fig:sed:example8} shows NVSS~J051042$-$094813 \citep{ccg+98}, which also has a non-power law spectrum, though we do not explore these sources further here.} {With the addition of the GLEAM-300 data point,} the gap in spectral coverage between GLEAM and RACS (231--888\,MHz) is now more accurately filled-in across the whole Southern Sky.

\subsection{Astrometry}\label{sec:astrometry}

For assessment of the astrometry, we also cross match the GLEAM-300 catalogue to the SUMSS and the NVSS. We also compare to the other GLEAM catalogues, TGSS ADR1, and {VCSS1} again. We consider isolated and compact sources (in each catalogue), and calculate the astrometric offsets as GLEAM-300 $-$ external survey for sources above $50\sigma_\text{rms}$ in the GLEAM-300 catalogue. Figure~\ref{fig:astrometry} shows the positional offsets in $\alpha_\text{J2000}$ and $\delta_\text{J2000}$ for the survey comparisons.

As GLEAM is {largely} used for phase calibration and post-imaging astrometric corrections, we see bulk $\alpha_\text{J2000},\delta_\text{J2000}$ offsets of {$-0\farcs06 \pm 1\farcs40$} and  {$+0\farcs52 \pm 1\farcs52$}, noting that the smallest pixel size in the mosaic images is {$17\farcs9$} (with a maximum pixel size of {44\farcs6}). Other surveys tend towards a bulk offset in declination{, up to a few arcsec}, with median bulk offsets in $\delta_\text{J2000}$ of {$+1\farcs55 \pm 3\farcs29$} for GLEAM-X DR2, {$+0\farcs60 \pm 3\farcs52$} for {VCSS1}, {$+1\farcs99 \pm 3\farcs30$} for the TGSS ADR1, {$+2\farcs19 \pm 3\farcs07$} for SUMSS, and {$+1\farcs30 \pm 3\farcs11$} for the NVSS. The GLEAM SGP on the other hand features {the largest} offset in $\alpha_\text{J2000}$, with {$\Delta\alpha\cos\delta = -1\farcs39 \pm 2\farcs32$} along with a marginal offset in $\delta_\text{J2000}$. GLEAM-X DR2 uses both SUMSS and NVSS for its astrometric corrections so we expect the offsets against SUMSS and NVSS seen by \citet{gleamx-dr2} should be present here too. Of particular note is elongation of the offset distributions towards the lower right of each plot, except in the SUMSS comparison. Figure~\ref{fig:astrometry:dec} shows the median-binned declination offsets as a function of declination for the GLEAM, GLEAM-X DR2, and TGSS{~ADR1} cross-matches. The GLEAM cross-matches deviate in the SCP region {and towards the Northern Hemisphere}, whereas both the GLEAM-X DR2 and TGSS offsets show similar structure as a function of $\delta_\text{J2000}$. The GLEAM-300 catalogue has largely inherited an inherent astrometric uncertainty from GLEAM. 

\section{Applications and future work}\label{sec:future}

\subsection{Diffuse and extended radio emission}

The GLEAM-300 catalogue provides crucial frequency coverage for understanding the spectral evolution of diffuse non-thermal phenomena in several environments. Both the so-called `remnants' or `fossils' associated with switched-off AGN \citep[e.g.][]{Murgia2011,Quici2025} and the steep-spectrum radio sources associated with cosmic structures---groups and clusters of galaxies \citep[see e.g.][respectively]{Eckert2021_groups_review,vda+19}---require broad-band spectral information to understand the physics of the particle acceleration mechanisms at work.

Oftentimes these spectra show significant curvature, such as the remnant radio galaxy NGC~1534 \citep{Duchesne2019_NGC1534}, the `ultra-steep spectrum jellyfish' associated with Abell~2877 \citep{Hodgson2021_Jellyfish}, or the ultra-steep spectrum fossil in Abell~3266 \citep{Duchesne2022,Riseley2022_A3266}. The latter object in particular shows a clear break in the spectrum between 216~MHz (measured by the MWA) and 943~MHz (measured by ASKAP) and constraining the break frequency, as well as the type of break---whether the spectrum is truly curved or shows a sharp break in the injection index, for example---would provide insights into the ageing and evolution of these sources.

Similarly, in cosmic structures such as galaxy groups, bulk motions and interactions in the environment can re-distribute and re-accelerate fossil plasma onto different scales \cite[e.g.][]{Brienza2022_NGC507,Candini2023_NGC6086} but the mechanisms by which this occurs are still poorly understood. The GLEAM-300 measurements occupy a critical frequency regime that has been historically served only down to $\delta_\text{J2000} \approx -53\degr$ (the southern limit of the GMRT); the catalogue published herein significantly broadens the scope of such studies.

Furthermore, the frequency coverage provided by the GLEAM-300 catalogue is essential in supernova remnants \citep[see][for a recent review]{Dubner2015} studies which exhibit a variety of spectral behaviours connected to their particle acceleration mechanisms, interactions with the surroundings and energy losses. The GLEAM-300 frequency band is beneficial for identifying turnovers in the spectra at low radio frequencies, which can signify absorption mechanisms such as free-free absorption in ionised gas {\citep[e.g.][]{Castelletti2021}, and in particular the 300-MHz data can provide an anchor for the unabsorbed power-law portion of the spectrum at a frequency where interstellar medium (ISM) absorption has not yet taken effect.}

The GLEAM-300 data provide a unique opportunity to probe these effects by extending the spectral coverage into an under-explored range. This dataset effectively bridges the very low frequency of GLEAM (72--231\,MHz) and the higher frequencies provided by surveys like the Evolutionary Map of the Universe \citep[EMU;][]{Norris2021,Hopkins2025} at 944\,MHz while awaiting the new SKA infrastructure. Expanding studies of supernova remnants with this catalogue will help constrain the physical processes governing particle acceleration and evolution in these sources.

\subsection{Transients and variability}

\begin{figure}[t]
    \centering
    \includegraphics[width=1\linewidth]{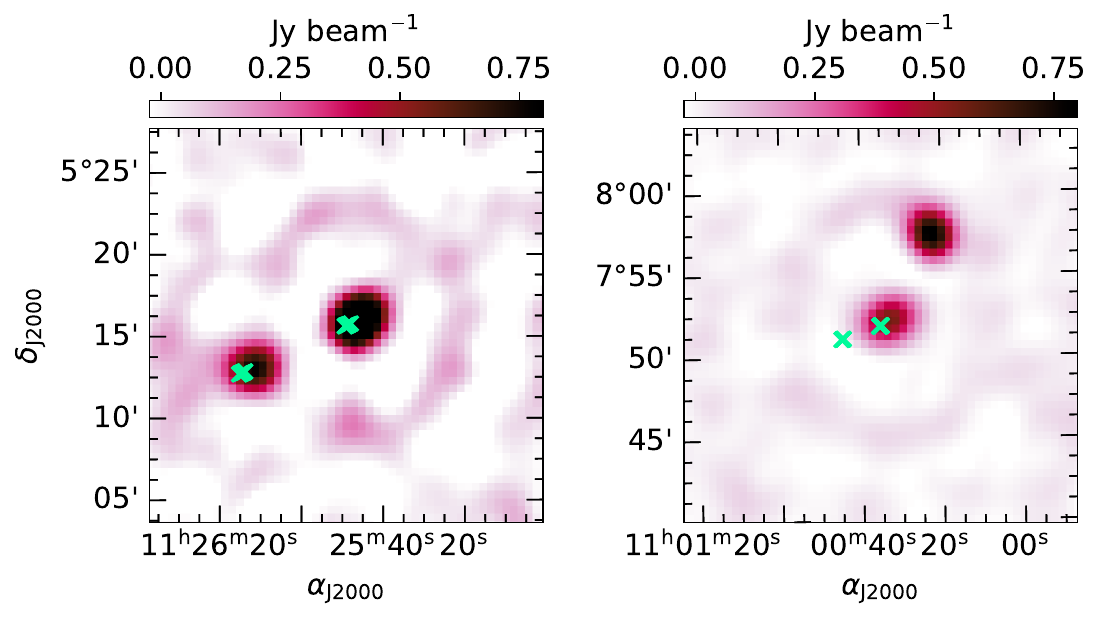}
    \caption{\label{fig:jupiter} {Jupiter detections in the mosaic {images}. The positions of Jupiter in the individual snapshots are indicated by green crosses, and sources detected at those locations are the weighted-average detection in the mosaic. In the left panel, both radio sources at the marked positions are Jupiter detections. In the right panel, only one source is a Jupiter detection.}}
\end{figure}

Over the last few years, widefield imaging surveys have become an increasingly popular tool for finding sources that vary on minute- to hour-long time scales, a hitherto relatively unexplored parameter space \citep[e.g.][]{2007ApJ...660L.121H,2013PASA...30....6M,Murphy2021,2022Natur.601..526H}. Such surveys can also potentially detect shorter duration (sub-second) events that are sufficient bright \citep[e.g.][]{2021ApJ...920...45W,2023PASA...40....3S,McSweeney2025}. The sensitivity of a given survey towards different classes of transient sources is a non-trivial function of frequency. At lower radio frequencies ($\lesssim 200$\,MHz), detection can be inhibited by higher sky temperatures, intrinsic low-frequency turnovers, as well as the temporal smearing effects of interstellar dispersion and scattering. On the other hand, transient sources' typically steep spectra ($-4 \lesssim \alpha \lesssim -1$) make detection difficult at higher frequencies ($\gtrsim 1$\,GHz). This survey, at 300\,MHz, may therefore be sensitive to transients that, for one reason or another, are difficult to detect at other frequencies. The individual snapshots have a median noise of $\sigma \approx {68}$\,mJy\,beam$^{-1}$; we expect that they will be sensitive to transient events whose brightness (integrated over the two-minute snapshot) is $S_\text{300} > 5\sigma \approx {340}$\,mJy\,beam$^{-1}$. In particular, this survey is likely to detect new long period transients whose single pulses can last anywhere from a few tens of seconds \citep{deRuiter2025,2024ApJ...976L..21H} to several minutes \citep{2023Natur.619..487H,2025arXiv250109133L}. To enable searches of such transients, the individual source lists for each two-minute snapshot are being made available.

{We note that Jupiter is detected in snapshots taken on 20--21 February 2016 and 25--26 April 2016 and appears in the mosaic images. The movement across sky over the two nights in February is seen as two separate sources, whereas only a single radio source is seen over the two nights in April. We show the mosaic detections in Figure~\ref{fig:jupiter}, with green markers indicating the direction of Jupiter in the individual snapshots. Note that Jupiter is not included in the CLEAN mask so has noticeable PSF sidelobes around it. Three sources in the final catalogue are co-located with Jupiter's positions: GLEAM-300~J112611$+$051306, GLEAM-300~J112546$+$051601, and GLEAM-300~J110033$+$075214.}

\subsection{General data improvements and other additions}

\begin{figure*}[p]
    \begin{subfigure}[b]{1\linewidth}
        \includegraphics[width=1\linewidth]{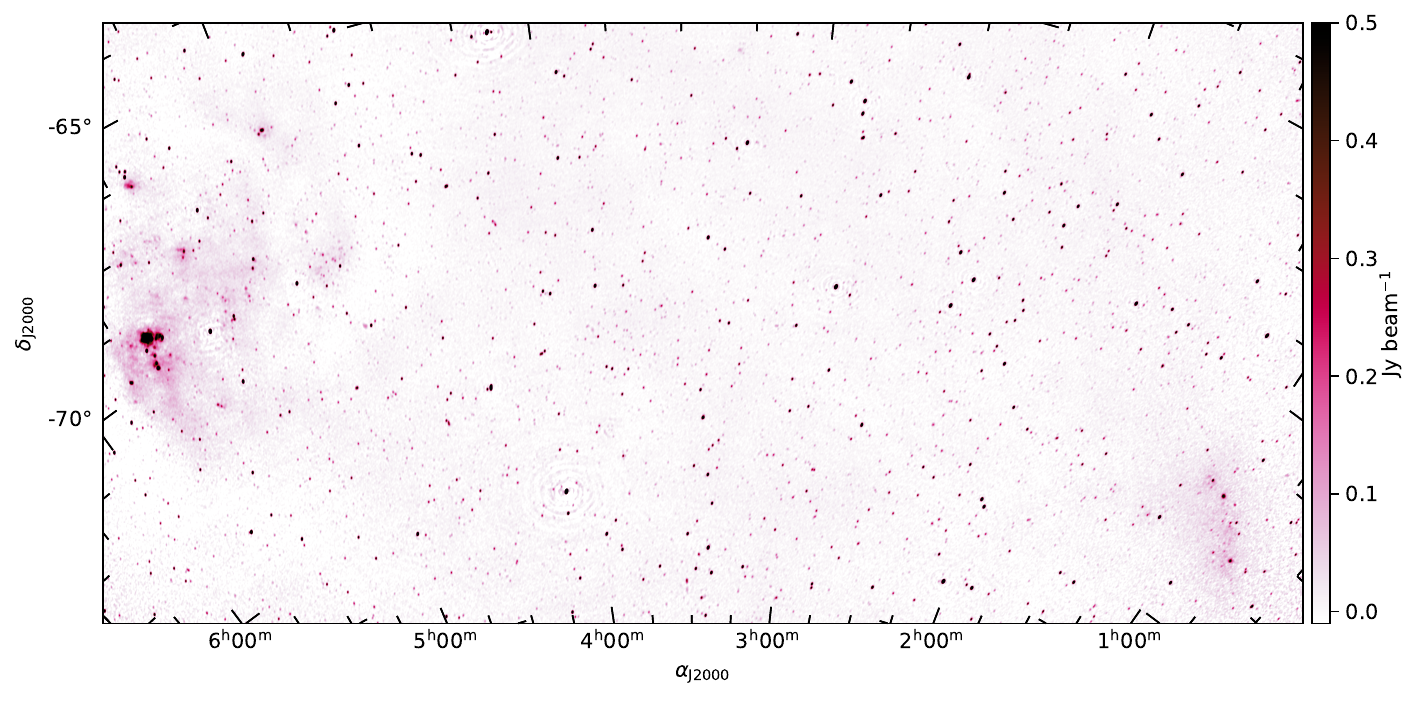}
        \caption{\label{fig:singledish:original} GLEAM-300 image.}
    \end{subfigure}\\%
    \begin{subfigure}[b]{1\linewidth}
        \includegraphics[width=1\linewidth]{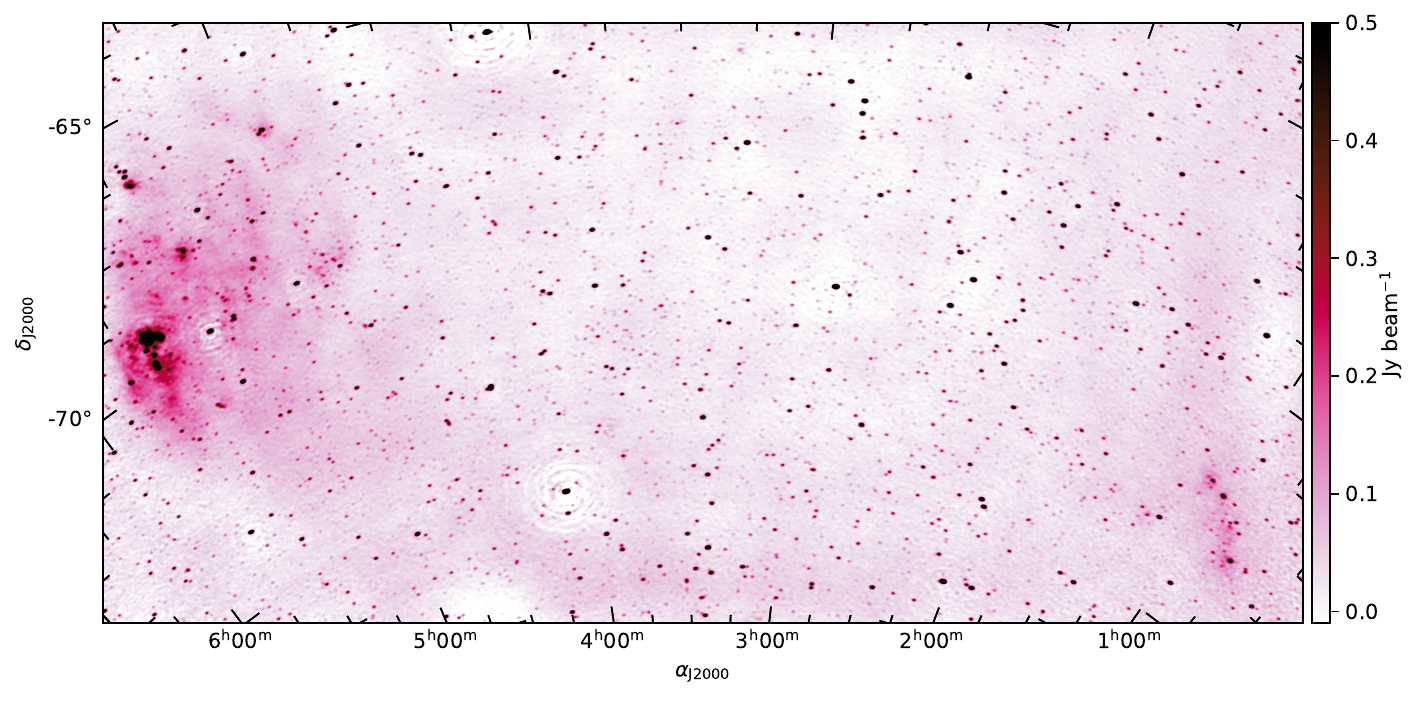}
        \caption{\label{fig:singledish:feather} GLEAM-300 image feathered with the Murriyang data.}
    \end{subfigure}
    \caption{\label{fig:singledish} Example mosaic region containing both the LMC and SMC, comparing the GLEAM-300 image by itself \subref{fig:singledish:original} and the GLEAM-300 image feathered with an equivalent image from GMIMS-LBS \citep{Wolleben2019} \subref{fig:singledish:feather}. The brightness scaling is the same in both panels, and the feathered image highlights additional extended emission within/around both the LMC and SMC as well as Galactic emission in this region.}
\end{figure*}

One of the limitations of our current GLEAM-300 processing strategy is the lack of consistent in-field calibration. \citet{Franzen2021a} highlighted that self-calibration (after initial calibration from other methods) can be used in place of in-field calibration, providing similar results. We find through MWA processing experience that the level of improvement will also depend on the direction being observed and what is in the sky model. Either more consistent in-field calibration or a successful self-calibration procedure would improve the overall sensitivity of the GLEAM-300 survey and reliability at low SNR. At present it is not clear why these processes do not work consistently, though it is likely the combination of inaccurate source models, certain primary beam pointings being more difficult due to attenuated bright sources at the edges of the primary beam lobes, as well as other time/observation-dependent effects such as ionospheric activity.

An interesting avenue for further work in this area will be to develop a linear polarization counterpart to GLEAM-300, akin to the POGS effort at lower MWA frequencies \citep{Riseley2018,Riseley2020}. A detailed understanding of the magnetoionic properties and environments of radio sources requires broadband linear polarization measurements, and the primary figure of merit is complete sampling of a broad range of wavelength-squared ($\lambda^2$) coverage. GLEAM-300 now provides well-understood visibility data from which the required data products---full Stokes continuum cubes---can be produced and analysed. This is however a non-trivial exercise because additional calibration effects require attention, such as XY phase differences, ionospheric Faraday rotation measure corrections, and characterisation and correction of polarization leakage. But the gains can be significant, as a polarization counterpart of GLEAM-300 would help to fill a key gap in $\lambda^2$ coverage between POGS and the ASKAP SPICE-RACS catalogue \citep{spiceracs} at considerably higher frequency. These intermediate-$\lambda^2$ measurements are particularly valuable to understand the details of depolarization effects \citep[see, e.g.,][]{2011AJ....141..191F} that dramatically reduce the density of linearly polarized sources seen at MWA frequencies compared to GHz-regime observations like those with ASKAP \citep[e.g.,][]{2023MNRAS.519.5723O,2024A&A...687A.267P}.

Whilst the focus of this work is on producing a compact source catalogue at 300\,MHz, the short baselines of the MWA provide excellent sensitivity to extended diffuse emission from the Milky Way and nearby galaxies. The MWA in the Phase I compact configuration was sensitive to scales of up to $\sim7.4$\,deg based on the minimum baseline of 7.7\,m at 300\,MHz. As described in the introduction, GMIMS-LBS~\citep{Wolleben2019} covers the entire GLEAM-300 band and almost the entire survey area (up to $\delta_\text{J2000}=+20\degr$). Having been observed with Murriyang, a 64\,m single dish telescope, GMIMS-LBS can provide the missing `zero spacings' in GLEAM-300 and provide sensitivity in spatial scales across the entire sky.

By way of example we have combined the GLEAM-300 mosaic of the region surround the LMC and SMC (field `J0326-72') with GMIMS-LBS using the `feathering' algorithm \citep{Weiss2001}\footnote{Implemented in \url{https://github.com/AlecThomson/FeatherPy}.}. We show the original and `feathered' image in Figure~\ref{fig:singledish}, highlighting the additional recovered signal from the large-scale features in the LMC in particular. We intend to combine the entirety of GLEAM-300 with GMIMS-LBS as an enhanced data product in future work.

\section{Summary}

This work presents a new widefield radio survey conducted by the MWA at the heretofore unexplored frequency band of 300 MHz. The survey, part of the Galactic and Extragalactic MWA All-sky (GLEAM) project and titled GLEAM-300, covers the whole Southern Sky and extends to $\delta_\text{J2000} \approx {+40}\degr$. The survey was conducted similarly to other GLEAM surveys, with 2-min snapshot observations taken in declination strips and processed semi-independently prior to mosaicking to increase sensitivity. While no observations directly covered the South Celestial Pole, imaging primary beam grating lobes allowed us to fill in the whole Southern Hemisphere.

The resultant mosaic images have a median angular resolution of $\psfMedianA \times \psfMedianB$ and median rms noise of \rmsMedian\,mJy\,beam$^{-1}$. The brightness scale of the images is reliable to {12--21\%}. A catalogue is constructed from the mosaic images resulting in \nsources\ unique Gaussian components covering the whole survey region. Due to corrections applied to snapshot images, the astrometric accuracy is similar to the original GLEAM, with some bulk offsets up to $\approx 2\arcsec$ in declination seen with respect to other surveys, as a function of declination. The overall astrometric accuracy is $\approx 2$--$4\arcsec$.  

We also cross-matched a subset of the catalogue to the second data release of GLEAM-X and the three RACS catalogues to illustrate where the 300-MHz measurements sit within the current ecosystem of Southern Sky widefield surveys, and show the overall agreement with the existing data. The 300-MHz datapoint in particular provides a unique measurement in the large spectral gap between previously-published MWA and ASKAP data. This is particularly timely given the ongoing construction of SKA-Low, which will observe from $50$--$350$\,MHz. Together with GLEAM and GLEAM-X, this $300$\,MHz catalogue will provide crucial information for initial calibration and validation of SKA-Low image products in this frequency range. For example, the catalogue is well suited to contribute to the construction of a broadband calibration sky model for SKA-Low, and provides a reference epoch for forthcoming variability studies using SKA-Low data. We release the catalogue and mosaic images to the astronomy community for use in science and data validation work, available at the CSIRO Data Access Portal: \url{https://doi.org/10.25919/z80f-5t62}.

\section*{Data availability}

The GLEAM-300 catalogue and mosaic images are available at the CSIRO Data Access Portal: \url{https://doi.org/10.25919/z80f-5t62}. 

Much of the code/scripts required to generate plots and other miscellaneous analysis tasks for the GLEAM-300 survey and this manuscript are collected in a \texttt{python} package hosted here: \url{https://gitlab.com/Sunmish/gleam300}. 

\begin{acknowledgement}

The authors would like to thank the anonymous referee for their in-depth reading of the manuscript. Their comments helped improve the work and in particular motivated us to revisit the bias in our original flux density measurements described in Appendix~\ref{app:faint}.

This scientific work uses data obtained from Inyarrimanha Ilgari Bundara / the Murchison Radio-astronomy Observatory. We acknowledge the Wajarri Yamaji People as the Traditional Owners and native title holders of the Observatory site. CSIRO’s ASKAP radio telescope is part of the Australia Telescope National Facility (\url{https://ror.org/05qajvd42}). Operation of ASKAP is funded by the Australian Government with support from the National Collaborative Research Infrastructure Strategy. ASKAP uses the resources of the Pawsey Supercomputing Research Centre. Establishment of ASKAP, Inyarrimanha Ilgari Bundara, the CSIRO Murchison Radio-astronomy Observatory and the Pawsey Supercomputing Research Centre are initiatives of the Australian Government, with support from the Government of Western Australia and the Science and Industry Endowment Fund. 

N.H.-W. is the recipient of an Australian Research Council Future Fellowship (project number FT190100231). CJR acknowledges financial support from the German Science Foundation DFG, via the Collaborative Research Center SFB1491 `Cosmic Interacting Matters – From Source to Signal'.

This research has made use of the VizieR catalogue access tool, CDS, Strasbourg, France (DOI: 10.26093/cds/vizier). The original description of the VizieR service was published in \citet{vizier}. We used a range of \texttt{python} software packages during this work and the production of this manuscript, including \texttt{astropy} \citep{astropy:2018}, \texttt{matplotlib} \citep{Hunter2007}, \texttt{numpy} \citep{numpy2020}, \texttt{scipy} \citep{scipy}, and \texttt{healpy} \citep{healpy}. We make use of \texttt{ds9} \citep{ds9}, \texttt{topcat} \citep{topcat}, and the Cube Analysis and Rendering Tool for Astronomy \citep[CARTA;][]{Comrie2021} for visualisation, as well as the ``Aladin sky atlas'' developed at CDS, Strasbourg Observatory, France \citep{aladin1,aladin2} for obtaining catalogue data. For precision rounding we used \texttt{to-precision}: \url{https://bitbucket.org/william_rusnack/to-precision/src/master/}. We make use of a selection of colourmaps provided by the \texttt{cmasher} package \citep{cmasher}, and use \url{https://github.com/eltos/gradient} for some colour selections.
\end{acknowledgement}


\bibliography{bib}

\appendix

\renewcommand{\thesubfigure}{(\roman{subfigure})}
\renewcommand\thefigure{A\arabic{figure}}   
\renewcommand\thetable{A\arabic{table}}
\setcounter{figure}{0}
\setcounter{table}{0}

\section{Faint source measurement bias}\label{app:faint}

In initial processing and subsequent assessment of the brightness scale of the GLEAM-300 catalogue, we noticed low-SNR sources were systematically fainter when compared to other catalogues. We first cross-matched the source-lists from each 300-MHz mosaic to the GLEAM catalogue, considering isolated and near-compact sources to assess the individual mosaics prior to combining the source list to make the contiguous catalogue. Despite having similar sensitivities, we also corrected for Eddington bias \citep{Eddington1913} for the GLEAM-300 and GLEAM catalogue measurements, following equation~4 from \citet{Hogg1998}. We found an offset in the lowest SNR bins of $\approx 29$\,mJy, though it varied as a function of source brightness. This offset was also present for comparisons with any surveys that reach or exceed the sensitivity of GLEAM-300 so pointed to an intrinsic bias within our data. We explore this issue and detail the resolution in the following sections.

\subsection{The effect of shallow CLEANing?}\label{app:shallow_clean}

\begin{figure}[t]
    \centering
    \includegraphics[width=1\linewidth]{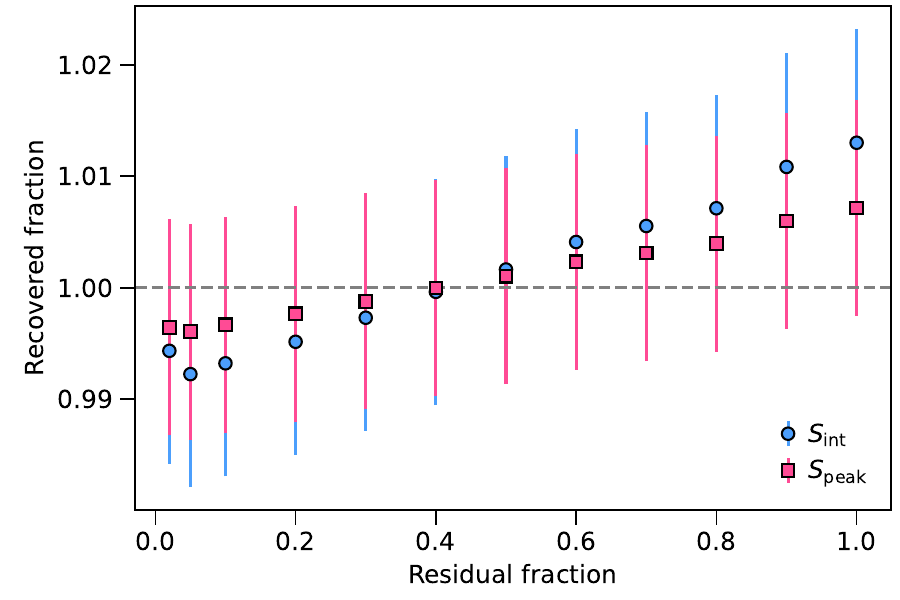}
    \caption{\label{fig:clean} Recovered integrated ($S_\text{int}$) and peak flux densities ($S_\text{peak}$) as measured by \texttt{aegean} for a simulated source at varying CLEAN depths. The y-axis shows the recovered fraction of the measurement and the x-axis shows the fraction of the measured emission that is composed of residual, `un-CLEAN` emission. The uncertainties are those reported by \texttt{aegean} from model fitting only. A residual fraction of 1 indicates no CLEANing done.}
\end{figure}

\begin{figure*}[p]
    \centering
    \includegraphics[width=1\linewidth]{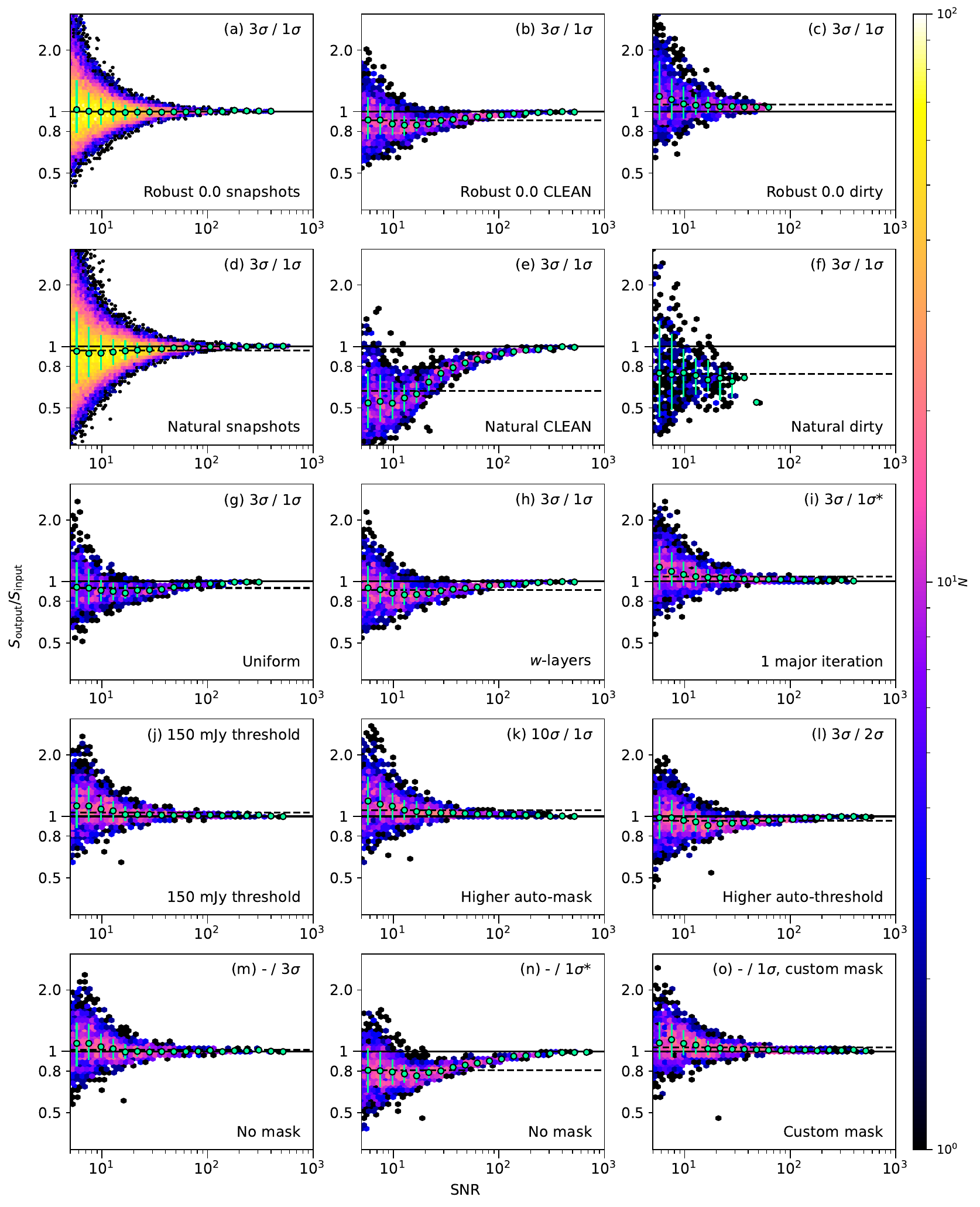}
    \caption{\label{fig:app:sim} Recovered flux density (as $S_\text{output} / S_\text{input}$) as a function of SNR ($S_\text{output,peak} / \sigma_\text{rms}$) in simulated data for the J1234$-$27 mosaic for a range imaging setups. Green markers indicate medians within SNR bins with associated 16-th and 84-th percentiles. The solid black line indicates a ratio of 1, and the dashed black line indicates the overall median for a given test. CLEAN depth stopping thresholds supplied to \texttt{WSClean} are recorded in the top right of each panel (either a fixed value or as auto-mask / auto-threshold). A `*' next to a threshold label indicates the real stopping threshold is larger than this value due to major or minor iteration limits. Note the difference in sensitivity between snapshots and mosaics is approximately a factor of ten.}
\end{figure*}

\begin{figure}[t]
    \centering
    \begin{subfigure}[b]{1\linewidth}
    \includegraphics[width=1\linewidth]{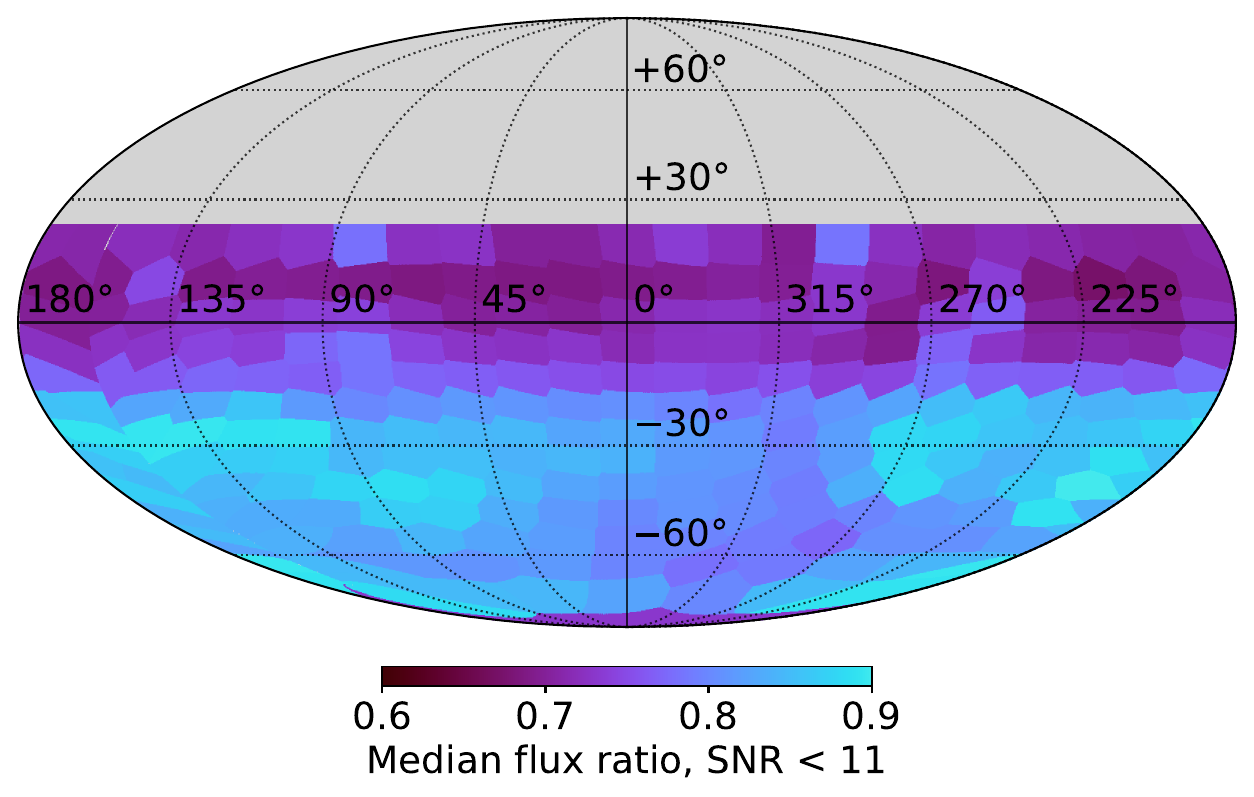}
    \caption{\label{fig:app:sim:pos:11}}
    \end{subfigure}\\%
    \begin{subfigure}[b]{1\linewidth}
    \includegraphics[width=1\linewidth]{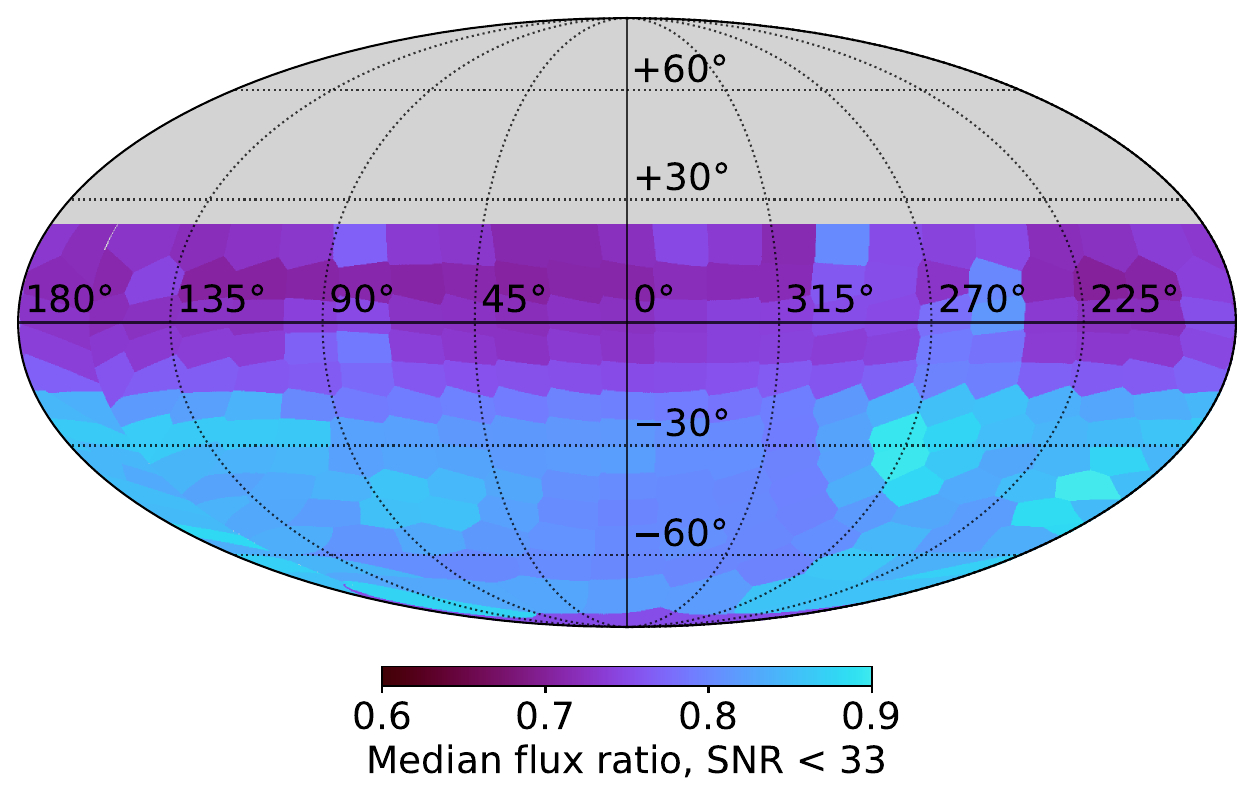}
    \caption{\label{fig:app:sim:pos:33}}
    \end{subfigure}\\%
    \caption{\label{fig:app:sim:pos} Interpolated, HEALPix-binned maps showing the median flux density ratio ($S_\text{output} / S_\text{input}$ for simulated mosaics after imaging with automatic masking and thresholding of 3 and 1, respectively. We show the medians for sources with SNR $<11$ \subref{fig:app:sim:pos:11} and sources with SNR $<33$~\subref{fig:app:sim:pos:33}, corresponding to the two approximate CLEAN depths.}
\end{figure}

\begin{table}[t]
    \centering
    \begin{threeparttable}
    \caption{List of sources subtracted when outside an image FoV.\label{tab:subtraction}}
    \begin{tabular}{l c c c}\toprule
        Source & $\alpha_\text{J2000}$ & $\delta_\text{J2000}$ & $S_\text{300\,MHz}$ \tnote{a}  \\
        & (hh:mm:ss) & (dd:mm:ss.s) & (Jy) \\\midrule
3C~123 & 04:37:04 & +29:40:13.8 & 140 \\
3C~161 & 06:27:10 & -05:53:05.0 & 59 \\
3C~353 & 17:20:28 & -00:58:47.0 & 200 \\
3C~409 & 20:14:28 & +23:34:53.0 & 69 \\
3C~444 & 22:14:32 & -17:01:39.0 & 49 \\
Cassiopeia~A & 23:23:28 & +58:48:42.0 & 9900 \\
Centaurus~A & 13:25:14 & -43:12:59.2 & 3200 \\
Centaurus~B & 13:46:49 & -60:24:12.6 & 110 \\
Cygnus~A & 19:59:24 & +40:43:52.1 & 3700 \\
Fornax~A & 03:22:45 & -37:12:43.9 & 460 \\
Hercules~A & 16:51:17 & +04:59:23.5 & 200 \\
Hydra~A & 09:18:05 & -12:03:45.8 & 160 \\
Pictor~A & 05:19:50 & -45:45:50.9 & 200 \\
Taurus~A & 05:34:33 & +22:01:37.8 & 1000 \\
Virgo~A & 12:30:48 & +12:22:01.9 & 280 \\
NGC~253 & 00:47:37 & -25:17:09.4 & 13 \\
PKS~0408$-$65 & 04:08:20 & -65:45:09.1 & 57 \\
PKS~0410$-$75 & 04:08:48 & -75:07:19.3 & 31 \\
PKS~0420$-$62 & 04:20:56 & -62:23:33.5 & 17 \\
PKS~0442$-$28 & 04:44:37 & -28:09:47.2 & 26 \\
PKS~0521$-$36 & 05:22:58 & -36:27:30.8 & 47 \\
PKS~1610$-$60 & 16:15:05 & -60:56:13.9 & 100 \\
PKS~1932$-$46 & 19:35:56 & -46:20:38.7 & 52 \\
PKS~2153$-$69 & 21:57:06 & -69:41:24.0 & 88 \\
PKS~2331$-$41 & 23:34:26 & -41:25:24.0 & 20 \\
PKS~2356$-$61 & 23:59:03 & -60:55:11.0 & 62 \\
PMN~J1607$-$6331 & 16:06:39 & -63:29:56.2 & 10 \\
\corrs{Galactic Centre} & 17:45:43 & -28:58:04.5 & 390 \\\bottomrule
    \end{tabular}
    \begin{tablenotes}
    {
    \footnotesize \item[a] Expected flux density at 300\,MHz based on the input models.
    }
    \end{tablenotes}
    \end{threeparttable}
\end{table}

Snapshot images are cleaned to a depth of, at best, $1\sigma_\text{rms}$, where $\sigma_\text{rms}$ is a position-independent measure. The typical $\sigma_\text{rms}$ in the 2-min snapshots is $\approx 68$\,mJy\,beam$^{-1}$ (Table~\ref{tab:strips}). The final mosaic images approach a median $\sigma_\text{rms} \approx 9$\,mJy\,beam$^{-1}$ (see Section~\ref{sec:noise}). This results in residual `un-CLEANed' emission in the mosaic images that becomes significant and does not have the same PSF as the restored images. When there is a significant difference between the area of the fitted 2-D Gaussian restoring beam and the original PSF, the measured flux densities of low-SNR sources can be incorrect \citep[e.g.][]{Jorsater1995,Ott2001,Duchesne2021c,Radcliffe2024}. 

To check this effect on our data, we perform a test on a simulated arbitrary brightness point source injected into an example snapshot dataset with typical $(u,v)$ coverage, and perform CLEANing to various depths (including no CLEANing at all, but assuming a 2-D Gaussian fitted beam). We measure the final integrated and peak flux densities using \texttt{aegean}, with resulting measurements shown in Figure~\ref{fig:clean}. We see variation in the measured integrated and peak flux densities of up to $\approx 2$\%, though within the fitting uncertainties reported by \texttt{aegean} and so are not likely to be a significant effect for these data.  

\subsection{The effect of \emph{deep} CLEANing?}

\citet{Duchesne2020} describes a bias observed in stacked MWA images that reduces the flux density of sources as a function of signal-to-noise ratio (SNR). This effect is only noticeable in the final stacked images, and presents similar to the CLEAN bias described by \citet{Becker1995} \citep[see also][]{White1997,ccg+98}, and \citet{Duchesne2020} assumed it was the same effect. CLEAN bias results in a reduction of peak flux density by a constant amount for all sources, which therefore affects fainter sources more severely. To assess any technical issues arising from CLEAN, imaging, or stacking (and to avoid any inherent biases within external catalogues such as GLEAM), we created simulated snapshot images and mosaics using a simple sky model based on the real 300 MHz catalogue. 

\subsubsection{Simulating snapshot images} 

We use \texttt{mwa\_hyperdrive} \footnote{\url{https://github.com/MWATelescope/mwa_hyperdrive}.} \citep{Jordan2025} to simulate the visibilities, which makes use of the snapshot metadata file to simulate the visibilities as they were observed. The snapshots are simulated at their original time resolution (0.5\,s) and then averaged to 4\,s after injecting the local sky model. 

For the sky model we use our initial 300-MHz catalogue. We remove sources if they have neighbours within 240~arcsec (corresponding to approximately the largest PSF in the catalogue), and for the purpose of the simulation assume all sources are point sources---this is to aid in cross-matching to the model catalogue once the images are made and to avoid other issues relating to source-finding/characterisation of extended sources. We construct a local apparent brightness sky model for each snapshot observation, restricted to sources within the output image bounds. The source brightness is attenuated across four specific frequencies within the band that are modelled by the FEE primary beam model. 

We also inject noise into the visibilities, which is assumed to be Gaussian and to have a standard deviation of 600\,Jy for the real and imaginary parts of the complex visibility. This generates $\approx 50$\,mJy\,beam$^{-1}$ rms noise in the output snapshot images which approximately corresponds to the  median rms noise we see in the real snapshots. We note this does not include flagging, real calibration errors, baseline/antenna-dependent noise, the effects of the ionosphere, or other real noise sources in the signal chain. 

The snapshots are then imaged using our original imaging parameters. Namely, we use \texttt{WSClean} with a auto-masking threshold of $3\sigma_\text{rms}$ and an auto-threshold of $1\sigma_\text{rms}$. The images are then primary-beam corrected and we use \texttt{aegean} for source-finding on each snapshot with a detection threshold of $5\sigma_\text{rms}$. Simulated (and real) snapshots usually have around 200 sources detected above $5\sigma_\text{rms}$, but approximately 4\,000--5\,000 are injected into each visibility dataset.

\subsubsection{Testing the snapshots and mosaics}

For the initial set of tests, we focus on a mosaic region centered on $(\alpha_\text{J2000}, \delta_\text{J2000}) = (12^\text{h}34^\text{m}, -27^\text{d}00^\text{m})$, using 343 snapshots taken near zenith.  We create the simulated mosaic using the simulated snapshot images following the process used for the real data. We run source-finding on each simulated snapshot and the resultant mosaic and cross-match to our input model and look at the ratio of the measured flux density in the simulated mosaic ($S_\text{output}$) compared to the input model ($S_\text{input}$). We also repeat the mosaicking and source-finding on the images \textit{without} deconvolution (i.e.\ the `dirty' images), assuming the same restoring beam in the final images \footnote{As noted in Section~\ref{app:shallow_clean}, we do not expect the integrated volume of the synthesised beam to differ much from the fitted 2-D Gaussian beam in the restored/CLEANed images for robust 0.0 image weighting, though for different image weightings with different PSFs this may be the case. In particular, the naturally weighted data shown in Figure~\ref{fig:app:sim} features an additional offset in the flux density recovery for sources that are not deconvolved and may be a result of PSF area differences.}. 

Figure~\ref{fig:app:sim}(a)--(c) shows the results of the simulated dataset with the original imaging parameters, with the flux density ratios ($S_\text{output} / S_\text{input}$) as a function of SNR for the snapshot source lists [Figure~\ref{fig:app:sim}(a)], the simulated mosaic [Figure~\ref{fig:app:sim}(b)], and the simulated dirty mosaic [Figure~\ref{fig:app:sim}(c)]. We see that the snapshot source lists (for sources above $5\sigma_\text{rms}$) do not immediately show any discrepancy. The simulated mosaic shows the bias, and the simulated dirty mosaic conversely shows the opposite with $S_\text{output} / S_\text{input} > 1$, but with larger scatter. We apply a correction for Eddington bias, so we do not expect an increase in $S_\text{output}$ for low-SNR sources. We note there is no significant difference between integrated and peak flux densities in the simulated data. For the first epoch VCSS data, \citet{vcss-memo} found a difference in the integrated and peak flux density measurements as a function of SNR that they note could be a result of blurring from astrometric offsets between images when making mosaics. That is not likely a significant effect here. 

As well as the normal imaging and mosaicking setup, to aid in understanding where the bias arises we perform a series of similar simulations with some modification to imaging or mosaicking parameters. These test are summarised as follows: \begin{enumerate}
    \item \emph{Primary beam as weight.} We normally use a position-dependent rms noise map for mosaic weighting. We construct a mosaic using primary beam weights instead, though this produces the same results as expected as the smoothly varying noise maps largely trace the primary beam attenuation pattern.
    \item \emph{Natural and uniform image weighting.} We repeat the imaging and mosaicking process for all snapshots natural weighting [Figure~\ref{fig:app:sim}(d)--(f)] and with uniform weighting [panel (g)]. For natural weighting, the effect becomes more significant, and we begin to see the effect in the individual snapshots as well. Natural weighting may also show the effect described in Appendix~\ref{app:shallow_clean}.  Conversely, uniform weighting reduces the effect somewhat.
    \item \emph{$w$-snapshot gridding.} We test the original widefield gridder implemented in \texttt{WSClean}. The result for the CLEAN mosaic is shown in Figure~\ref{fig:app:sim}(h), and is identical to the $w$-gridder case.  
    \item \emph{One major iteration during deconvolution.} We restrict CLEAN to one major iteration, which in practice CLEANs to $\approx 3\sigma_\text{rms}$. In this case bright sources are well-recovered, with fainter sources tracing the dirty mosaic (due to lack of deconvolution). While snapshots look the same with less CLEANing, the mosaic image quality is noticeably worse with obvious residual sidelobes around sources. The result is shown in Figure~\ref{fig:app:sim}(i).
    \item \emph{Fixed 150-mJy threshold.} We use a fixed 150-mJy threshold ($\approx 3\sigma_\text{rms}$) which returns the same result as using one major iteration as they CLEAN to similar depths. The result is shown on Figure~\ref{fig:app:sim}(j).
    \item \emph{Varying auto-mask and auto-threshold levels.} We test $10\sigma_\text{rms}$ / $1\sigma_\text{rms}$ [Figure~\ref{fig:app:sim}(k)] and  $3\sigma_\text{rms}$ / $2\sigma_\text{rms}$ [Figure~\ref{fig:app:sim}(l)] auto-mask / auto-threshold levels. The higher auto-mask level again returns the same result as the fixed threshold and single major iteration. $3\sigma_\text{rms}$ / $2\sigma_\text{rms}$ begins to show the bias. 
    \item \emph{No CLEAN masking.} We test $3\sigma_\text{rms}$ [Figure~\ref{fig:app:sim}(m)] and $1\sigma_\text{rms}$ [Figure~\ref{fig:app:sim}(n)] automatic thresholds without masks. $3\sigma_\text{rms}$ is effectively repeating previous tests with almost identical results. In the $1\sigma_\text{rms}$ case, the effective CLEAN depth is only $\approx 2\sigma_\text{rms}$ as the minor iteration limit (250\,000) is reached first. The $1\sigma_\text{rms}$ threshold yields the worst results, highlighting the importance of the mask when CLEANing into the noise.
    \item \emph{Custom mask around known sources.} Finally, we construct a custom CLEAN mask using $3\times 3$ pixel boxes placed at the location of simulated sources and CLEAN to an automatic $1\sigma_\text{rms}$ threshold. The result is shown on Figure~\ref{fig:app:sim}(o)---this removes the bias, suggesting the normal auto-masking process is not adequate for our use-case and in combination with other imaging parameters. 
\end{enumerate}

To assess the effect over the whole survey, we repeated the initial simulation process for all of the original mosaics, including $\approx 14\,000$ snapshots. Figure~\ref{fig:app:sim:pos} shows the median ratio of $S_\text{output}/S_\text{input}$ after interpolation and HEALPix binning for the 242 mosaic regions. We show this for sources with SNR $<11$ [Figure~\ref{fig:app:sim:pos:11}] and SNR $<33$ [Figure~\ref{fig:app:sim:pos:33}], separately, corresponding to the two CLEAN thresholds used by \texttt{WSClean}. In general, there is further reduction of flux density recovery beyond SNR $=33$, but there is some flattening of the reduction beyond this initial CLEAN masking threshold. We note also clear variation as a function of declination, with $\delta_\text{J2000} > -26.7\degr$ regions showing the largest reduction in recovered flux density, generally. Some variation as a function of position can be attributed to the difference in the numbers of sources for each mosaic region, as we only simulate sources that are reported in the final catalogue. While the PSF tends to be larger at high declination and for the SCP region, the flux density does not appear to reduce as a function of the overall Gaussian PSF size. Similarly, source density does not appear to correlate with flux density bias in the simulated mosaics.

The result of the tests highlights the importance of a carefully constructed mask when CLEANing below $3\sigma_\text{rms}$ (i.e.\,into the noise). We conclude that the auto-masking with our other imaging parameters (including iteration limits and rms calculations for thresholding) are resulting in over-CLEANing. While our initial attempt at resolving this issue was to derive corrections based on these simulated data, because the effect varies significantly as a function of position we found the post-imaging corrections to not be as effective as simply re-imaging with modified imaging parameters. To reduce this effect, we suggest either higher CLEAN thresholds or more careful masking. We opted for both, since imaging stages for these data are relatively cheap computationally compared to other processing stages---e.g.\ flagging and off-axis source subtraction take a majority of the processing time and needs to be redone when re-imaging as we do not store calibrated MeasurementSets.

\section{Subtracted sources}\label{app:subtraction}

{As described in Section~\ref{sec:imaging}, we subtract or peel a selection of bright sources when they appear above 2.3\,Jy and 77\,Jy in apparent brightness, respectively, in a given observation when outside of the imaged FoV. As they are only subtracted when outside of the imaged FoV, they remain in the mosaics and are therefore included in the catalogue. The complete collection of subtracted sources is summarised in Table~\ref{tab:subtraction} with model absolute 300-MHz total flux densities. Models are a compilation from those used by general MWA continuum processing \citep[e.g.][]{HurleyWalker2017,Duchesne2020,DuchesnePhD,Franzen2021a}, \corrs{including a new model of a portion of the Galactic Centre which was constructed by running \texttt{PyBDSF} \citep{pybdsf} on initial GLEAM-300 mosaic images to generate 2-D Gaussian components.}

\end{document}